\documentclass[twocolumn,preprint]{aastex631}

\usepackage{hyperref}
\usepackage{xcolor}

\newcommand{\msun}{M_\odot}

\newcommand{\NII}{\hbox{{\rm [N}~{\sc ii}{\rm ]}}}

\newcommand{\Ha}{\hbox{{\rm H}$\alpha$}}

\shorttitle{The $M_{\rm BH}-M_{\star}$ Relationship at $3<z<7$}
\shortauthors{Jones et al.}
\graphicspath{{./}{figures/}}

\defcitealias{Reines_Volonteri_2015}{RV15}

\begin{document}

\title{\large \bf The $M_{\rm BH}-M_{\star}$ Relationship at $3<z<7$: Big Black Holes in Little Red Dots}

\suppressAffiliations

\author[0000-0002-8360-3880]{Brenda L. Jones}
\affiliation{Department of Physics and Astronomy, University of Maine, Orono, ME 04469, USA}

\author[0000-0002-8360-3880]{Dale D. Kocevski}
\affiliation{Department of Physics and Astronomy, Colby College, Waterville, ME 04901, USA}

\author[0000-0001-9879-7780]{Fabio Pacucci}
\affiliation{Center for Astrophysics $\vert$ Harvard \& Smithsonian, Cambridge, MA 02138, USA}
\affiliation{Black Hole Initiative, Harvard University, Cambridge, MA 02138, USA}

\author[0000-0003-1282-7454]{Anthony J. Taylor}
\affiliation{Department of Astronomy, The University of Texas at Austin, Austin, TX, USA}

\author[0000-0001-8519-1130]{Steven L. Finkelstein}
\affiliation{Department of Astronomy, The University of Texas at Austin, 2515 Speedway, Stop C1400, Austin, TX 78712, USA}

\author[0000-0003-0426-6634]{Johannes Buchner}
\affiliation{Max Planck Institute for Extraterrestrial Physics, Giessenbachstrasse, 85741 Garching, Germany}

\author[0000-0002-1410-0470]{Jonathan R. Trump}
\affiliation{Department of Physics, 196 Auditorium Road, Unit 3046, University of Connecticut, Storrs, CT 06269, USA}

\author[0000-0002-6748-6821]{Rachel S. Somerville}
\affiliation{Center for Computational Astrophysics, Flatiron Institute, 162 5th Avenue, New York, NY 10010, USA}

\author[0000-0002-3301-3321]{Michaela Hirschmann}
\affiliation{Institute of Physics, Laboratory of Galaxy Evolution, Ecole Polytechnique Fédérale de Lausanne (EPFL), Observatoire de Sauverny, 1290 Versoix, Switzerland}

\author[0000-0003-3466-035X]{{L. Y. Aaron} {Yung}}
\affiliation{Space Telescope Science Institute, 3700 San Martin Drive, Baltimore, MD 21218, USA}

\author[0000-0002-0786-7307]{Guillermo Barro}
\affiliation{Department of Physics, University of the Pacific, Stockton, CA 90340 USA}

\author[0000-0002-5564-9873]{Eric F.\ Bell}
\affiliation{Department of Astronomy, University of Michigan, 1085 S. University Ave, Ann Arbor, MI 48109-1107, USA}

\author[0000-0003-0492-4924]{Laura Bisigello}
\affiliation{Dipartimento di Fisica e Astronomia "G.Galilei", Universit\'a di Padova, Via Marzolo 8, I-35131 Padova, Italy}
\affiliation{INAF--Osservatorio Astronomico di Padova, Vicolo dell'Osservatorio 5, I-35122, Padova, Italy}

\author[0000-0003-2536-1614]{Antonello Calabr{\`o}}
\affiliation{Osservatorio Astronomico di Roma, via Frascati 33, Monte Porzio Catone, Italy}

\author[0000-0001-7151-009X]{Nikko J. Cleri}
\affiliation{Department of Astronomy and Astrophysics, The Pennsylvania State University, University Park, PA 16802, USA}
\affiliation{Institute for Computational \& Data Sciences, The Pennsylvania State University, University Park, PA 16802, USA}
\affiliation{Institute for Gravitation and the Cosmos, The Pennsylvania State University, University Park, PA 16802, USA}

\author[0000-0003-4174-0374]{Avishai Dekel}
\affiliation{Racah Institute of Physics, The Hebrew University, Jerusalem 91904, Israel}
\affiliation{SCIPP, University of California, Santa Cruz, CA 95064, USA}

\author[0000-0001-5414-5131]{Mark Dickinson}
\affiliation{NSF's National Optical-Infrared Astronomy Research Laboratory, 950 N. Cherry Ave., Tucson, AZ 85719, USA}

\author[0000-0003-3248-5666]{Giovanni Gandolfi}
\affiliation{Dipartimento di Fisica e Astronomia "G. Galilei", Università di Padova, Vicolo dell'Osservatorio 3, 35122 Padova, Italy}
\affiliation{INAF, Osservatorio Astronomico di Padova, Vicolo dell’Osservatorio 5, 35122 Padova, Italy}

\author[0000-0002-7831-8751]{{Mauro} {Giavalisco}}
\affiliation{University of Massachusetts Amherst, 710 North Pleasant Street, Amherst, MA 01003-9305, USA}

\author[0000-0001-9440-8872]{Norman A. Grogin}
\affiliation{Space Telescope Science Institute, Baltimore, MD, USA}

\author[0000-0001-9840-4959]{Kohei Inayoshi}
\affiliation{Kavli Institute for Astronomy and Astrophysics, Peking University, Beijing 100871, China}

\author[0000-0001-9187-3605]{Jeyhan S. Kartaltepe}
\affiliation{Laboratory for Multiwavelength Astrophysics, School of Physics and Astronomy, Rochester Institute of Technology, 84 Lomb Memorial Drive, Rochester, NY 14623, USA}

\author[0000-0002-6610-2048]{Anton M. Koekemoer}
\affiliation{Space Telescope Science Institute, 3700 San Martin Dr., Baltimore, MD 21218, USA}

\author[0000-0002-8951-4408]{Lorenzo Napolitano}
\affiliation{INAF – Osservatorio Astronomico di Roma, via Frascati 33, 00078, Monteporzio Catone, Italy}
\affiliation{Dipartimento di Fisica, Università di Roma Sapienza, Città Universitaria di Roma - Sapienza, Piazzale Aldo Moro, 2, 00185, Roma, Italy}

\author[0000-0003-2984-6803]{Masafusa Onoue}
\affiliation{Kavli Institute for the Physics and Mathematics of the Universe (Kavli IPMU, WPI), The University of Tokyo, Chiba 277-8583, Japan}

\author[0000-0002-5269-6527]{Swara Ravindranath}
\affiliation{Astrophysics Science Division, NASA Goddard Space Flight Center, 8800 Greenbelt Road, Greenbelt, MD 20771, USA}
\affiliation{Center for Research and Exploration in Space Science and Technology II, Department of Physics, Catholic University of America, 620}

\author[0000-0002-9415-2296]{Giulia Rodighiero}
\affiliation{Dipartimento di Fisica e Astronomia "G. Galilei", Universit\`a di Padova, Vicolo dell'Osservatorio 3, 35122 Padova, Italy}
\affiliation{INAF, Osservatorio Astronomico di Padova, Vicolo dell’Osservatorio $5$, 35122, Padova, Italy}

\author[0000-0003-3903-6935]{Stephen M.~Wilkins} %
\affiliation{Astronomy Centre, University of Sussex, Falmer, Brighton BN1 9QH, UK}
\affiliation{Institute of Space Sciences and Astronomy, University of Malta, Msida MSD 2080, Malta}

\affiliation{}

\begin{abstract}

JWST has identified a large population of faint, broad-line active galactic nuclei (AGN) in the early universe that are powered by black holes (BHs) that often appear overmassive relative to their host galaxies.  In this study, we examine the relationship between BH mass and galaxy stellar mass at $3<z<7$ using a sample of 70 broad-line AGN identified using NIRSpec/G395M spectroscopy from the CEERS, JADES, and RUBIES surveys.  Roughly half (43\%) of our sample appear heavily reddened and are classified as little red dots (LRDs).  We estimate BH masses ($M_{\rm BH}$) using single-epoch virial techniques, while host stellar masses ($M_{\star}$) are inferred using a combination of two-dimensional surface brightness profile fitting and spectral energy distribution modeling.  We find that a majority of our sources (50/70) have $M_{\rm BH}/M_{\star}$ ratios that are 1-2 dex higher than that observed in AGN locally.  Using a forward-modeling Bayesian framework that accounts for uncertainties, intrinsic scatter, and selection effects, we infer a $M_{\rm BH}$--$M_{\star}$ relationship that is $>3\sigma$ above the relationship measured for local broad-line AGN.  We derive an intrinsic scatter in this relationship of $0.9$ dex, which does not vary over the redshift range of our sample.  We also find that the $M_{\rm BH}/M_{\star}$ ratio increases by $2.3$ dex from $z = 3.5$ and $z = 6.5$ with a confidence level of $ > 3\sigma$.  We attribute this trend with the increasing fraction of LRDs in our sample at $z>4$ as their host masses are $\sim1$ dex lower than the non-LRD AGN in our sample.  These results support a picture in which the BHs powering JWST's broad-line AGN are  genuinely overmassive and become increasingly so with redshift. We discuss the implications of our findings on early BH growth relative to that of their host galaxies and the constraints it places on BH seeding models.

\end{abstract}


\section{Introduction}

Studies over the past three decades have found that supermassive black holes (SMBHs) are universally found at the centers of massive galaxies and that their growth is connected to that of their hosts.  The latter is evidenced by scaling relations between the mass of the central black hole (BH) and a variety of galaxy properties, such as their velocity dispersion and bulge mass \citep{Kormendy1995,Magorrian98, Gebhardt00, Ferrarese00, McConnell13, Kormendy_Ho_2013,Sun15}, as well as similarities between the cosmic star formation rate density and the BH accretion rate density with redshift \citep{Boyle_Terlevich98, Heckman04, Silverman09, Madau_Dickinson14, Aird15}.  How and when the scaling relationships observed in the local universe are put in place and whether they evolve with redshift are still very much open questions.

In recent years, JWST has provided an extraordinary view of BHs in the early universe by pushing the BH horizon to higher redshifts ($z=9-11$; \citealp{Maiolino24_gnz11,Bogdan24,Napolitano25,Taylor25b}) and lower masses ($10^6 < M_\odot < 10^8$; \citealp{onoue23,Kocevski23b, Harikane23, Matthee_2023, Maiolino23}) than previously possible.  These lower-mass BHs, in particular, are more representative of the underlying BH population than bright quasars and are potentially the key to 
constraining models of BH seeding \citep{Pacucci_2017, Pacucci_2019, Li_Inayoshi23, Fragione_2023} and the early coevolution of galaxies and BHs \citep{Yung2021, Habouzit22, Inayoshi_2022,Pacucci_2023_overmassive}.

Roughly a third of the broad-line AGN identified with JWST appear heavily reddened in the rest-frame optical, yet have relatively blue colors in the rest-frame ultraviolet (UV) \citep{Kocevski23b, Harikane23, Furtak24, Matthee_2023,Greene_2023,Killi23, Hviding25}.  Sources with this ``v-shaped”, red plus blue spectral energy distribution (SED) have come to be known as “little red dots” (LRDs) in the literature \citep{Matthee_2023}.  LRDs are now thought to be a previously-unknown population of gas-enshrouded AGN \citep{Inayoshi_Maiolino25, Naidu25, Taylor25b} that are ubiquitous at high redshifts, but whose number density declines rapidly below $z\sim4$ \citep{Kocevski25, Ma25, Zhuang_2025, Bisigello25}.  

A key property of the broad-line AGN identified with JWST at high redshifts, including the LRDs, is an elevated BH to galaxy stellar mass ratio.  Several studies have reported BH masses that are 1-10\% of their host masses \citep{Kocevski23b, Harikane23, Ubler23, Furtak24, Maiolino23, Kokorev23, Yue24, Kocevski25, Juodžbalis_25}, which is higher than the 0.1\% commonly found in local AGN (see, e.g., \citealt{Reines_Volonteri_2015}).
In some exceptional cases, the BH mass is estimated to be even higher than the galaxy stellar mass; for instance, \citealt{Juodzbalis_2025} reports the detection of a LRD at $z \approx 7$ with $M_{\rm BH}/M_\star  > 2$.

There is significant debate in the literature as to whether these findings signal an evolution of the $M_{\rm BH}-M_{\star}$ relationship towards overmassive BHs at higher redshifts \citep{Pacucci_2023_overmassive, Pacucci2024, Durodola_2025} or if these sources are simply the massive, biased tail of a larger population of AGN that lie below the detection threshold of JWST \citep[][Brooks et al., in preparation]{Sun2024, LiSilverman2025, Silverman25}.

Several models have pointed out that the $M_{\rm BH}-M_{\star}$ relationship should evolve with redshift based on gas availability or self-regulation arguments \citep{Wyithe_Loeb03, Caplar18, Yang18, Inayoshi24, Pacucci_Loeb_2025}; however, observational studies out to $z\sim3$ have produced mixed results \citep[e.g.,][]{Peng06, Jahnke09, Decarli10, Suh20, Tanaka25}.  Furthermore, there is no consensus from cosmological simulations as to whether the $M_{\rm BH}/M_{\star}$ ratio is higher or lower at higher redshifts compared to the local Universe, however most fail to reproduce the overmassive BH population observed with JWST \citep{Habouzit22}.   There is also the possibility that the scatter of the high-redshift relationship is higher than that observed locally due to the effects of merger-averaging, which reduces the intrinsic scatter at lower redshifts \citep{Peng07, Hirschmann11, Jahnke11}.

In this paper, we make use of the largest sample of JWST-identified broad-line AGN yet compiled in the redshift range $3<z<7$ to statistically investigate the nature of the $M_{\rm BH}-M_{\star}$ relationship in the early universe and its potential evolution with redshift.  We make use of a forward-modeling Bayesian framework that accounts for selection biases to infer the intrinsic $M_{\rm BH}-M_{\star}$ relationship at these redshifts \citep{Pacucci_2023_overmassive}.  

We present our analysis as follows.  In Section \ref{sec:obs_data}, we describe the {\it JWST} imaging and spectroscopic data used for this study, while Section \ref{sec:sample} describes our sample selection and the properties of the broad-line AGN used in this work.  Section \ref{sec:method} describes the methodology used to measure our BH and galaxy stellar masses, as well as the forward modeling approach used in determining the intrinsic $M_{\rm BH}-M_{\star}$ scaling relationship.  We present our results in Section \ref{sec:result} and discuss the implications of our findings in Section \ref{sec:discussion}.  Throughout this paper we use the following cosmological parameters: $H_{0} = 70~{\rm km~s^{-1}~Mpc^{-1}; \Omega_{tot}, \Omega_{\Lambda}, \Omega_{m} = 1, 0.7, 0.3}$.

\section{Data Description \& Reduction} \label{sec:obs_data} 

\subsection{NIRCam Imaging}

In this study, we make use of {\it JWST} NIRCam imaging from the Cosmic Evolution Early Release Science Survey (CEERS; \citet{Finkelstein25}), the JWST Advanced Deep Extragalactic Survey (JADES; \citet{Eisenstein_JADES_2023}), and the Public Release IMaging for Extragalactic Research survey \citep[PRIMER][]{Dunlop21}. For the JADES survey, we make use of the public NIRCam mosaics made available as part of the JADES second data release\footnote{https://archive.stsci.edu/hlsp/jades}.  
The CEERS and PRIMER NIRCam data were processed using the {\it JWST} Calibration Pipeline\footnote{\url{http://jwst-pipeline.readthedocs.io/en/latest/}} (versions 1.8.5 and 1.10.2, respectively) with custom modifications described in \citet{Finkelstein25} and \citet{bagley22}. The resulting images were registered to the same World Coordinate System reference frame (based on Gaia DR1.2; \citealt{Gaia16}) and combined into a single mosaic for each field using the drizzle algorithm with an inverse variance map weighting \citep{fruchter02,Casertano00} via the Resample step in the pipeline.  The final mosaics in all fields have pixel scales of 0\farcs03/pixel. 

Source detection and photometry on the NIRCam mosaics were computed on PSF-matched images using \texttt{SExtractor} \citep{bertin96} version~2.25.0 in two-image mode, with an inverse-variance weighted combination of the PSF-matched F277W and F356W images as the detection image.  Photometry was measured in all of the available NIRCam bands in each field, as well as the F606W and F814W HST bands using public data from the CANDELS and 3D-HST surveys \citep{grogin11, koekemoer11,brammer12,Momcheva16}.  Two runs of \texttt{SExtractor} were conducted in a hot+cold setup similar to the process described in \citet{Galametz13}.  
Sources are first selected with conservative (cold) detection parameters, designed not to split up large, bright galaxies, followed by a more aggressive (hot) run, which is performed
to identify fainter objects.  Objects from the hot catalog that fall outside the segmentation map from the cold run are added to the final photometry catalog.  Our analysis of the NIRCam imaging data is similar to that described in \citet{Finkelstein24};
hence, we refer the reader there for additional details.

\subsection{NIRSpec Spectroscopy}

The NIRSpec observations used in this study consist of data from the Cycle 1 CEERS survey (JWST ERS \#1345, PI Finkelstein \citealt{Finkelstein_2025}; the Cycle 2 RUBIES survey (JWST GO \#4233, PIs de Graaff and Brammer, \citealt{RUBIES_2025}); and the Cycle 1 JADES survey \citep{Eisenstein_JADES_2023}.  The CEERS data consists of six MSA pointings in the EGS field, while the RUBIES data consists of six MSA pointings in EGS and 12 pointings in the UKIDSS Ultra Deep Survey (UDS) field.  For this study, we only make use of NIRSpec observations taken with the G395M/F290LP $R\simeq 1000$ grating/filter pair.

The CEERS observations used a three-nod dither pattern and the NRSIRS2 readout mode with 14 groups per nod, resulting in a total exposure time of 3107~s.  The RUBIES observations used a three-nod dither pattern with the NRSIRS2RAPID readout mode with 65 groups per nod, resulting in a total exposure time of 2889~s.  The details of the JADES observations, which used multiple configurations, can be found in \cite{Bunker23} and \cite{Maiolino23}.

For the JADES spectroscopic data, we made use of the reduced 1D spectra made available as part of the JADES third data release.  The CEERS and RUBIES data were independently reduced as described in \cite{Taylor25} using version~1.13.4 of the JWST Science Calibration Pipeline with the Calibration Reference Data System (CRDS) mapping 1215, starting from the Level~1 uncalibrated data products (``\_uncal.fits'' files) available on MAST. Custom parameters were used for the \texttt{jump} step at the detector-level calibration for a better treatment of the ``snowballs''\footnote{\url{https://jwst-docs.stsci.edu/data-artifacts-and-features/snowballs-and-shower-artifacts}} produced by high-energy cosmic ray events, and a nodded background subtraction was adopted.

The reduced two-dimensional (2D) spectra (``s2d'') have a rectified trace with a flat slope. To best optimize the extraction of one-dimensional (1D) spectra from the 2D spectra, we perform a weighted extraction based on the methodology of \cite{Horne86}. Briefly, for a given spectrum, we take the median of the 2D spectrum along the spectral direction to produce a spatial profile for the source. We then identify the central peak of this profile, which corresponds to the source's spectral trace. We then set all pixels in the spatial profile that are not a part of this central feature to zero and normalize the area under this masked spatial profile to one. We then use the normalized profile as the variable \textbf{P} in Table~1 of \cite{Horne86} and follow the prescription given therein to extract an optimized 1D spectrum. 

\begin{table*}
\renewcommand\thetable{1} 
\caption{Properties of Our Sample of Broad-line AGN}\vspace{0mm}
\begin{center}
\footnotesize
\begin{tabular}{ccccccccc}
\hline
\hline
ID & RA & Dec & $z_{\rm spec}$ &  F$_{{\rm H}\alpha, {\rm broad}}$  & FWHM$_{{\rm H}\alpha, {\rm broad}}$  & $\log\,(M_{\rm BH}/{\rm \msun})$ & $\log\,(M_{\rm *}/{\rm \msun})$ & LRD  \\
  & (J2000) & (J2000) & & (10$^{-18}$ erg s$^{-1}$ cm$^{-2}$) & (km s$^{-1}$) & &   \\
\hline
\multicolumn{9}{c}{Kocevski et al.~(2023)} \\
\hline
CEERS 746         & 214.809142 & 52.868484 & 5.624  & 4.52${\pm 0.45}$  & 1800${\pm 200}$  & 7.19$^{+0.11}_{-0.13}$  & 7.99${\pm 0.37}$  & 1 \\
CEERS 2782        & 214.823453 & 52.830281 & 5.241  & 5.62${\pm 0.72}$  & 2060${\pm 290}$  & 7.32$^{+0.14}_{-0.16}$  & $<$ 9.53  & 0 \\
\hline
\multicolumn{9}{c}{Harikane et al.~(2023)} \\
\hline
CEERS  397 & 214.836197 & 52.882693 & 6.000 & $4.27\pm 0.39$  & 2091${\pm 158}$  & 7.34$^{+0.08}_{-0.09}$  & $9.21\pm 0.33$  & 0 \\
CEERS  672 & 214.889677 & 52.832977 & 5.666 & 2.76${\pm 0.24}$  & 1504${\pm 193}$  & 6.93$^{+0.13}_{-0.14}$  & 8.33${\pm 0.24}$  & 1 \\
CEERS 1236 & 215.145298 & 52.967279 & 4.484 & 2.38${\pm 0.24}$  & 2959${\pm 0.24}$  & 7.39$^{+0.13}_{-0.14}$  & 9.04${\pm 0.24}$  & 0 \\
\hline
\multicolumn{9}{c}{Maiolino et al.~(2023)} \\
\hline
JADES-GN      954 & 189.15197 & 62.25964  & 6.760 & 1.73${\pm 0.07}$  & 1911${\pm 126}$  & 7.60$^{+0.07}_{-0.07}$  & 8.92${\pm 0.76}$  & 1 \\
JADES-GN     1093 & 189.17974 & 62.22463  & 5.595 & 1.27${\pm 0.30}$  & 1591${\pm 252}$  & 6.81$^{+0.18}_{-0.21}$  & 7.82${\pm 0.32}$  & 1 \\
JADES-GS     8083 &  53.13284 & -27.80186 & 4.648 & 3.35${\pm 0.21}$  & 1750${\pm 130}$  & 7.01$^{+0.08}_{-0.08}$  & 8.50${\pm 0.24}$  & 0 \\
JADES-GN    11836 & 189.22059 & 62.26368  & 4.409 & 4.09${\pm 0.35}$  & 1518${\pm 171}$  & 6.90$^{+0.11}_{-0.13}$  & 8.54${\pm 0.72}$  & 0 \\
JADES-GN    20621 & 189.12252 & 62.29285  & 4.681 & 3.36${\pm 0.40}$  & 1670${\pm 475}$  & 6.97$^{+0.25}_{-0.33}$  & $<$ 9.08  & 0 \\
JADES-GN    61888 & 189.16802 & 62.21701  & 5.875 & 3.31${\pm 0.25}$  &  1320${\pm 133}$ & 6.86$^{+0.10}_{-0.11}$  & 8.24${\pm 0.32}$  & 1 \\
JADES-GN    62309 & 189.24898 & 62.21835  & 5.172 & 1.22${\pm 0.24}$  & 986.8${\pm 144}$  & 6.34$^{+0.16}_{-0.19}$  & 8.34${\pm 0.31}$  & 0 \\
JADES-GN    77652 & 189.29323 & 62.199    & 5.229 & 3.59${\pm 0.47}$  & 922.4${\pm 204}$  & 6.51$^{+0.20}_{-0.25}$  & 8.04${\pm 0.89}$  & 0 \\
\hline 
\multicolumn{9}{c}{Kocevski et al.~(2025)} \\
\hline
RUBIES-EGS 37124    &   214.99098  &   52.916523  &  5.684  & 9.00${\pm 0.37}$  & 1520${\pm 115}$  & 7.18$^{+0.07}_{-0.08}$  & 8.42${\pm 0.37}$  & 1 \\
RUBIES-EGS 42046    &   214.79537  &   52.788848  &  5.280  & 1.97${\pm 0.01}$  & 3300${\pm 60.0}$  & 8.47$^{+0.02}_{-0.02}$  & 8.96${\pm 0.46}$  & 1 \\
RUBIES-EGS 42232    &   214.88680  &   52.855376  &  4.955  & 3.76${\pm 0.06}$  & 1850${\pm 50.0}$  & 7.58$^{+0.03}_{-0.03}$  & 8.71${\pm 0.46}$  & 1 \\
RUBIES-EGS 49140    &   214.89225  &   52.877406  &  6.687  & 4.54${\pm 0.47}$  & 5420${\pm 370}$  & 8.72$^{+0.08}_{-0.09}$  & 9.01${\pm 0.35}$  & 1 \\
RUBIES-EGS 55604    &   214.98304  &   52.956006  &  6.986  & 1.88${\pm 0.11}$  & 4180${\pm 480}$ & 8.55$^{+0.11}_{-0.12}$  & 9.05${\pm 0.52}$  & 1 \\
RUBIES-EGS 60935    &   214.92338  &   52.925588  &  5.288  & 1.93${\pm 0.03}$  & 1670${\pm 60.0}$  & 7.39$^{+0.04}_{-0.04}$  & 8.49${\pm 0.29}$ & 1 \\
RUBIES-EGS 61496    &   214.97245  &   52.962196  &  5.079  & 2.31${\pm 0.17}$  & 1800${\pm 300}$  & 7.00$^{+0.15}_{-0.18}$  & 8.68${\pm 0.46}$  & 1 \\
RUBIES-EGS 926125   &   215.13706  &   52.988557  &  5.286  & 4.48${\pm 0.13}$  & 1690${\pm 70.0}$  & 7.10$^{\pm 0.04}$  & 8.32${\pm 0.36}$  & 1 \\
RUBIES-EGS 927271   &   215.07826  &   52.948504  &  6.786  & 9.60${\pm 1.9}$  & 1410${\pm 200}$  & 6.74$^{+0.16}_{-0.18}$  & 8.70${\pm 0.36}$  & 1 \\
RUBIES-UDS 40579    &   34.244190  &   -5.245834  &  3.103  & 1.49${\pm 0.08}$  & 2700${\pm 130}$  & 8.29$^{+0.05}_{-0.06}$  & 8.09${\pm 0.51}$  & 1\\
RUBIES-UDS 50716    &   34.313132  &   -5.226765  &  6.17   & 1.83${\pm 0.27}$  & 2280${\pm 220}$  & 7.26$^{+0.10}_{-0.12}$  & 8.16${\pm 0.43}$  & 1 \\
RUBIES-UDS 59971    &   34.260537  &   -5.209120  &  5.365  & 1.12${\pm 0.18}$  & 1540${\pm 280}$  & 6.74$^{+0.18}_{-0.21}$  & 8.61${\pm 0.48}$  & 1 \\
RUBIES-UDS 63166    &   34.312143  &   -5.202546  &  6.518  & 3.08${\pm 0.50}$  & 2200${\pm 400}$  & 7.36$^{+0.18}_{-0.22}$  & $<$ 9.92  & 0 \\
\hline 
\multicolumn{9}{c}{Taylor et al.~(2025)} \\
\hline
RUBIES-EGS 920  &  214.052344  & 52.884268  &  3.616 & 1.13${\pm 0.076}$  & 831.7${\pm 196}$   & 6.21$^{+0.23}_{-0.29}$   & 8.86${\pm 0.19}$  & 0 \\
RUBIES-EGS 6411   &  215.109185  & 52.939770  &  4.880 & 1.15${\pm 0.14}$  & 1118${\pm 132}$  & 6.41$^{+0.12}_{-0.14}$  & 8.45${\pm 0.26}$  & 0 \\  
RUBIES-EGS 13872  &  215.132933  & 52.970705  &  5.261 & 1.18${\pm 0.32}$  & 1510${\pm 165}$  & 6.72$^{+0.14}_{-0.17}$  & 8.14${\pm 0.49}$ & 0 \\
RUBIES-EGS 15825  &  215.079264  & 52.934252  &  3.666 & 64.8${\pm 2.4}$  & 3343${\pm 179}$  & 8.08$^{+0.05}_{-0.06}$   & $<$ 10.4  & 0 \\
RUBIES-EGS 19174  &  214.860840  & 52.784773  &  3.774 & 8.15${\pm 2.0}$  & 672.588${\pm 79.4}$   & 6.23$^{+0.14}_{-0.17}$  & 10.8${\pm 0.73}$  & 0 \\
RUBIES-EGS 28812  &  214.924149  & 52.849050  &  4.223 & 9.36${\pm 0.20}$  & 2191${\pm 63.0}$  & 7.37$^{+0.03}_{-0.03}$   & 9.01${\pm 0.43}$  & 1 \\    
RUBIES-EGS 29489  &  215.022071  & 52.920786  &  4.543 & 24.4${\pm 1.0}$  & 2091${\pm 161}$  & 7.56$^{+0.08}_{-0.08}$  & 8.65${\pm 0.39}$  & 1 \\ 
RUBIES-EGS 34978  &  214.861690  & 52.818438  &  3.772 & 19.8${\pm 1.2}$  & 1357${\pm 77.6}$   & 7.04$^{+0.06}_{-0.06}$  & 9.81${\pm 0.19}$  & 0 \\   
RUBIES-EGS 37032  &  214.849388  & 52.811824  &  3.850 & 11.9${\pm 0.58}$  & 1969${\pm 167}$   & 7.28$^{+0.08}_{-0.09}$  & 8.67${\pm 0.33}$  & 1 \\       
RUBIES-EGS 46985  &  214.805654  & 52.809497  &  4.963 & 6.81${\pm 0.33}$  & 1389${\pm 151}$  & 6.98$^{+0.10}_{-0.11}$  & 9.04${\pm 0.70}$  & 0 \\    
RUBIES-EGS 50052  &  214.823454  & 52.830277  &  5.240 & 14.4${\pm 0.27}$  & 2092${\pm 60.3}$  & 7.52$^{+0.03}_{-0.03}$  & 8.28${\pm 0.36}$  & 0 \\    
RUBIES-EGS 50522  &  214.855980  & 52.854661  &  3.614 & 1.09${\pm 0.03}$  & 1515${\pm 424}$   & 6.53$^{+0.27}_{-0.35}$   & 10.3${\pm 0.12}$  & 0 \\
RUBIES-EGS 50812  &  214.845487  & 52.848281  &  3.519 & 1.13${\pm 0.076}$  & 1850${\pm 266}$   & 7.17$^{+0.13}_{-0.15}$   & 8.22${\pm 0.38}$  & 1 \\
RUBIES-EGS 58237  &  214.850571  & 52.866030  &  3.651 & 7.66${\pm 0.27}$  & 2002${\pm 46.9}$  & 7.65$^{+0.03}_{-0.03}$   & $<$ 10.9  & 0 \\
\hline
\end{tabular}
\normalsize
\end{center}
\label{tab:MBH_results}
\vspace{-0.15in}
\tablecomments{All line widths are reported after correcting for instrument broadening.  The relationship of \citet{reines13} used to derived our BH masses has a scatter of $\sim0.5$ dex, which will dominate over our measurement uncertainties.}
\end{table*}

\begin{table*}
\renewcommand\thetable{1} 
\caption{Properties of Our Sample of Broad-line AGN (continued)}\vspace{0mm}
\begin{center}
\footnotesize
\begin{tabular}{ccccccccc}
\hline
\hline
ID & RA & Dec & $z_{\rm spec}$ &  F$_{{\rm H}\alpha, {\rm broad}}$  & FWHM$_{{\rm H}\alpha, {\rm broad}}$  & $\log\,(M_{\rm BH}/{\rm \msun})$ & $\log\,(M_{\rm *}/{\rm \msun})$ & LRD \\
  & (J2000) & (J2000) & & (10$^{-18}$ erg s$^{-1}$ cm$^{-2}$) & (km s$^{-1}$) & &   \\
\hline
\multicolumn{9}{c}{Taylor et al.~2025} \\
\hline
RUBIES-UDS 5496   &   34.405872  &  -5.312951 &  3.655 & 7.14${\pm 0.98}$  & 1810${\pm 407}$  & 7.08$^{+0.21}_{-0.26}$  & 9.94${\pm 0.31}$  & 0 \\  
RUBIES-UDS 8895   &   34.363041  &  -5.306108 &  3.982 & 19.2${\pm 1.3}$  & 1517${\pm 57.9}$  & 7.16$^{+0.05}_{-0.05}$  & 11.0${\pm 0.08}$  & 0 \\
RUBIES-UDS 10036  &   34.381671  &  -5.303742 &  3.806 & 10.9${\pm 0.54}$  & 1205${\pm 64.5}$  & 6.82$^{+0.06}_{-0.06}$  & $<$ 9.65  & 0 \\    
RUBIES-UDS 11721  &   34.411039  &  -5.300780 &  3.978 & 1.65$^{\pm 0.23}$  & 1689${\pm 110}$  & 6.76$^{+0.08}_{-0.09}$    & 9.82${\pm 0.31}$  & 0 \\  
RUBIES-UDS 18302   &   34.233628  &  -5.283850 &  3.698 & 2.73${\pm 0.31}$  & 1482${\pm 260}$  & 6.71$^{+0.17}_{-0.20}$  & 10.4${\pm 0.76}$  & 0 \\
RUBIES-UDS 19484  &    34.232426 &  -5.280654 &  4.656 & 1.21${\pm 0.16}$  & 1288${\pm 229}$  & 6.53$^{+0.17}_{-0.20}$  & $<$ 9.80  & 0 \\  
RUBIES-UDS 19521  &    34.383672 &  -5.287732 &  5.669 & 2.82${\pm 0.30}$  & 1682${\pm 211}$  & 7.03$^{+0.13}_{-0.14}$  & 8.14${\pm 0.92}$  & 1 \\ 
RUBIES-UDS 21944  &   34.469218  &  -5.283563 &  3.526 & 7.49${\pm 0.79}$  & 1739${\pm 161}$  & 7.03$^{+0.01}_{-0.11}$  & 10.3${\pm 0.23}$  & 0 \\ 
RUBIES-UDS 22304  &   34.399170  &  -5.283007 &  3.906 & 2.73${\pm 7.4}$  & 747.4${\pm 89.2}$  & 6.122$^{+0.15}_{-0.18}$  & 8.48${\pm 0.20}$  & 0 \\ 
RUBIES-UDS 29813  &    34.453355 &  -5.270717 &  5.440 & 8.15${\pm 0.36}$  & 2329${\pm 168}$  & 7.52$^{+0.07}_{-0.08}$  & 8.29${\pm 0.35}$  & 1 \\ 
RUBIES-UDS 30969  &   34.296356  &  -5.268672 &  4.000 & 11.8${\pm 1.2}$  & 1018${\pm 44.5}$  & 6.71$^{+0.06}_{-0.06}$  & 9.73${\pm 0.23}$  & 0 \\
RUBIES-UDS 35974  &    34.331644 &  -5.260593 &  4.367 & 5.76${\pm 0.68}$  & 1101${\pm 141}$  & 6.67$^{+0.13}_{-0.15}$  & 10.1${\pm 0.16}$  & 0 \\ 
RUBIES-UDS 46885  &   34.291666  &  -5.233788 &  3.730 & 7.83${\pm 0.51}$  & 781.0${\pm 26.9}$  & 6.35$^{+0.21}_{-0.26}$  & 7.97${\pm 0.25}$  & 0 \\ 
RUBIES-UDS 48507 &    34.284578 &  -5.230702 &  4.468 & 4.69${\pm 0.38}$  & 765.5${\pm 37.3}$  & 6.32$^{+0.06}_{-0.06}$  & 7.87${\pm 0.52}$  &  0 \\ 
RUBIES-UDS 61627  &   34.238394  &  -5.205775 &  3.654 & 3.06${\pm 0.57}$  & 1586${\pm 114}$  & 6.79$^{+0.10}_{-0.11}$  & 10.5${\pm 0.41}$  & 0 \\ 
RUBIES-UDS 63139  &    34.230848 &  -5.202607 &  4.435 & 2.06${\pm 0.30}$  & 1240${\pm 105}$  & 6.58$^{+0.10}_{-0.11}$  & 8.00${\pm 0.62}$  & 0 \\ 
RUBIES-UDS 119957  &    34.268908 &  -5.176722 &  4.149 & 8.61${\pm 0.28}$  & 1908${\pm 94.0}$  & 7.22$^{+0.05}_{-0.05}$  & 8.07${\pm 0.45}$  & 1 \\
RUBIES-UDS 139709  &    34.296002 &  -5.149895 &  5.685 & 36.7${\pm 1.5}$  & 2366${\pm 138}$  & 7.86$^{+0.06}_{-0.06}$  & 8.65${\pm 0.41}$  & 1 \\
RUBIES-UDS 143683  &    34.316389 &  -5.144678 &  4.226 & 3.90${\pm 0.28}$  & 1351${\pm 203}$  & 6.76$^{+0.14}_{-0.16}$  & 8.74${\pm 0.85}$  & 0 \\
RUBIES-UDS 146995  &    34.331043 &  -5.139963 &  3.732 & 25.2${\pm 1.5}$  & 1403${\pm 87.0}$  & 7.12$^{+0.07}_{-0.07}$  & $<$ 10.8  & 0 \\
RUBIES-UDS 147411  &    34.360718 &  -5.139081 &  3.966 & 4.16${\pm 0.31}$  & 1667${\pm 298}$  & 6.93$^{+0.16}_{-0.19}$  & $<$ 8.78  & 0 \\
RUBIES-UDS 150323  &    34.417822 &  -5.5287732 &  3.618 & 9.20${\pm 0.32}$  & 1804${\pm 109}$  & 7.12$^{+0.06}_{-0.06}$  & 8.94${\pm 0.54}$  & 1 \\
RUBIES-UDS 153207 &   34.493112  &  -5.130999 &  3.597 & 177${\pm 3.6}$  & 1287${\pm 27.8}$  & 7.42$^{+0.02}_{-0.02}$  & 10.2${\pm 0.25}$  & 0 \\ 
RUBIES-UDS 155916  &    34.317031 &  -5.127611 &  4.098 & 2.46${\pm 0.20}$  & 1595${\pm 154}$  & 6.80$^{+0.10}_{-0.11}$  & 10.7${\pm 0.58}$  & 0 \\
RUBIES-UDS 172350  &    34.368951 &  -5.103941 &  5.580 & 32.0${\pm 0.71}$  & 1884${\pm 71.7}$  & 7.63$^{+0.04}_{-0.04}$  & 8.23${\pm 0.47}$  & 1 \\
RUBIES-UDS 174752  &    34.205808 &  -5.100500 &  6.039 & 4.39${\pm 0.25}$  & 1496${\pm 110}$  & 
7.18$^{+0.06}_{-0.06}$  & 8.94${\pm 0.64}$  & 0 \\
RUBIES-UDS 182791  &    34.213813 &  -5.087050 &  4.718 & 31.4${\pm 0.41}$  & 2936${\pm 203}$  & 7.93$^{+0.02}_{-0.02}$  & 9.00${\pm 0.58}$  & 1 \\
RUBIES-UDS 807469 &    34.376139 &  -5.310366 &  6.778 & 4.14${\pm 0.49}$  & 1667${\pm 197}$  & 7.19$^{+0.12}_{-0.14}$  & 8.34${\pm 0.53}$  & 1 \\
RUBIES-UDS 970351  &    34.261900 &  -5.105205 &  5.282 & 5.57${\pm 0.36}$  & 1731${\pm 251}$  & 7.16$^{+0.13}_{-0.15}$  & $<$ 9.95  & 0 \\

\hline
\end{tabular}
\normalsize
\end{center}
\vspace{-0.15in}
\tablecomments{All line widths are reported after correcting for instrument broadening. The relationship of \citet{reines13} used to derived our BH masses has a scatter of $\sim0.5$ dex, which will dominate over our measurement uncertainties.}
\end{table*}

\section{Sample Description}\label{sec:sample}

For this study, we make use of 70 broad-line AGN at $3<z<7$ that were previously identified with NIRSpec by five studies: \cite{Kocevski23b}, \cite{Harikane23}, \cite{Maiolino23}, \cite{Kocevski25}, and \cite{Taylor25}.  The coordinates and redshifts of these AGN are listed in Table \ref{tab:MBH_results}.  Five sources reported in \cite{Kocevski23b} and \cite{Harikane23} were identified using the CEERS dataset, eight sources from \cite{Maiolino23} were found in the JADES dataset, and the remaining 57 sources were found using the RUBIES dataset.  

\begin{figure}[t]
\centering
\includegraphics[width=\linewidth]{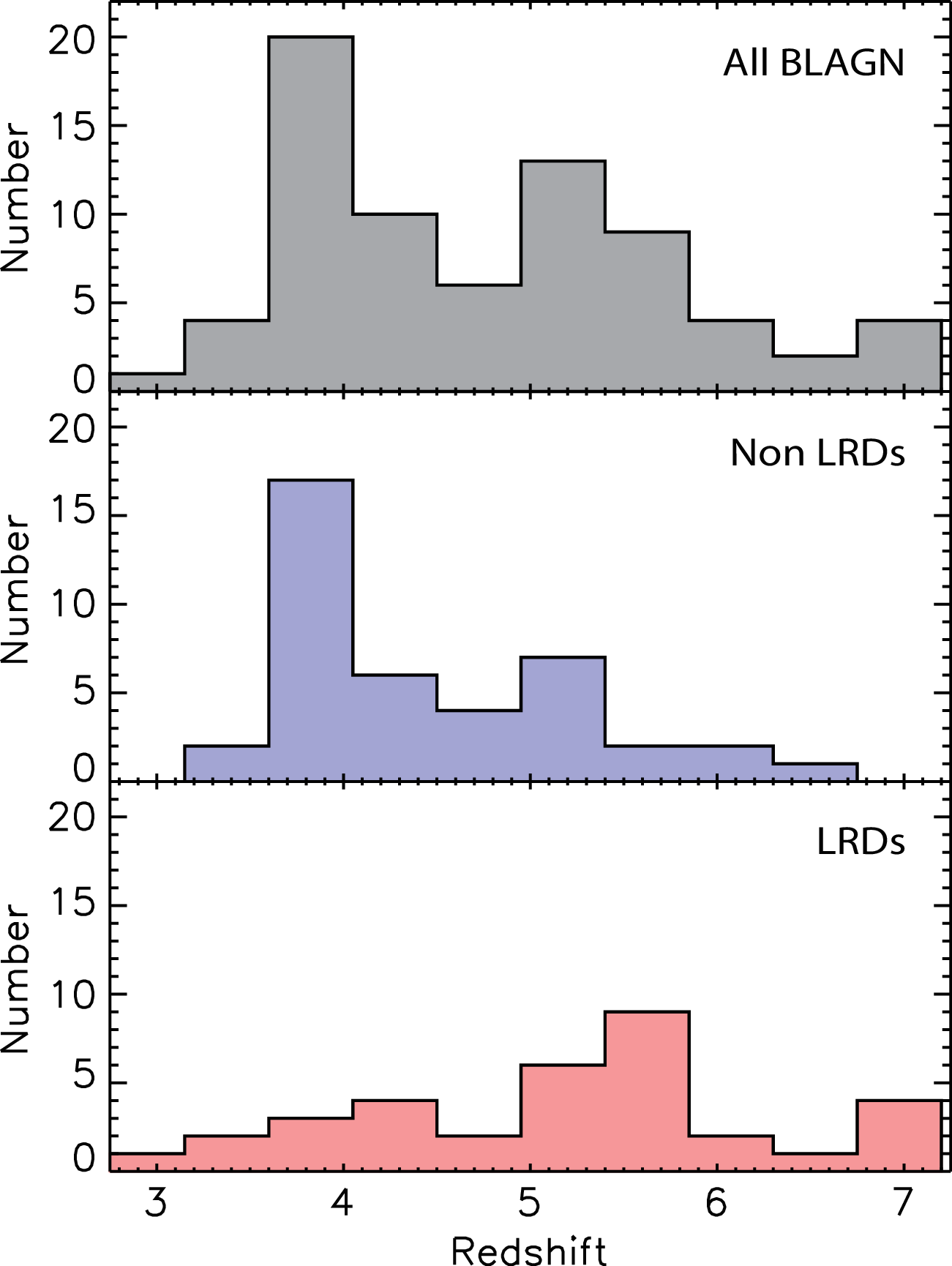}
\caption{The redshift distribution of the broad-line AGN used in this study. The top panel shows the distribution of our entire sample, while the bottom and middle panels show the distribution for sources that are and are not classified as LRDs based on the criteria of \cite{Kocevski25}.  The redshift distribution of the non-LRDs peaks at $z<4$, while that of the LRDs peaks at $z>5$. \label{fig:zhist_plot}}
\end{figure}

The redshift distribution of our broad-line AGN sample is shown in Figure \ref{fig:zhist_plot}. Roughly 43\% (30 sources) of our sample appear heavily reddened and are classified as LRDs, based on the criteria developed in \citet{Kocevski25}; i.e., they have a compact morphology and a ``v-shaped" SED such that they are red in the rest-optical and blue in the rest-UV.  

While both LRD and non-LRD sources are found over the entire redshift range probed by our sample, their redshift distributions skew in different directions.  The LRDs are preferentially found at higher redshifts: 4 are located at $z<4$ compared to 27 at $z>4$.  On the other hand, the redshift distribution of the non-LRDs peaks at lower redshifts: 19 are located at $z<4$ and 20 at $z>4$.  We discuss how this redshift distribution might impact our findings in \S\ref{sec:discussion}.  

\begin{figure}[t]
\centering
\includegraphics[width=\linewidth]{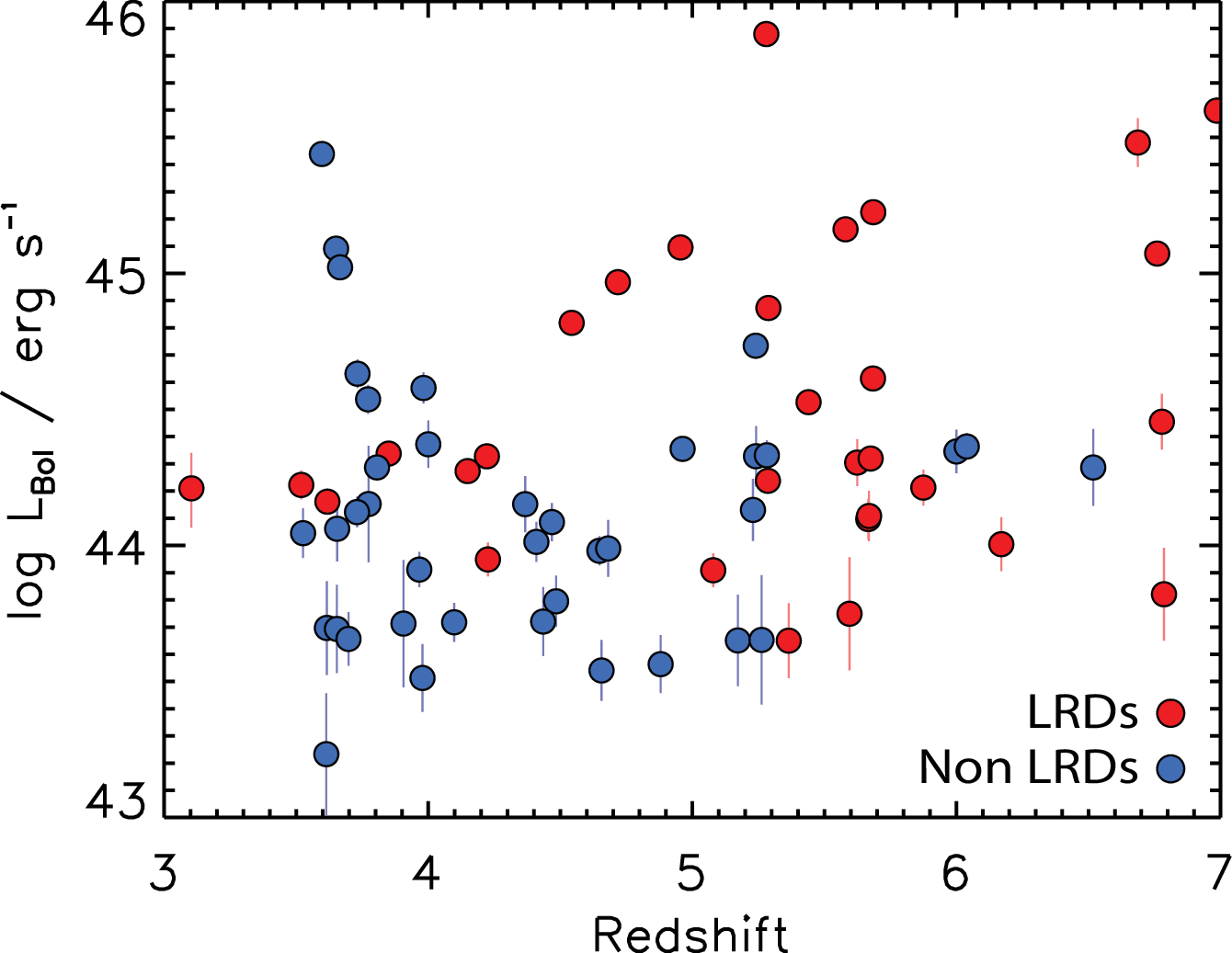}
\caption{Bolometric luminosity versus redshift for the broad-line AGN used in this study. The red (blue) points denote sources that are (are not) classified as LRDs based on the criteria of \cite{Kocevski25}.  \label{fig:lbol_plot}}
\end{figure}

Finally, the bolometric luminosities, $L_{\rm Bol}$, of the sample are shown in Figure \ref{fig:lbol_plot}.  Here we derive $L_{\rm Bol}$ from the broad H$\alpha$ line luminosity ($L_{\rm H\alpha}$) as measured in Section 4.1 using the $L_{\rm H\alpha}$-$L_{\rm Bol}$ relation from \citet{Stern_Laor12}. The bolometric luminosities range from 10$^{43}$ to 10$^{46}$ erg s$^{-1}$, with a median luminosity of $1.7\times10^{44}$ erg s$^{-1}$.  The median luminosity of the LRDs in our sample is $2.1\times10^{44}$ erg s$^{-1}$, while that of the non-LRDs is $1.2\times10^{44}$ erg s$^{-1}$.

\section{Methodology}\label{sec:method}
In this Section, we detail the methodologies used for the measurement of the BH mass (Sec. \ref{sec:BH_mass_measurements}), the morphology of the hosts (Sec. \ref{sec:galfit}), and the measurement of the stellar mass (Sec. \ref{sec:sed_fitting}). We conclude with a summary description of the methodology used to infer the relation between $M_{\rm BH}$ and $M_\star$ as a function of redshift (Sec. \ref{subsec:MCMC}).

\subsection{Black Hole Mass Measurements}\label{sec:BH_mass_measurements}
  
We estimate the virial BH masses of our AGN sample assuming that their broadened emission lines trace the kinematics of gas in the broad-line region (see, e.g., \citealt{Greene_2005}).  We measure line fluxes and widths in the available NIRSpec G395M/F290LP spectra using a Levenberg-Marquardt least-squares method implemented by the \texttt{mpfit} IDL code\footnote{https://pages.physics.wisc.edu/$\sim$craigm/idl/fitting.html} as described in \citet{Kocevski23b}.
We simultaneously fit multiple Gaussians for features in the Balmer and Paschen line regions, depending on the redshift of the source.  To account for broad components, we fit lines with two Gaussians: one narrow with width $<350$~km~s$^{-1}$ and one broad with width $>350$~km~s$^{-1}$.  The line centers, widths, and fluxes are all free parameters and the broad and narrow components can be kinematically offset from each other.  When fitting \Ha\, we also include additional Gaussian components for the \NII~$\lambda\lambda$6550, 6585 doublet. While the \Ha\ and \NII~$\lambda$6585 lines are separated by roughly three times the resolution limit, the lines will blend together in the presence of a sufficiently broad \Ha\ component.  We account for this by constraining the line widths and relative line centers of the \NII\ doublet to that of narrow \Ha.  Four sources in our sample also show blue-shifted absorption in their Balmer and Paschen lines similar to the ones reported in \citet{Matthee_2023} and \citet{Kocevski25}.  To account for this, we add an additional absorption component to our fits of these sources.  We find the absorption features have a full-width at half maximum (FWHM) in the range 250-500 km s$^{-1}$.  

\begin{figure*}[ht]
\centering
\includegraphics[width=\linewidth]{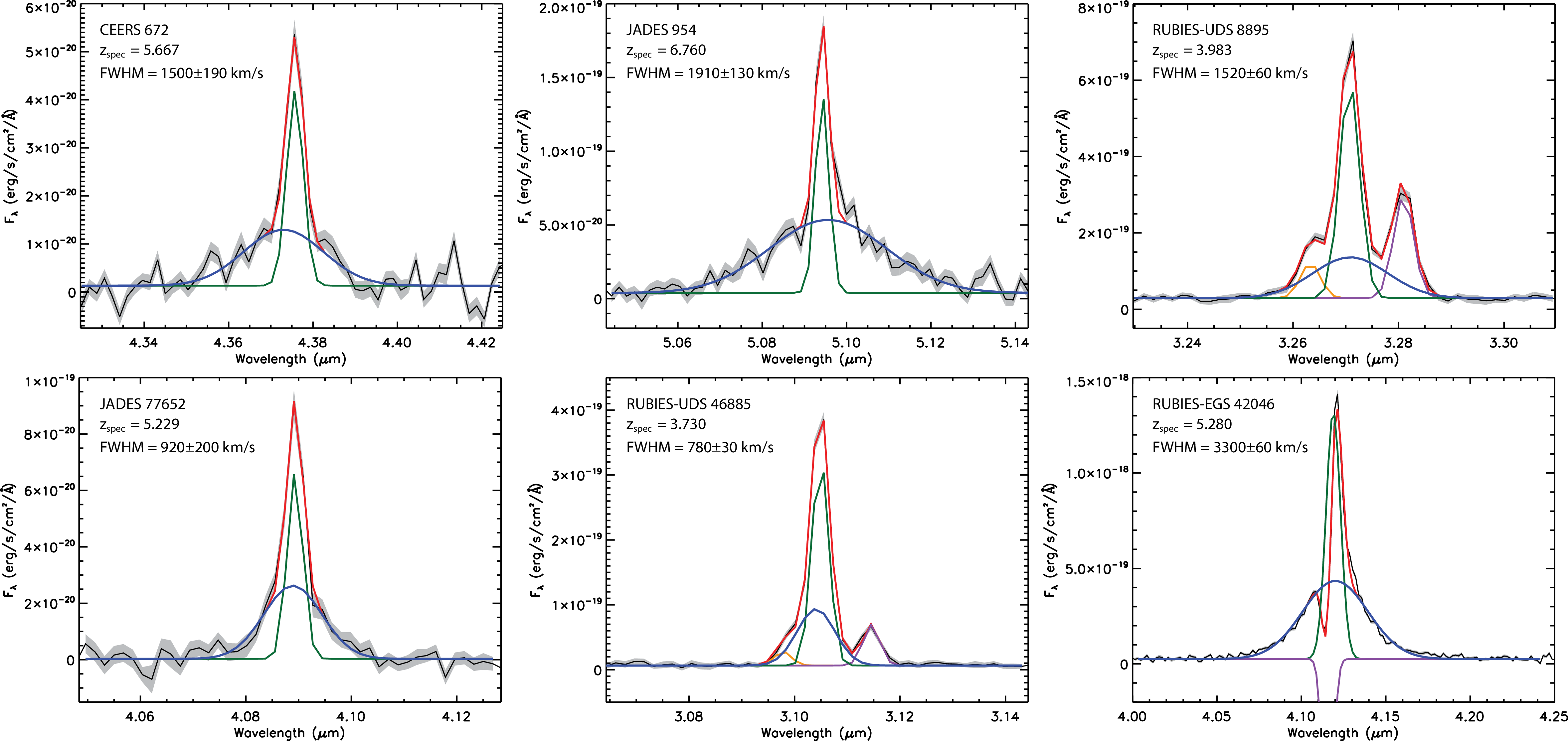}
\caption{Examples of our broad-line fits to the NIRSpec G395M spectra of six sources in our sample from the CEERS, JADES, and RUBIES datasets. Green lines show the best-fit Gaussian for the narrow emission line component, blue lines show the best-fit broad component, purple lines show our best-fit absorption components, and red lines show the best overall (narrow plus broad) fit to the observed emission line (black line and shaded area).  The FWHM of the broad component, corrected for instrument broadening, is shown in the upper left of each panel. \label{fig:line_fits}}
\end{figure*}

Examples of our line fits for sources over the full range of redshifts probed by our sample are shown in Figure \ref{fig:line_fits}.  The broad-line widths and line fluxes that we measure are listed in Table \ref{tab:MBH_results}.  Uncertainties on our line measurements are derived using a Monte Carlo approach.  For each spectral element, we perform 1000 random draws from a Gaussian distribution whose mean is set to the measured flux at that wavelength and whose standard deviation is set to the error in that measurement.  Our line fitting is then repeated on all 1000 mock spectra, and standard deviations are calculated from the resulting distributions.

\begin{figure*}
\centering
\includegraphics[width=4.5in]{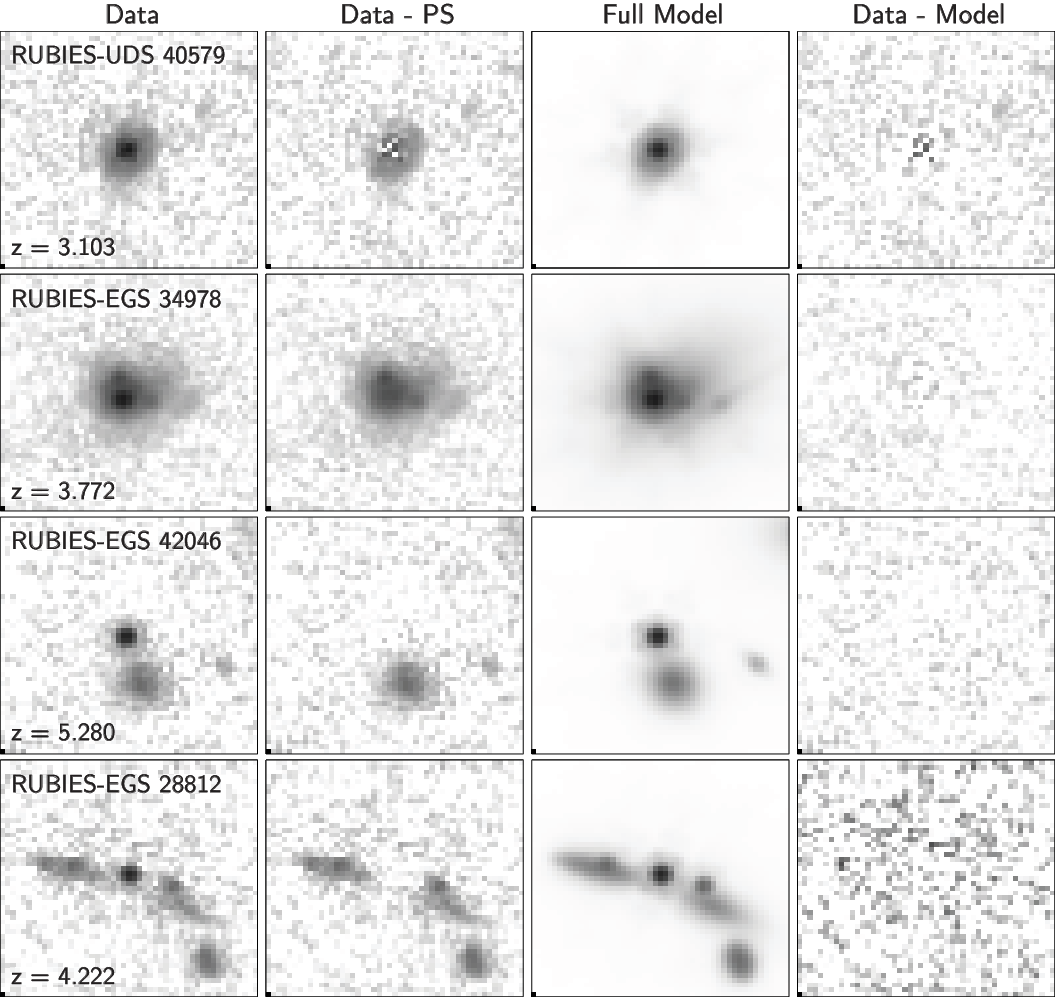}
\caption{Examples of our two-dimensional surface brightness profile fitting. From left to right, the columns show F200W images of several broad-line AGN, residual images after subtracting off our best-fit point source only model, our full, best-fit \texttt{GALFIT} model, and residual images after subtracting our full, best-fit model.  Images are $1.5^{\prime\prime}\times~1.5^{\prime\prime}$ in size. Sources RUBIES-UDS 40579, RUBIES-EGS 42046, and RUBIES-EGS 28812 are LRDs. RUBIES-UDS 40579 and RUBIES-EGS 34978 both have extended morphologies, while RUBIES-EGS 42046 and RUBIES-EGS 28812 are best-fit using a point source only model. \label{fig:galfit_plot}}
\end{figure*}

\begin{table}
\centering
\caption{GRAHSP modules and their parameters.}
\begin{tabular}{lc}
     Parameter & Prior \\
     \hline
     \hline
     Galaxy components: & \\
     \hline
    \texttt{stellar\_mass}  & log-uniform between $10^5$ and $10^{15} M_\odot$ \\ [2mm]
     \textbf{\texttt{[sfhdelayed]}} & \\
     \texttt{tau\_main} & 50, 100, 200, 500, 1000, 3000, \\
     & 5000, 7000, 10000 Myr \\
     \texttt{age} & 50 to 10000  Myr in 20 log-uniform steps \\
     \texttt{sfr\_A} & 1 \\ [2mm]
     \textbf{\texttt{[bc03]}} & \\
     \texttt{imf} & 1 (Chabrier) \\
     \texttt{metallicity} & 0.02 (solar) \\
     \texttt{separation\_age} & 10 Myr \\[ 2mm]
     \textbf{\texttt{[nebular]}} & \\
     \texttt{logU}  & -2.0 \\
     \texttt{f\_esc} & 0.0 \\
     \texttt{f\_dust} & 0.0 \\
     \texttt{lines\_width} & 300 km/s \\ [2mm]
     \textbf{\texttt{[galdale]}} & \\
     \texttt{alpha} &  0.75 to 2.75 in 32 uniform steps \\
     \hline
     \hline
    AGN components & \\
     \hline
     \textbf{\texttt{[activate]}} & \\
     \texttt{L\_AGN} & log-uniform between $10^{38}$ and $10^{50}$ erg/s \\[2mm]
     \textbf{\texttt{[activatelines]}} & \\
     \texttt{AFeII} & 0.6 to 32 in 10 log-uniform steps \\
     \texttt{Alines} & 0.3, 0.5, 0.7, 1, 1.5, 2, 4, 10, 20 \\
     \texttt{linewidth} & 10000 km/s  \\ [2mm]
     \textbf{\texttt{[activategtorus]}} & \\
     \texttt{fcov} & 0.05 to 0.95 in 18 uniform steps \\
     \texttt{Si} & -4 to +4 in 40 uniform steps\\
     \texttt{COOLlam} & 10 to 30 in 20 uniform steps \\
     \texttt{COOLwidth} & 0.2 to 0.65 in 10 uniform steps \\
     \texttt{HOTfcov} & 0.04, 0.1, 0.2, 0.4, 0.6, 0.8, 1.0 \\ 
        & 1.2, 1.4, 1.6, 2.0, 2.5, 3, 5, 10 \\
     \texttt{HOTlam} & 1 to 5.5 in 450 uniform steps of 0.01 \\
     \texttt{HOTwidth} & 0.2 to 0.65 in 10 uniform steps \\[2mm]
     \textbf{\texttt{[activatepl]}} & \\
     \texttt{uvslope} & 0 \\
     \texttt{plslope} & -2.7 to -1 in 170 uniform steps \\
     \texttt{plbendloc} & 50.  80.  90. 100. 120. 150. nm \\
     \texttt{plbendwidth} & 0.1 to 10 in 10 log-uniform steps  \\ [2mm]
     \textbf{\texttt{[biattenuation]}} & \\
     \texttt{E(B-V)}  & 0.01 to 10 in 80 log-uniform steps \\
     \texttt{E(B-V)\_AGN} & 0.01 to 10 in 80 log-uniform steps\\ 
     \hline
     \hline
\end{tabular}
\label{tab:GRAHSP}
\end{table}

To calculate BH masses, we follow the methodology of \citet{Taylor25} and make use of the single-epoch scaling relationship presented in \citet{reines13}:  
 \begin{equation} \label{eq: GH05}
 M_{\rm BH} = 3.7 \times 10^6 \left( \frac{L_{\rm H\alpha}}{10^{42}\ {\rm erg\ s^{-1}}}\right)^{0.47}  \left(\frac{{\rm FWHM_{\rm H\alpha}}}{10^3\ {\rm km\ s^{-1}}} \right)^{2.06} M_\odot.
 \end{equation} 
Here $L_{\rm H\alpha}$ and ${\rm FWHM_{\rm H\alpha}}$ are the luminosity and FWHM of the broad H$\alpha$ line.  For RUBIES-EGS 55604 and RUBIES-UDS 40579, where ${\rm H\alpha}$ is not visible, we make use of the H$\beta$ and P$\delta$ line widths and luminosities, respectively, assuming a H$\beta$-\Ha~line ratio of 2.86 and a P$\delta$-\Ha~line ratio of 46.7, consistent with Case B recombination.  The \citet{reines13} relationship has an uncertainty of roughly 0.5 dex \citep{Reines_Volonteri_2015} which will dominate over our measurement uncertainties.  However, since the \citet{reines13} relationship was derived for AGN at $z<0.06$, these uncertainties may increase further by applying it to higher redshift sources.

We note that, in calculating our BH masses, we do not correct our line luminosities for dust attenuation.  There has been considerable debate in the literature as to whether LRDs are as dust-obscured as their red colors suggest.  Instead, it has been proposed that they are enshrouded in dense gas clouds \citep{Inayoshi_Maiolino25, Naidu25, Taylor25b}, which would help explain their general lack of hot dust emission in the near and mid infrared \citep{Williams23, Pablo24, Akins24, Leung24}.  Given this uncertainty and the fact that LRDs comprise a significant portion of our sample, we have opted not to correct their line luminosities for dust extinction.  If our broad-line AGN have significant dust attenuation ($A_{\rm V} > 2$), our reported BH masses may be underestimated, such that the BHs would be even more overmassive. However, \citet{Taylor25} note that the effect of reddening on BH masses computed this way is likely small ($\sim$0.19 dex) compared to the $\sim$0.5 dex uncertainty from using the locally-calibrated prescription from \citet{reines13}.

\begin{figure*}
\centering
\includegraphics[width=\linewidth]{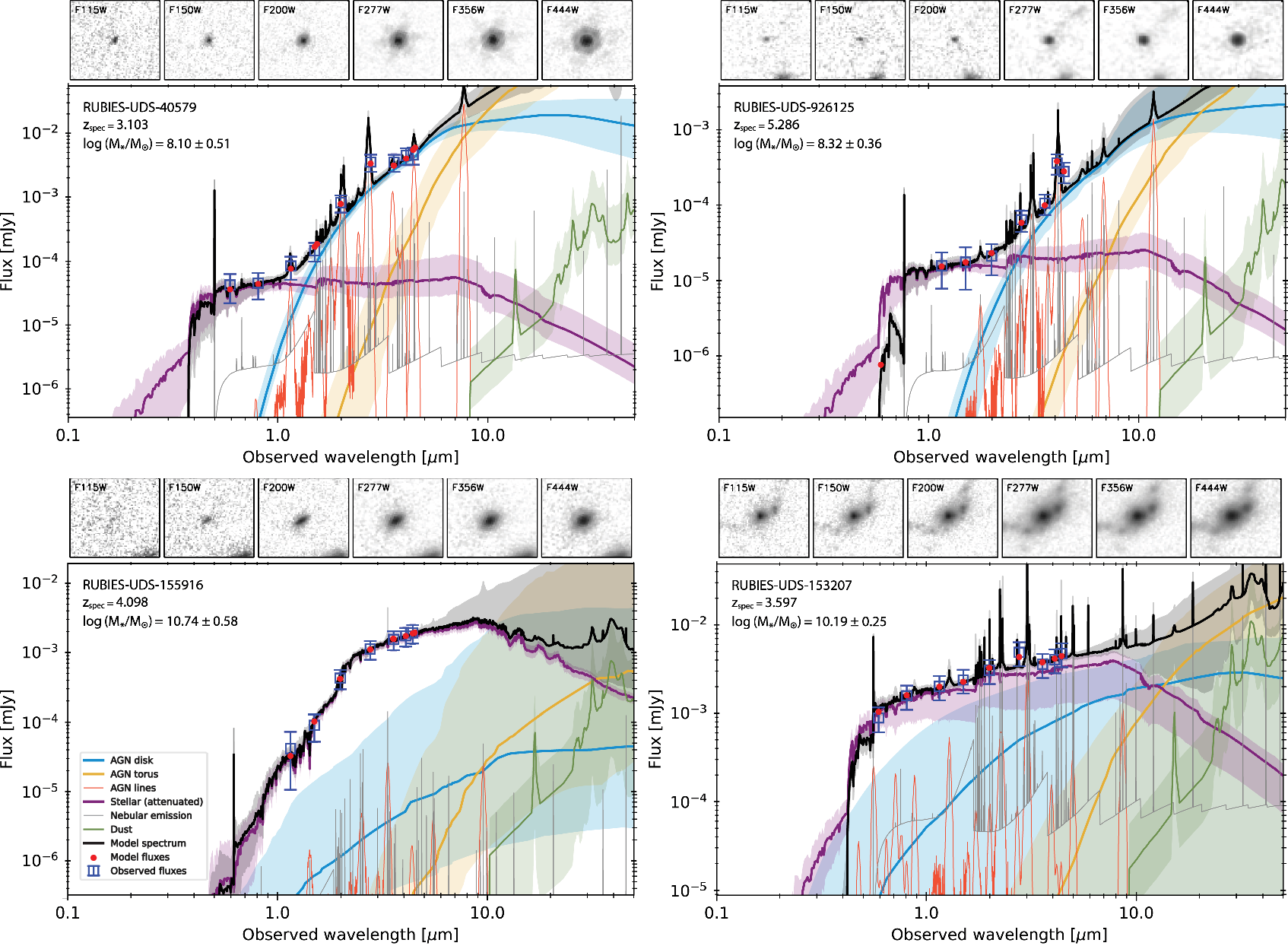}
\caption{Examples of our SED fits with GRAHSP. The total model is shown in black, while individual galaxy and AGN components are represented by the various colored lines (see legend for details).  The posterior mean (solid line) and 2 sigma equivalent uncertainties (shaded areas) are shown for each component.  Observed fluxes are shown as blue squares with $3\sigma$ error bars and predicted model fluxes are shown as red points. NIRCam images on the upper panel are $2.0^{\prime\prime}\times~2.0^{\prime\prime}$ in size.  The top two sources (RUBIES-UDS 40579 and RUBIES-UDS 926125) are LRDs with their distinctive ``v-shaped" SEDs. \label{fig:sedfits_plot}}
\end{figure*}

\subsection{Host Galaxy Morphologies}\label{sec:galfit}

We assess the underlying morphology of the AGN host galaxies using two-dimensional surface brightness profile fitting with the \texttt{GALFIT} software \citep{peng02}.  We carry out these fits in the NIRCam F115W, F150W, F200W, F277W, F356W, and F444W imaging (at their native resolution) to determine if an extended host component is visible and use this knowledge to help inform our SED fitting.  For these fits, we provide \texttt{GALFIT} with empirical PSFs constructed from the NIRCam imaging in each field and noise images that account for both the intrinsic image noise (e.g., background noise and readout noise) as well as added Poisson noise due to the objects themselves.  We fit each source with a single point source component in each band and examine the residual image for underlying light from the host galaxy.  When extended emission is visible, additional S\'ersic components are added to help minimize the flux in our residual images.  All neighboring objects down to three magnitudes fainter than the primary source in the fitting band are fit simultaneously using single S\'{e}rsic profiles.  

Examples of our \texttt{GALFIT} analysis in the F200W filter for two point-like and two extended sources are shown in Figure \ref{fig:galfit_plot}.  The second column from the left shows the residual image after subtracting off a point-source only model, which can help reveal if any extended component is present, while the last column on the right shows the residual image after subtracting our full, best-fit model.  

Roughly half (52.9\%) of our sample appears point-like and is best-fit with a point-source only model.  The LRDs in our sample are more likely to have a point-like morphology compared to the non-LRD sources. Roughly 87\% of the LRDs show no extended emission when fit with a point-source only model.  On the other hand, extended host emission is visible for 74\% of the non-LRDs.

Although the compact size of LRDs is one of their key characteristics, we find four LRDs in our sample that show signs of a spatially extended host galaxy.  These are JADES-GN-954, RUBIES-UDS-40579, RUBIES-EGS-49140, and RUBIES-EGS-50812.   Our Galfit model fit for RUBIES-UDS-40579 at z=3.103 is shown in Figure \ref{fig:galfit_plot}.  In all four cases, we find that the unresolved point source component begins to dominate our two-component fits at wavelengths longer than F277W, indicating that nuclear light from the AGN is responsible for the bulk of their emission at rest-optical wavelengths.

\subsection{Host Galaxy Stellar Mass Measurements}\label{sec:sed_fitting}

To determine the host stellar masses of our broad-line AGN sample, we model their SEDs using \texttt{GRAHSP} \citep{Buchner24}.  \texttt{GRAHSP} combines galaxy stellar population and nebular emission models with a flexible, empirically validated AGN model consisting of power-law continuum emission from an accretion disk, broad and narrow emission lines, and torus emission in the infrared. \texttt{GRAHSP} employs a Bayesian approach to constrain model parameters using a nested sampling inference procedure.

For the host stellar population, \texttt{GRAHSP} makes use of several modules originally developed for \texttt{CIGALE} \citep{Boquien19,Yang22}.  For this study, we adopt a delayed-$\tau$ star formation history (the \texttt{sfhdelayed} module), a \citet{chabrier03} initial mass function with a solar metallicity ($Z = 0.02$), and \citet{BC03} for the simple stellar population (SSP) module.  We add nebular emission using the {\tt nebular} module \citep{VillaV21} and account for dust extinction using the {\tt biattenuation} module.  The latter allows for independent attenuation of the galaxy and AGN components using an SMC-like dust attenuation model.

For the AGN components, we use the {\tt activatepl} module to model the power-law continuum emission from the accretion disk, the {\tt activatelines} module to account for both narrow and broad AGN emission lines, and the {\tt activategtorus} module to add infrared emission associated with reprocessing of UV light by a dust torus.  Attenuation of the resulting AGN emission is handled by the aforementioned {\tt biattenuation} module. 

Table \ref{tab:GRAHSP} summarizes the modules used and their parameters.  For the galaxy components, these follow previous CIGALE-based fits by \cite{Ciesla_2015} and \cite{Yang_2020}, while the AGN components generally follow the ranges found in \citep{Buchner24}.  \texttt{GRAHSP} uses the \texttt{Ultranest} package to sample from the posterior distribution.  Our reported host masses and uncertainties are computed using the median and standard deviation of 1000 equally weighted samples of the posterior probability distribution returned by \texttt{Ultranest}.

All sources are nominally fit with both galaxy and AGN components activated. However, if our \texttt{Galfit} analysis finds the morphology of a source is best fit with a pure point source model and no extended host emission is detected, then we run \texttt{GRAHSP} with only the galaxy modules activated and report the resulting mass as an upper limit.  This is similar to the approach used by \cite{Kocevski23b} and \cite{Harikane23}.  

An exception to this approach is made for the LRD population.  There has been substantial debate in the literature as to the origin of the rest-frame blue UV and red optical emission from these sources, but the prevailing theory is that the UV excess originates from the host galaxy, while the optical emission originates from the AGN.  As discussed in \ref{sec:galfit}, 
we find evidence to support this scenario in the form of extended host emission in the bluest NIRCam bands for four LRDs in our sample, two of which are at $z<4$ (see also \citealt{Rinaldi_2025}, which present a spatially extended LRD at $z \approx 2$, and \citealt{Billand_2025} a sample of LRD descendants at $z < 4$).  Our SED fit for one of these sources, RUBIES-UDS-40579 at $z=3.103$, is shown in Figure \ref{fig:sedfits_plot}.  At higher redshift, LRDs are routinely point-like in NIRCam imaging, yet their SEDs are virtually identical to their more extended, lower redshift counterparts.  As a result, we fit all LRDS with both galaxy and AGN modules in \texttt{GRAHSP} and report their host masses and uncertainties even when their morphologies are point-like.  Examples of our SED fits for both LRDs and non-LRDs are shown in Figure \ref{fig:sedfits_plot}.

\subsection{Inferring the High-z $M_{\rm BH}-M_{\star}$ Relation}
\label{subsec:MCMC}

The black hole and stellar mass measurements, along with their associated uncertainties, can be used to execute a statistical inference of the evolution of the $M_{BH}-M_\star$ relation in the redshift range of our sample.

Local scaling relations (see, e.g., \citealt{Kormendy_Ho_2013, Reines_Volonteri_2015}) indicate that, typically, $M_{BH} \sim 10^{-3} \times M_{\star}$, although with an intrinsic scatter of $\sim 0.5$ dex \citep{Reines_Volonteri_2015}.
Using early JWST data from \cite{Maiolino23}, \cite{Kocevski23a}, \cite{Harikane23} and \cite{Uebler23}, \cite{Pacucci_2023_overmassive} found that galaxies hosting broad-line AGN in the redshift range $4<z<7$ violate the local $M_{BH}-M_\star$ at $>3\sigma$, with black holes being overmassive by factors of $10-100$.

In this paper, we make use of the same Markov-Chain Monte Carlo (MCMC) method developed in \cite{Pacucci_2023_overmassive} to execute a statistical analysis of the relation between black hole mass and host's stellar mass. This method allows a careful treatment of the mass bias, i.e., the fact that observing overmassive black holes is intrinsically easier because they tend to be more luminous. 

In the remainder of this Section, we describe the MCMC model in broad strokes. The interested reader is strongly encouraged to refer to the original study (i.e.,  \citealt{Pacucci_2023_overmassive}), where an in-depth description is provided. The interested reader is also referred to \cite{Hogg_2010}, where a detailed description of the likelihood function used in \cite{Pacucci_2023_overmassive} is provided, together with additional methodologies to fit models to data.

\begin{figure*}
\centering
\includegraphics[width=\linewidth]{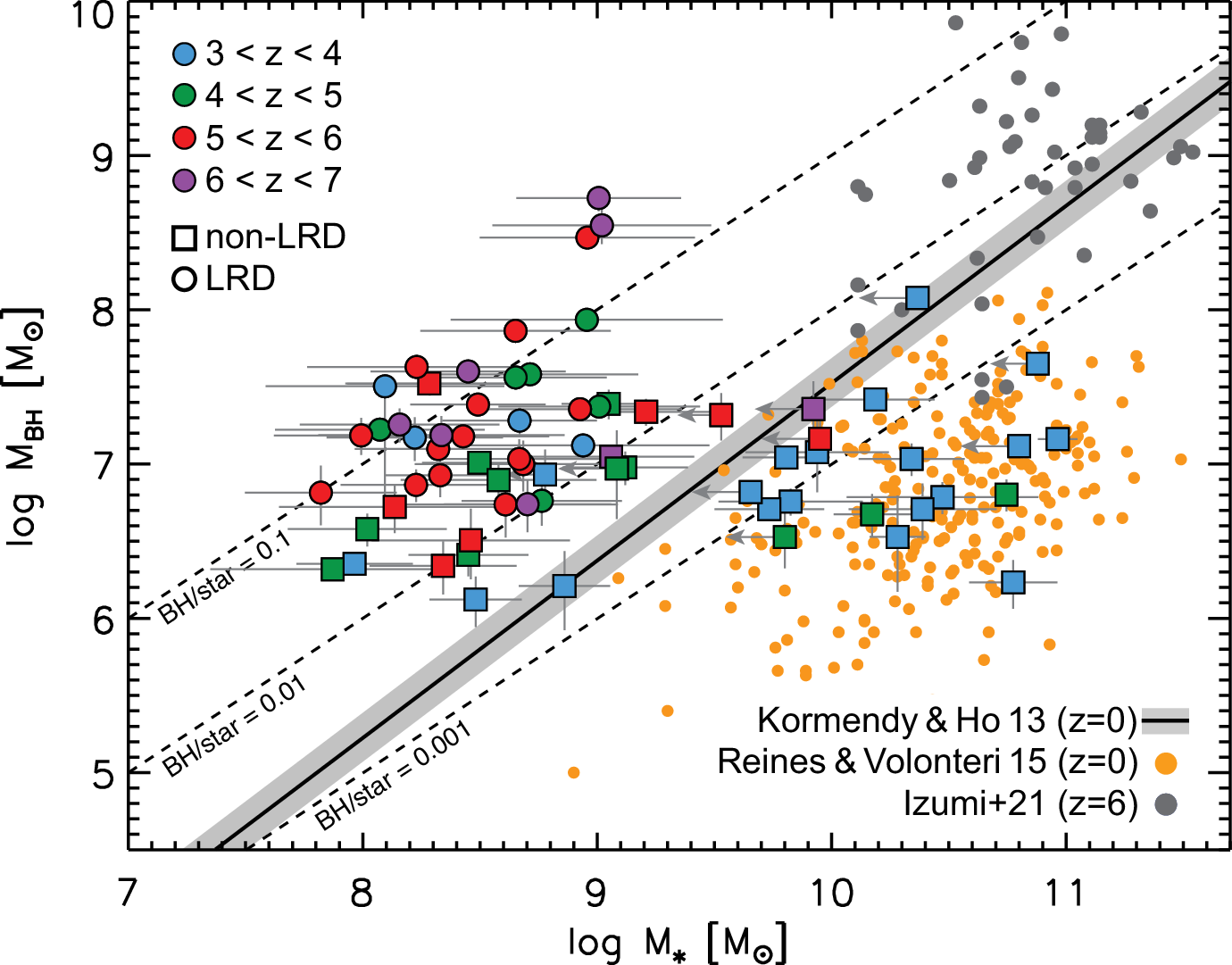}
\caption{Host stellar mass versus BH mass for our sample of broad-line AGN at $3<z<7$, color coded by redshift. Arrows indicate the upper limit on the stellar masses. Grey circles show the $z >6$ quasar samples compiled by \cite{Izumi_2021}. 
Orange circles are the observational samples in the local universe provided by \citet{Reines_Volonteri_2015}, while the gray line is the best-fit local relationship from \citet{Kormendy_Ho_2013}.
The diagonal dashed lines represent ratios of $M_{\rm BH}/M_\star= 0.1$, $0.01$, and $0.001$, as indicated. We find that our sources preferentially scatter above the local scaling relationship at higher redshifts ($z>4$). \label{fig:MM}}
\end{figure*}

The MCMC algorithm uses three main ingredients to find the best model that describes the given $M_{\rm BH}-M_\star$ data.
First, it uses the empirical local $M_{\rm BH}-M_\star$ relation from \cite{Reines_Volonteri_2015} as a prior, selected due to its relevance to the sample at hand, which includes galaxies with similar BH mass. Of course, the redshifts of the galaxies in our sample are much higher, while \cite{Reines_Volonteri_2015} used galaxies with $z < 0.055$. The adopted relation is:
\begin{equation}
\log \left(\frac{M_{BH}}{M_\odot} \right) = \alpha + \beta \log \left(\frac{M_\star}{10^{11} M_\odot} \right) \, ,
\end{equation}
where $\alpha = 7.45 \pm 0.08$ and $\beta = 1.05 \pm 0.11$. 
Second, it uses the JWST sensitivity in H$\alpha$ for a resolution element of the spectrum, which is then translated into a minimum black hole mass that is thus detectable.
Third, it uses the high-$z$ stellar mass function \citep{Song_2016} to account for the fact that low-mass galaxies are intrinsically more numerous than high-mass galaxies.

These ingredients are then fed into the likelihood function, which is central to modeling the high-$z$ $M_{\rm BH}-M_\star$ relation while taking into account (severe) observational biases. A detailed description of the likelihood function is provided in \cite{Pacucci_2023_overmassive}, to which the interested reader is referred.

The algorithm, when applied to a data set of values ($M_{\rm BH},M_\star$), together with their associated uncertainties, provides the posterior distribution functions for three quantities, characterizing the high-$z$ $M_{\rm BH}-M_\star$ relation: (i) the slope $m$; (ii) the intercept $b$; (iii) the orthogonal variance $\nu$, which is related to the standard scatter via $\sqrt{\nu}\sec\theta$, where $\theta$ is the angle between the $M_{\rm BH}-M_\star$ relation and the x-axis.   Hence, the intrinsic scatter is derived from the data, and not fixed a priori. However, \cite{Guia_2024_scatter} showed that there is no significant redshift evolution in the intrinsic scatter of the original set of overmassive black holes used in \cite{Pacucci_2023_overmassive}.

\begin{figure}[ht]
\centering
\includegraphics[width=\linewidth]{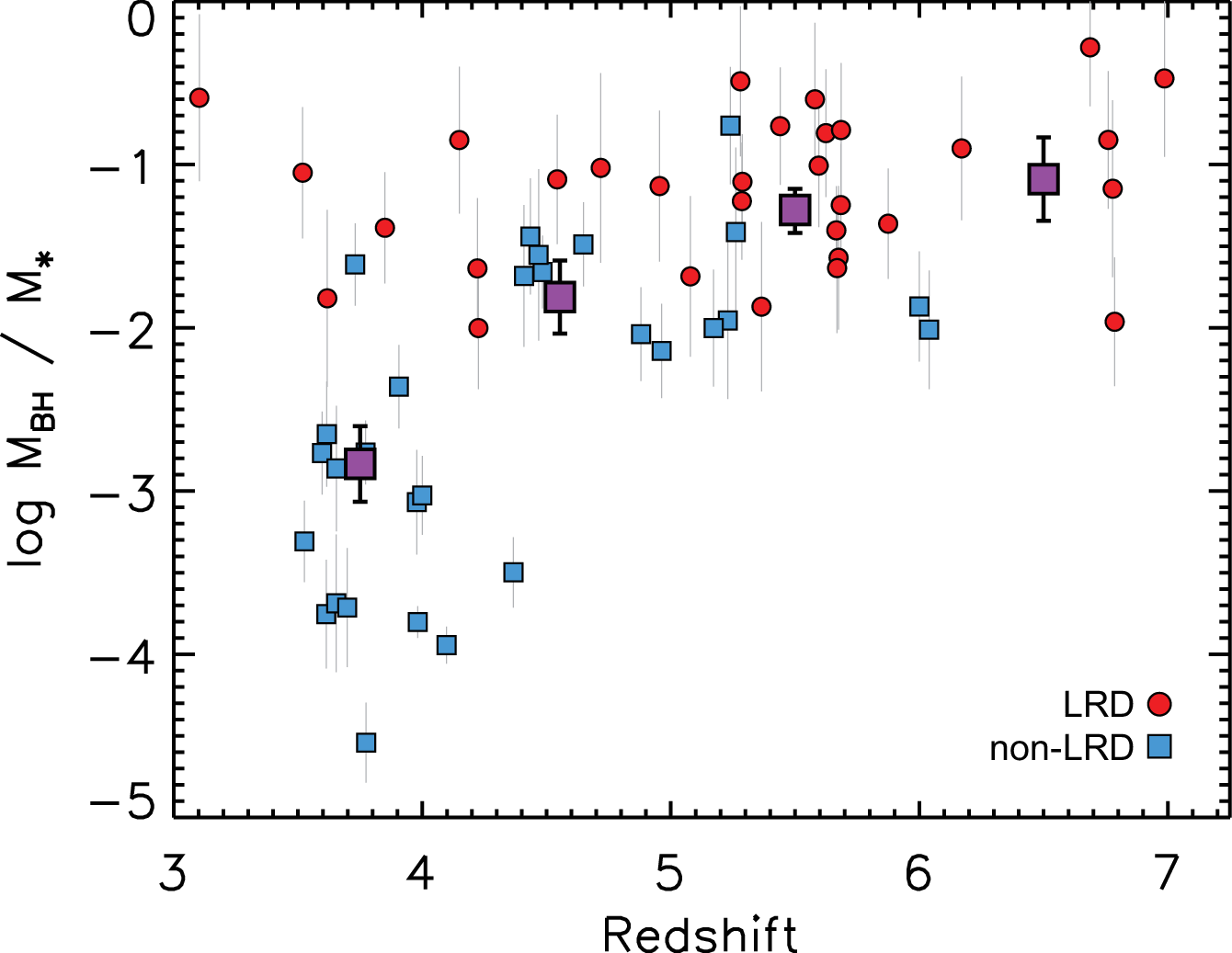}
\caption{Redshift evolution of the ratio between BH mass and stellar mass for the sample of broad-line AGN presented in this study. The colored symbols show individual sources and their $1\sigma$ uncertainties. Red circles (blue squares) denote sources that are (are not) classified as LRDs based on their ``V-shaped" SEDs.  These data points are then grouped into four redshift bins: $3<z<4$, $4<z<5$, $5<z<6$, and $6<z<7$. Their average and propagated $1\sigma$ errors are displayed with purple squares. A redshift evolution is evident, with the ratio $M_{\rm BH}/M_\star$ evolving steadily from the local value of $\sim 0.1\%$ to an overmassive value of $\sim 10\%$ at $z \sim 6.5$.  However, the mass ratio for the LRD sample is elevated over the entire redshift range.
\label{fig:mass_ratio_plot}}
\end{figure}

\section{Results}\label{sec:result}

In Figure \ref{fig:MM}, we show our derived host galaxy stellar masses plotted against the BH masses of our broad-line AGN sample.  Also shown are the local scaling relationships of \citet[hereafter KH13]{Kormendy_Ho_2013} and \citet[hereafter RV15]{Reines_Volonteri_2015}, which are derived for low-redshift, bulge-dominated quiescent galaxies and broad-line AGN, respectively. We use the RV15 relationship as our local benchmark because it was empirically estimated from a sample of broad-line AGN using broad H$\alpha$ lines to measure BH masses and because the BH mass distribution of the sample is comparable to our own.  Consistent with several previous studies, we find that the majority (50/70) of our sources scatter above the local scaling relationship of RV15 and have BH to stellar mass ratios that are 1-2 dex higher than observed their local AGN sample.

We find a redshift trend such that host masses decrease with increasing redshift among sources with the same $M_{\rm BH}$ (and hence $L_{\rm bol}$, assuming a similar accretion regime).  As a result, sources at $z<4$ preferentially sit near the KH13 relationship, while those at higher redshifts scatter above it.  This is reflected in the BH to stellar mass ratio, which can be seen as a function of redshift in Figure \ref{fig:mass_ratio_plot}.   We find that sources at $z<4$ have a median $M_{\rm BH}/M_{\rm \star}$ ratio of 0.15\%, while this increases to 4\% for sources at $z\sim 5.5$. The ratio in our highest redshift bin ($6<z<7$) rises to 7\%. 

Our findings suggest that the higher-redshift, broad-line AGN in our sample are powered by BHs that are increasingly overmassive relative to their local counterparts.  While it is expected that more luminous BHs will preferentially be selected at higher redshifts in a flux-limited survey due to observational biases, an evolving $M_{\rm BH}/M_{\rm \star}$ ratio is supported by our forward modeling  that is designed to take such biases into account (see \S5.1).

Instead we propose the change in the $M_{\rm BH}/M_{\rm \star}$ ratio with redshift is likely related to the increasing fraction of LRDs in our sample at $z>4$.  Of the sources at $z<4$, 83\% (19 sources) are galaxy-dominated, non-LRDs with extended host morphologies.  Only 17\% (4 sources) are LRDs.  On the other hand, at $z>4$, 55\% are LRDs (26 sources) with point-like morphologies.  LRDs are generally faint in the UV, which is the emission that we attribute to their host galaxies.  As a result, their average host mass is $\sim1$ dex lower than that of the non-LRD AGN in our sample.  That said, even the non-LRDs do show signs of an increasing BH to stellar mass ratio with redshift, as also found by \cite{Durodola_2025}.  When we exclude the LRDs, we find that the mass ratio of the non-LRDs at $z>4$ increases by $\sim1$ dex compared to their counterparts at $3<z<4$.  We discuss further how the varying composition of our broad-line AGN sample as a function of redshift impacts our findings in \S\ref{sec:discussion}.

\subsection{Evolution of the $M_{\rm BH}-M_{\rm \star}$ Relation with Redshift}
\label{subsec:evolution}

\begin{figure*}
\centering
\includegraphics[width=\linewidth]{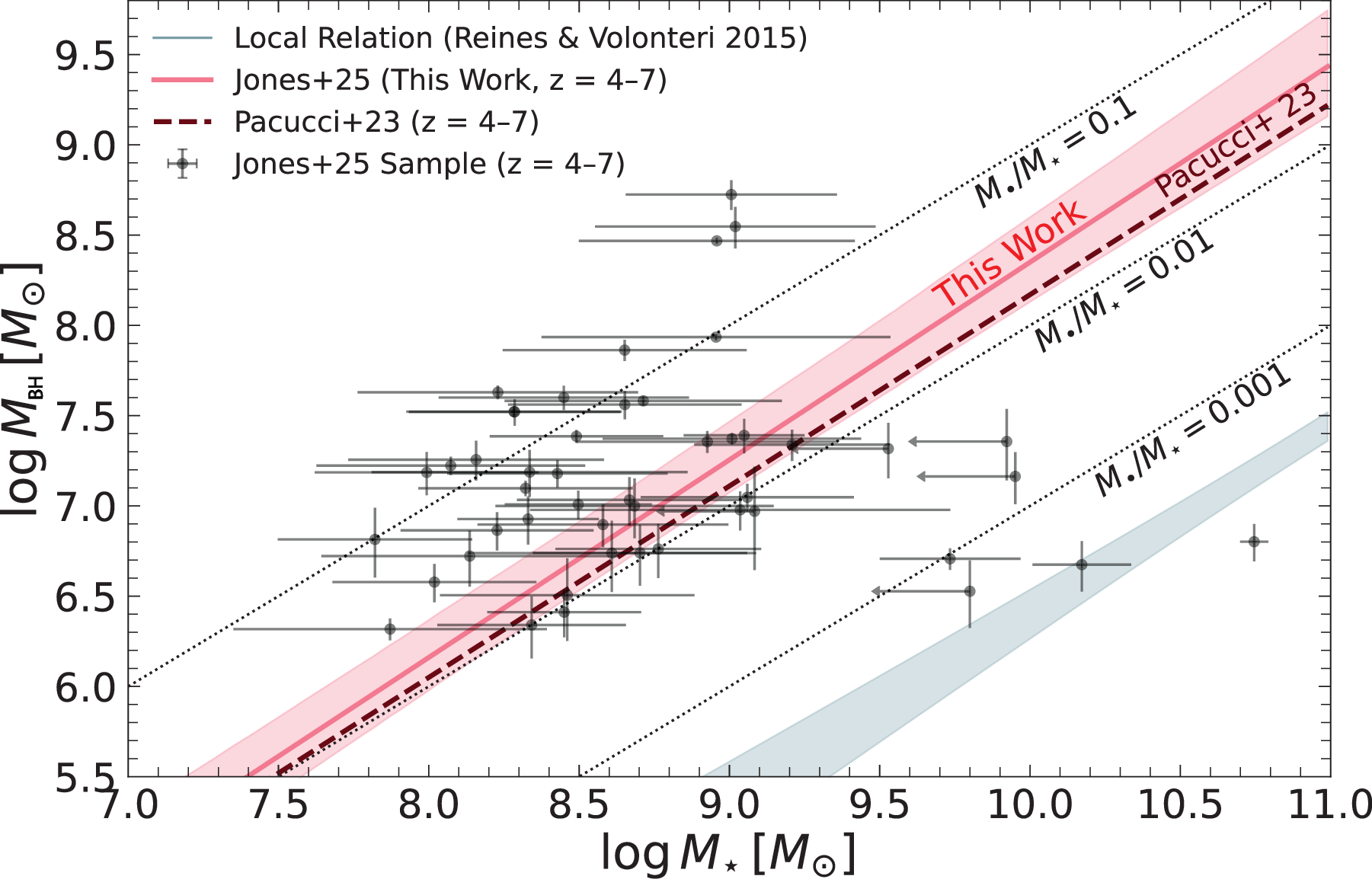}
\caption{The inferred high–$z$ $M_{\rm BH}$–$M_{\star}$ relation (solid red line and $1\sigma$ band) for our sample at $4<z<7$, compared to the local relation from \cite{Reines_Volonteri_2015} (solid teal line with shaded $1\sigma$ scatter) and the original high–$z$ relation from \cite{Pacucci_2023_overmassive} (dashed brown line). We recover a best‐fit intercept $b=-2.50\pm0.75$ and slope $m=1.10\pm0.09$, confirming a $\gtrsim3\sigma$ offset above the local relation. The new relation inferred is practically indistinguishable from the original relation found by \cite{Pacucci_2023_overmassive}. Diagonal dashed black lines indicate constant ratios, for reference: $M_{\rm BH}/M_{\star}=0.1,\;0.01,\;0.001$. Data points show our sample measurements with $1\sigma$.}
\label{fig:MM_new}
\end{figure*}

In order to account for selection biases and quantify the redshift evolution of the relationship between $M_{\rm BH}$ and $M_{\star}$, we applied the same MCMC model described in \cite{Pacucci_2023_overmassive} to our broad-line AGN sample.  Because we observe a sharp increase in the $M_{\rm BH}/M_{\rm \star}$ ratio at $z>4$ (see Fig. \ref{fig:mass_ratio_plot}), we have chosen to limit our analysis to sources in the redshift range $4<z<7$.  When run on the AGN in this redshift range, we recover a best-fit relation of
\begin{equation}
\log\left(\frac{M_{\rm BH}}{M_\odot}\right) \;=\; b + m\,\log\left(\frac{M_\star}{10^{11}M_\odot}\right)
\end{equation}
with intercept $b = -2.50 \pm 0.75$ and slope $m = 1.10 \pm 0.09$.  Our best-fit relationship is shown in Figure~\ref{fig:MM_new}. This inferred relation lies more than $3\sigma$ above the local RV15 
relationship, and is in excellent agreement with the original \cite{Pacucci_2023_overmassive} high-$z$ result (also shown in Figure~\ref{fig:MM_new}). 

\begin{figure}
\centering
\includegraphics[width=\linewidth]{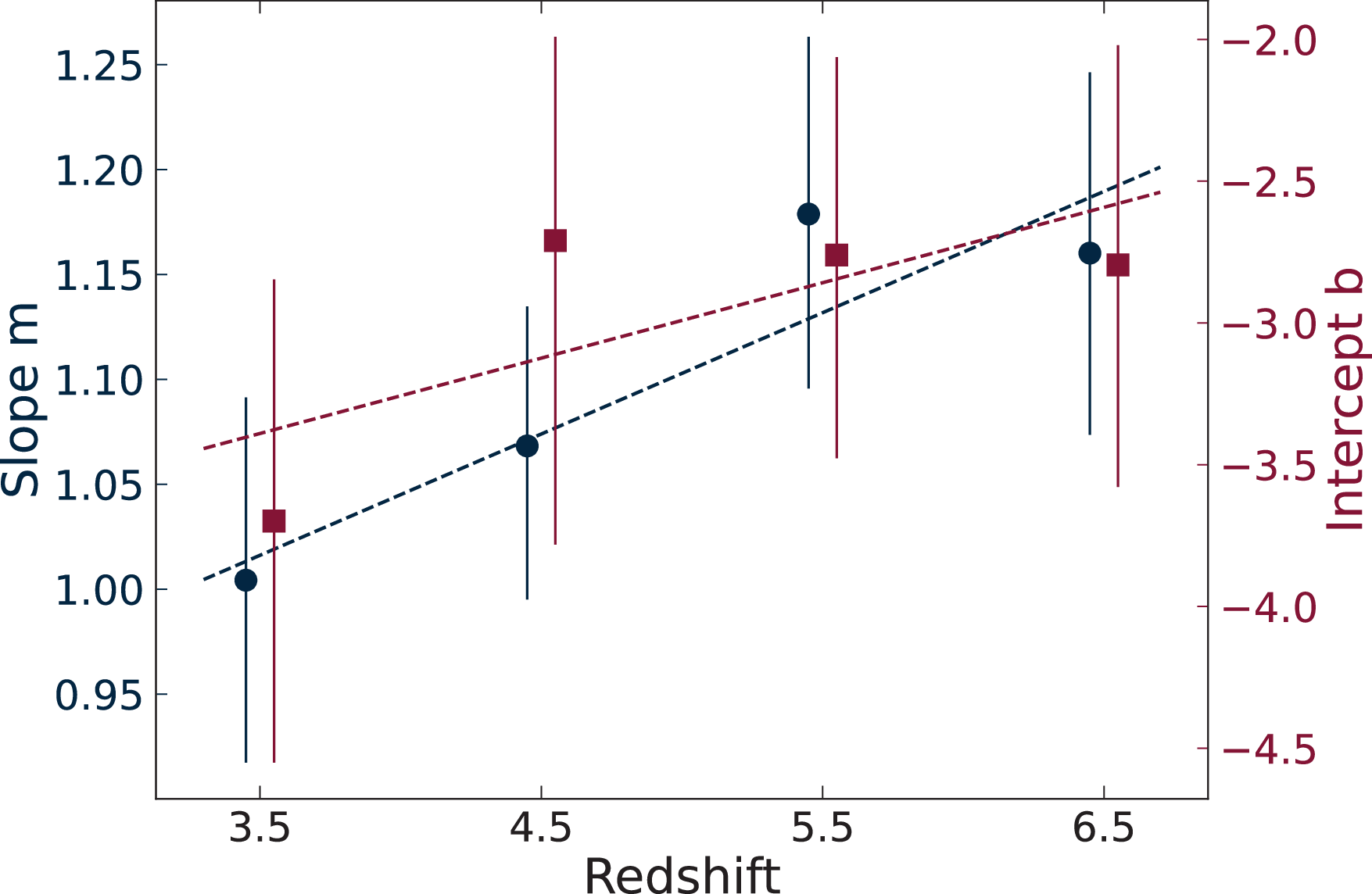}
\caption{Redshift evolution of the best‐fit $M_{\rm BH}$–$M_{\star}$ relation parameters.  Blue circles show the slope $m$ and red squares show the intercept $b$ in four redshift bins ($3<z<4$, $4<z<5$, $5<z<6$, $6<z<7$). Both quantities are plotted at slightly offset horizontal positions to avoid overlap ($\Delta z = \pm0.05$).  Error bars indicate the posterior $1\sigma$ uncertainties from our MCMC fits in each bin.  We find a $\sim1\sigma$–$2\sigma$ increase in both slope and intercept from $z\sim3.5$ to $z\sim6.5$, with probabilities of evolution $P(m_{\rm 6-7}>m_{\rm 3-4})=90\%$ and $P(b_{\rm 6-7}>b_{\rm 3-4})=80\%$. The dashed, colored lines indicate the respective best-fit (i.e., linear regressions) lines for the slope (blue) and intercept (red) trends.}
\label{fig:evolution}
\end{figure}

To investigate any redshift trends, we subdivide the sample into four bins ($3<z<4$, $4<z<5$, $5<z<6$, $6<z<7$) and rerun the MCMC inference on each bin independently.
In this case, we include the data points in the redshift range $3<z<4$ because they offer an average value of the ratio $M_{\rm BH}/M_{\star}$  that is closer to the local one. We extract the median and $1\sigma$ uncertainties of $b$ and $m$ in each bin, then plot these eight values as a function of bin center, as displayed in Figure~\ref{fig:evolution}.

To quantify the significance of any redshift evolution, we compare the posterior distributions of $m$ and $b$ in the lowest ($3<z<4$) and highest ($6<z<7$) bins.  Specifically, for each of 15,000 random draws from the two chains, we record whether the high-$z$ value exceeds the low-$z$ value, yielding:
\begin{equation}
P(m_{6-7} > m_{3-4}) \;=\; 0.896\,, 
\qquad
P(b_{6-7} > b_{3-4}) \;=\; 0.780\,.
\end{equation}
These probabilities correspond to roughly $1-2\sigma$ in a Gaussian sense, therefore we interpret this as only marginal evidence for evolution in both slope and intercept across $3<z<7$.

We can also use this same methodology to examine the evolution of the intrinsic $M_{\rm BH}/M_{\star}$ ratio over the redshift range of our AGN sample.  Calculated for the median $M_{\star}$ in each redshift bin in Figure \ref{fig:evolution}, we find that $M_{\rm BH}/M_{\star}$ increases on average by a factor of $\sim197$ (i.e., $2.294$ dex) between $z = 3.5$ to $z = 6.5$.  The uncertainty on this change is $0.751$ dex (calculated assuming redshift independence), which means this increase is measured with a confidence level of $ > 3\sigma$.

Finally, we can estimate the intrinsic scatter of our inferred $M_{\rm BH}$–$M_{\star}$ relationship based on the best-fit orthoginal variance.  We find a typical scatter of $0.9$ dex in all four redshift bins.  This scatter is larger than what was originally inferred in \cite{Pacucci_2023_overmassive} (i.e., $0.69$) and does not appear to significantly evolve with redshift, as already suggested in \cite{Guia_2024_scatter}.

To summarize, the key findings of this study are the following:
\begin{itemize}
  \item We find that the $4<z<7$ $M_{\rm BH}$–$M_{\star}$ relationship as probed by our sample of broad-line AGN is $>3\sigma$ above the local relationship of RV15. 
  \item We derive an intrinsic scatter in this relationship of $0.9$ dex, which does not vary over the redshift range of our sample. 
  \item We find that the $M_{\rm BH}/M_{\star}$ ratio increases on average by $2.294$ dex from $z = 3.5$ and $z = 6.5$ with a confidence level of $ > 3\sigma$. 
 \item We find only marginal evidence ($\sim$1–2$\sigma$) that both the slope and intercept of the $M_{\rm BH}$–$M_{\star}$ increases over the redshift range of our sample. 
\end{itemize}
These data suggest that the $M_{\rm BH}/M_{\star}$ ratio increases significantly with increasing redshift. At the same time, the intrinsic scatter remains constant, despite being larger than its typical value in the local Universe.

\section{Discussion}\label{sec:discussion}

Using a forward modeling approach that accounts for selection biases, we find that the faint, broad-line AGN identified by JWST at $z>4$ are overmassive by 1-2 dex relative to low-redshift AGN and the resulting $M_{\rm BH}$–$M_{\star}$ relationship is incompatible with the local relationship of RV15 with a confidence level of $> 3\sigma$.

A contributing factor to the increasing $M_{\rm BH}/M_{\star}$ ratio that we observe at higher redshifts is the increasing fraction of LRDs in our sample with increasing redshift.  While LRDs make up 17\% of our sample at $z<4$, they are the majority of the sample at $z>4$, consistent with the redshift distribution of photometrically-selected LRDs \citep[e.g.,][]{Kocevski25}, and with theoretically-derived redshift distributions \citep[e.g.,][]{Pacucci_Loeb_2025}. Due to their faint rest-frame UV emission, the average host mass of the LRDs in our sample is $\sim1$ dex lower than that of the non-LRD AGN.  

Without the LRDs included, the bulk of the remaining BLAGN in our sample have $M_{\rm BH}/M_{\star}$ ratios that are largely consistent with the RV15 relationship.  The bimodal distribution that we observe in the mass ratios of LRD and non-LRD BLAGN may indicate that LRDs are a specific evolutionary phase of SMBH evolution, where early BH growth has outpaced that of their host galaxies.  

That said, there is still significant debate as to the nature of LRDs and which part, if any, of their ``v-shaped" SEDs is due to stellar emission. As a result, their host masses may be underestimated. 
However, if we attribute their rest-optical light to stellar emission, which would make their host masses more consistent with the RV15 relationship, it would result in extreme stellar densities given their measured sizes \citep[e.g.,][]{Furtak22,Akins24}. Their stellar densities would be higher than the densities observed in any system at any redshift and, in the most extreme cases, would greatly exceed the density necessary for runaway stellar collisions \citep{Baggen24, Guia24, Pacucci_Hernquist_2025}.

Another important caveat is that the BH masses of LRDs may be overestimated for a number of reasons, including if the locally-calibrated virial mass relationship of \citet{reines13} is no longer applicable at high redshifts, if electron scattering contributes to the broadening of the Balmer lines \citep{Rusakov25}, or if the LRDs have lower bolometric luminosities than commonly assumed \citep[e.g.,][]{Greene25}.   That said, \citet{Juodzbalis_2025} recently reported that their direct measurement of the BH mass in an LRD at $z=7.01$ is in excellent agreement with the mass inferred from single-epoch virial estimates.  This suggests that these relationships may in fact be reliable for LRDs out to the Epoch of Reionization.

\subsection{Comparison to Previous Results}

Our finding of a statistically significant deviation from the $M_{\rm BH}$–$M_{\star}$ relation of RV15 differs from the recent results from \citet{Li25}, who examined a subset of the sources used in this study.  In addition to the larger number of LRDs in our sample, a key difference between the approach of \citet{Li25} and that of \citet{Pacucci_2023_overmassive}, which we use in this study, is the adoption of an Eddington ratio ($\lambda_{\rm Edd}$) distribution function derived for bright quasars at $z\sim6$ that has a mean of $\lambda_{\rm Edd}\sim0.1$.  Instead, our forward modeling approach translates $M_{\rm BH}$ directly into an observed luminosity, assuming an observationally-motivated value of 1000 km s$^{-1}$ for ${\rm FWHM_{H\alpha}}$.

Lower values of $\lambda_{\rm Edd}$ translate into lower luminosities for a given $M_{\rm BH}$, making lower-mass BHs harder to detect. However, there is growing evidence that LRDs may be accreting at higher Eddington ratios than the $10\%$ assumed in \citet{Li25}.  First, large collections of LRDs at $z > 4$ \citep{Harikane23, Maiolino23, Kokorev24} suggest a median $\lambda_{\rm Edd}$ ranging from $0.4$ to $1.1$, i.e., slightly super Eddington, and more than one order of magnitude higher than the assumption in \cite{Li25}.  Additionally, from a theoretical perspective, there is growing evidence that LRDs may accrete close to or above the Eddington limit, as this would help explain many of their observed properties, including their X-ray weakness and weak variability \citep{Pacucci24b, Inayoshi24b, Lupi_2024, Madau24, Lambrides_2024}. Early super-Eddington growth that transitions to a typical AGN phase at lower redshifts would naturally explain the high $M_{\rm BH}$–$M_{\star}$ ratios observed in the LRDs compared to lower redshift AGN \citep{Inayoshi25}. 

Ultimately, a better understanding of the intrinsic Eddington ratio distribution for these broad-line AGN, and LRDs in particular, is necessary to improve our handling of selection biases.  If these sources are accreting with the same Eddington ratios as bright quasars, the offset from the local relationship may indeed be minimal. However, if they are accreting at higher rates, then the completeness correction would be less severe and allow for a higher normalization in the scaling relationship at $z\sim5$.  

That said, a potential issue with the bias-only explanation presented in \citet{Li25} and \citet{Silverman25}  is that it would require the intrinsic high-$z$ 
$M_{\rm BH}$-$M_{\star}$ relation to have a nearly local normalization but a very large scatter, so that we preferentially detect only the bright, overmassive tail (see also \citealt{Jahnke_2011}).
For instance, \cite{Silverman25} explicitly infers $\sigma\simeq 0.8^{+0.23}_{-0.28}$ dex and a normalization consistent with the local relation, attributing the observed offsets to selection and dispersion rather than true evolution. In that picture, the systems we observe at $z \gtrsim 4$ that sit $\sim 1.8-1.9$ dex above the local mean (i.e., a factor of $\sim 70$ in BH mass, as initially reported by \citealt{Pacucci_2023_overmassive}) represent the extreme Gaussian tail, not the bulk population. 

For a Gaussian with $\sigma \approx 0.8$ dex, the fraction above $\sim 1.85$ dex is only $\sim 1\%$; i.e., the underlying parent population would have to be $\sim 100$ times more numerous than the already observed counts to ``hide'' the remainder.  However, the number density of LRDs at $z\sim5-7$ already accounts for a relatively high fraction ($\sim1-10\%$) of the total galaxy population at the bright end ($M_{UV}\sim-22$; \citealp{Greene_2023, Kokorev24, Kocevski25}).  The observed BH mass function at $z \sim 5-6$ \citep{Taylor25} also already agrees in normalization and shape with physically motivated models and independent AGN luminosity functions , not requiring a large, undetected reservoir. Inflating the true number density by a significant factor to accommodate a bias-only scenario would overproduce the BH mass function relative to those same models and observed luminosity functions.

In addition, \cite{Geris_2025} recently stacked the spectra of $\sim 600$ JADES galaxies in the redshift range $3<z<7$, finding that the faint broad-line AGN population that they recover is still significantly above the local $M_{\rm BH}$-$M_{\star}$ relation. Hence, these observations suggest that we are not missing a significant population of AGN that sit on the local relation.

\subsection{Comparison to Cosmological Simulations and Semi-Analytical Models}

To understand the implications of a large population of overmassive BHs at high redshifts, in this section we compare our results to what current large-volume cosmological simulations and semi-analytical models (SAMs) predict regarding the existence of overmassive BHs at high redshift, and the corresponding evolution of the $M_{\rm BH}-M_\star$ relation through cosmic time.

Many cosmological simulations have difficulty reproducing the overmassive BH population observed with JWST without explicitly invoking heavier BH seeds.  For example, \cite{Jeon_2025}, using the cosmological zoom-in simulation code GIZMO, find that stellar-remnant seeds with standard Bondi-Hoyle accretion fail to grow rapidly enough to reproduce the overmassive BHs identified by JWST at $z \gtrsim 7$, but heavy seeds ($\sim10^{5-6} \, M_\odot$) in dense, biased environments can reach $M_{\rm BH}\sim10^{7-8} \, M_\odot$ if they accrete efficiently at or near the Eddington limit. Under these conditions, the simulated systems naturally appear overmassive with $M_{\rm BH}/M_\star$ ratios well above the local relation and comparable to those inferred from our sample. The simulations predict that such overmassive states can persist until at least $z \sim 5$, after which additional stellar growth or mergers are required to converge onto the local relations.

In addition, \cite{Bhomwick_2024}, using the BRAHMA simulations with heavy seed models that emulate direct collapse black hole (DCBH) formation, find that their most physically motivated models result in BHs at $z \sim 4$ that fall below our inferred high-$z$ relation by roughly an order of magnitude.  Only when some of their most stringent criteria for DCBH formation are dropped and merger delays between central BHs are assumed to be short (i.e., $\lesssim 750$ Myr) does the simulation reproduce the overmassive relation observed.  In these simulations, BH growth at $z \gtrsim 4$ is primarily merger-driven; hence, producing many overmassive systems requires permissive heavy seeding, higher initial seed masses, and/or more efficient accretion than in the baseline model. These results imply that persistent, widespread overmassive BHs at high redshift require heavy seeding and that BH mergers proceed efficiently.

Other cosmological simulations, such as \textsc{Astrid} and \textsc{TNG} (see, e.g., \citealt{Weller_2023_overmassive, Ni_2024_astrid}), offer a complementary perspective. The redshift evolution of the mean $M_{\rm BH}-M_\star$ relation is governed by how gas is partitioned between star formation and BH fueling. In these simulations, sources that are overmassive at $z=4$ typically grow more slowly thereafter and drift toward the mean by $z\simeq2$, whereas undermassive systems catch up. Thus, large positive offsets are transient rather than a long-lived, ubiquitous population in these models.

Finally, comparisons of \textsc{Astrid} and TNG50 \citep{Dattathri_2025} demonstrate that simulations can host strongly overmassive BHs, but at low stellar masses these systems are rare and, at low redshift, their origin is primarily environmental (e.g., tidal stripping in overdense regions) rather than a direct imprint of seeding. At $z > 6$, overmassive BHs are a signature of heavy seeds (see, e.g., \citealt{Scoggins_2023}). However, at later times, they are more often linked to suppressed star formation and environmental interactions.

The \textsc{Dark Sage} SAM \citep{Dattathri_2025} displays substantially larger scatter in $M_{\rm BH}$ at fixed $M_\star$. This broader dispersion arises from SAM choices for BH fueling and quenching that the black hole accretion rate from the star formation rate more readily than in \textsc{Astrid} or \textsc{TNG}, allowing a more numerous population of outliers at early times. SAMs of this kind can therefore more easily accommodate the existence of high-$z$ overmassive BHs. However, the prevalence and persistence of such offsets depend sensitively on the assumed prescriptions for gas supply, triggering, and feedback.

A complementary theoretical perspective is provided by the feedback-free starbursts (FFB) scenario \citep{Dekel_2023}. In that framework, as evaluated in \citet{Dekel_2025_MBH}, massive BH seeds
form at cosmic dawn by rapid core collapse within the young, rotating clusters that constitute the starbursting FFB galaxies.   The high $M_{\rm BH}/M_{\star}$ ratio is maintained as the clusters and BHs inspiral by dynamical friction to the galactic-disk centers and merge there while overcoming gravitational-wave recoil. This pathway yields typical ratios of $\sim 0.01$, broadly consistent with the average values we infer at $z \gtrsim 4$. However, reaching the extreme $\sim 0.1$ ratios seen in some of our sources may be challenging in this framework.

\section{Conclusions}\label{sec:conclusion}

In this study, we examine the relationship between BH mass and host galaxy stellar mass as probed by faint, broad-line AGN at $3 < z < 7$.  Our sample consists of 70 AGN previously identified in the literature based on their broad emission lines in NIRSpec G395M/F290LP observations from the CEERS, JADES, and RUBIES surveys.  Our AGN sample includes both reddened sources classified as LRDs, as well as bluer, non-LRDs in roughly equal number, although the LRDs are preferentially found at higher redshifts and become the majority of the sample at $z>4$.  

We perform emission line fitting using a consistent methodology for the entire sample and infer BH masses using the single-epoch scaling relationship presented in Reines et al. (2013).  We assess the underlying morphology of the AGN host galaxies using two-dimensional surface brightness profile fitting and measure host stellar masses with SED fitting using a hybrid model that includes stellar continuum and nebular emission from the galaxy and power-law continuum emission and broad and narrow line emission from the AGN.

We find that the BLAGN at $3<z<4$ have a median $M_{\rm BH}/M_{\rm \star}$ ratio consistent with the local scaling relationship of RV15, however at $z>4$ the mass ratios are 1-2 dex higher than that observed in their low-$z$ broad-line AGN.  We attribute this redshift trend with the increasing fraction of LRDs in our sample at $z>4$ as their host masses are $\sim1$ dex lower than the non-LRD AGN in our sample.

Using a forward-modeling Bayesian framework that accounts for uncertainties, intrinsic scatter, and selection effects, we infer an $M_{\rm BH}-M_{\star}$ relationship from our sample at $4 < z < 7$ that is inconsistent with the local RV15 relationship at a confidence level of $>3\sigma$.  We derive an intrinsic scatter in this relationship of 0.9 dex, which is higher than the 0.5 dex found locally, but does not vary over the redshift range of our sample. Using this same methodology, we find that the intrinsic $M_{\rm BH}/M_{\star}$ ratio increases by over 2 dex with a confidence level  of $>3\sigma$ from $z=3.5$ to $z=6.5$.  

Our findings support a picture in which the BHs powering JWST’s broad-line AGN are
genuinely overmassive and become increasingly so with redshift.  The presence of a large population of overmassive BHs at high redshifts could be due to initial seeding conditions or efficient BH accretion that outpaces star formation in early galaxies, however most cosmological simulations and SAMs have difficulty recreating such a population without employing the use of a heavy BH seeds.

Given that our results are heavily influenced by the increasing fraction of LRDs in our sample with redshift, we caution that much remains to be determined about these sources.  We have worked under the assumption that the blue UV emission of LRDs is due to stellar emission and that the locally-calibrated, single-epoch scaling relationship used to determine our BH masses is valid for LRDs at high redshifts.  Ultimately, additional study of LRDs will be needed to determine if these assumptions are valid.  Several ongoing spectroscopic follow-up programs and dedicated monitoring campaigns of LRDs should provide a better understanding of their BH masses, accretion rates, and intrinsic luminosities in the near future.

\section{Acknowledgments}

We thank the JADES and RUBIES team for their work in designing and preparing the NIRSpec observations used in this study.  This work is supported by NASA grants JWST-AR-02446 and JWST-GO-5718 based on observations made with the NASA/ESA/CSA James Webb Space Telescope. The data were obtained from the Mikulski Archive for Space Telescopes
at the Space Telescope Science Institute, which is operated by the Association of Universities for Research in Astronomy,
Inc., under NASA contract NAS 5-03127 for JWST.


\begin{thebibliography}{}
\expandafter\ifx\csname natexlab\endcsname\relax\def\natexlab#1{#1}\fi
\providecommand{\url}[1]{\href{#1}{#1}}
\providecommand{\dodoi}[1]{doi:~\href{http://doi.org/#1}{\nolinkurl{#1}}}
\providecommand{\doeprint}[1]{\href{http://ascl.net/#1}{\nolinkurl{http://ascl.net/#1}}}
\providecommand{\doarXiv}[1]{\href{https://arxiv.org/abs/#1}{\nolinkurl{https://arxiv.org/abs/#1}}}

\bibitem[{{Aird} {et~al.}(2015){Aird}, {Coil}, {Georgakakis}, {Nandra}, {Barro}, \& {P{\'e}rez-Gonz{\'a}lez}}]{Aird15}
{Aird}, J., {Coil}, A.~L., {Georgakakis}, A., {et~al.} 2015, \mnras, 451, 1892, \dodoi{10.1093/mnras/stv1062}

\bibitem[{{Akins} {et~al.}(2024){Akins}, {Casey}, {Lambrides}, {Allen}, {Andika}, {Brinch}, {Champagne}, {Cooper}, {Ding}, {Drakos}, {Faisst}, {Finkelstein}, {Franco}, {Fujimoto}, {Gentile}, {Gillman}, {Gozaliasl}, {Harish}, {Hayward}, {Hirschmann}, {Ilbert}, {Kartaltepe}, {Kocevski}, {Koekemoer}, {Kokorev}, {Liu}, {Long}, {McCracken}, {McKinney}, {Onoue}, {Paquereau}, {Renzini}, {Rhodes}, {Robertson}, {Shuntov}, {Silverman}, {Tanaka}, {Toft}, {Trakhtenbrot}, {Valentino}, \& {Zavala}}]{Akins24}
{Akins}, H.~B., {Casey}, C.~M., {Lambrides}, E., {et~al.} 2024, arXiv e-prints, arXiv:2406.10341, \dodoi{10.48550/arXiv.2406.10341}

\bibitem[{{Baggen} {et~al.}(2024){Baggen}, {van Dokkum}, {Brammer}, {de Graaff}, {Franx}, {Greene}, {Labb{\'e}}, {Leja}, {Maseda}, {Nelson}, {Rix}, {Wang}, \& {Weibel}}]{Baggen24}
{Baggen}, J. F.~W., {van Dokkum}, P., {Brammer}, G., {et~al.} 2024, \apjl, 977, L13, \dodoi{10.3847/2041-8213/ad90b8}

\bibitem[{{Bagley} {et~al.}(2022){Bagley}, {Finkelstein}, {Koekemoer}, {Ferguson}, {Arrabal Haro}, {Dickinson}, {Kartaltepe}, {Papovich}, {P{\'e}rez-Gonz{\'a}lez}, {Pirzkal}, {Somerville}, {Willmer}, {Yang}, {Yung}, {Fontana}, {Grazian}, {Grogin}, {Hirschmann}, {Kewley}, {Kirkpatrick}, {Kocevski}, {Lotz}, {Medrano}, {Morales}, {Pentericci}, {Ravindranath}, {Trump}, {Wilkins}, {Calabr{\`o}}, {Cooper}, {Costantin}, {de la Vega}, {Hutchison}, {Lucas}, {McGrath}, {Wang}, \& {Wuyts}}]{bagley22}
{Bagley}, M.~B., {Finkelstein}, S.~L., {Koekemoer}, A.~M., {et~al.} 2022, arXiv e-prints, arXiv:2211.02495.
\newblock \doarXiv{2211.02495}

\bibitem[{{Bertin} \& {Arnouts}(1996)}]{bertin96}
{Bertin}, E., \& {Arnouts}, S. 1996, \aaps, 117, 393, \dodoi{10.1051/aas:1996164}

\bibitem[{{Bhowmick} {et~al.}(2024){Bhowmick}, {Blecha}, {Torrey}, {Somerville}, {Kelley}, {Vogelsberger}, {Weinberger}, {Hernquist}, \& {Sivasankaran}}]{Bhomwick_2024}
{Bhowmick}, A.~K., {Blecha}, L., {Torrey}, P., {et~al.} 2024, \mnras, 533, 1907, \dodoi{10.1093/mnras/stae1819}

\bibitem[{{Billand} {et~al.}(2025){Billand}, {Elbaz}, {Gentile}, {Tarrasse}, {Franco}, {Magnelli}, {Daddi}, {Lyu}, {Dekel}, {Pacucci}, {Sangalli}, {Dickinson}, {Giavalisco}, {Holwerda}, {Kocevski}, {Koekemoer}, {Kokorev}, {Lucas}, \& {P{\'e}rez-Gonz{\'a}lez}}]{Billand_2025}
{Billand}, J.-B., {Elbaz}, D., {Gentile}, F., {et~al.} 2025, arXiv e-prints, arXiv:2507.04011, \dodoi{10.48550/arXiv.2507.04011}

\bibitem[{{Bisigello} {et~al.}(2025){Bisigello}, {Euclid Collaboration}, {Rodighiero}, {Fotopoulou}, {Ricci}, {Jahnke}, {Feltre}, {Allevato}, {Shankar}, {Cassata}, {Dalla Bont{\`a}}, {Gandolfi}, {Girardi}, {Giulietti}, {Grazian}, {Lovell}, {Maiolino}, {Matamoro Zatarain}, {Mezcua}, {Prandoni}, {Roberts}, {Roster}, {Salvato}, {Siudek}, {Tarsitano}, {Toba}, {Vietri}, {Wang}, {Zamorani}, {Baes}, {Belladitta}, {Nersesian}, {Spinoglio}, {Lopez Lopez}, {Aghanim}, {Altieri}, {Amara}, {Andreon}, {Auricchio}, {Aussel}, {Baccigalupi}, {Baldi}, {Balestra}, {Bardelli}, {Basset}, {Battaglia}, {Bender}, {Biviano}, {Bonchi}, {Branchini}, {Brescia}, {Brinchmann}, {Camera}, {Ca{\~n}as-Herrera}, {Capobianco}, {Carbone}, {Carretero}, {Casas}, {Castellano}, {Castignani}, {Cavuoti}, {Chambers}, {Cimatti}, {Colodro-Conde}, {Congedo}, {Conselice}, {Conversi}, {Copin}, {Courbin}, {Courtois}, {Cropper}, {Da Silva}, {Degaudenzi}, {De Lucia}, {Di Giorgio}, {Dolding}, {Dole}, {Dubath}, {Duncan}, {Dupac}, {Dusini}, {Ealet}, {Escoffier},
  {Farina}, {Farinelli}, {Faustini}, {Ferriol}, {Finelli}, {Frailis}, {Franceschi}, {Galeotta}, {George}, {Gillard}, {Gillis}, {Giocoli}, {G{\'o}mez-Alvarez}, {Gracia-Carpio}, {Granett}, {Grupp}, {Gwyn}, {Haugan}, {Hoekstra}, {Holmes}, {Hook}, {Hormuth}, {Hornstrup}, {Hudelot}, {Jhabvala}, {Keih{\"a}nen}, {Kermiche}, {Kiessling}, {Kubik}, {K{\"u}mmel}, {Kunz}, {Kurki-Suonio}, {Le Boulc'h}, {Le Brun}, {Le Mignant}, {Liebing}, {Ligori}, {Lilje}, {Lindholm}, {Lloro}, {Mainetti}, {Maino}, {Maiorano}, {Mansutti}, {Marcin}, {Marggraf}, {Martinelli}, {Martinet}, {Marulli}, {Massey}, {Maurogordato}, {Medinaceli}, {Mei}, {Melchior}, {Mellier}, {Meneghetti}, {Merlin}, {Meylan}, {Mora}, {Moresco}, {Moscardini}, {Nakajima}, {Neissner}, {Niemi}, {Nightingale}, {Padilla}, {Paltani}, {Pasian}, {Pedersen}, {Percival}, {Pettorino}, {Pires}, {Polenta}, {Poncet}, {Popa}, {Pozzetti}, {Raison}, {Rebolo}, {Renzi}, {Rhodes}, {Riccio}, {Romelli}, {Roncarelli}, {Rossetti}, {Rottgering}, {Rusholme}, {Saglia}, {Sakr}, {Sapone},
  {Sartoris}, {Schewtschenko}, {Schirmer}, {Schneider}, {Schrabback}, {Scodeggio}, {Secroun}, {Seidel}, {Serrano}, {Simon}, {Sirignano}, {Sirri}, {Stanco}, {Steinwagner}, {Tallada-Cresp{\'\i}}, {Taylor}, {Teplitz}, {Tereno}, {Toft}, {Toledo-Moreo}, {Torradeflot}, {Tutusaus}, {Valenziano}, {Valiviita}, {Vassallo}, {Verdoes Kleijn}, {Veropalumbo}, \& {Wang}}]{Bisigello25}
{Bisigello}, L., {Euclid Collaboration}, {Rodighiero}, G., {et~al.} 2025, arXiv e-prints, arXiv:2503.15323, \dodoi{10.48550/arXiv.2503.15323}

\bibitem[{{Bogd{\'a}n} {et~al.}(2024){Bogd{\'a}n}, {Goulding}, {Natarajan}, {Kov{\'a}cs}, {Tremblay}, {Chadayammuri}, {Volonteri}, {Kraft}, {Forman}, {Jones}, {Churazov}, \& {Zhuravleva}}]{Bogdan24}
{Bogd{\'a}n}, {\'A}., {Goulding}, A.~D., {Natarajan}, P., {et~al.} 2024, Nature Astronomy, 8, 126, \dodoi{10.1038/s41550-023-02111-9}

\bibitem[{{Boquien} {et~al.}(2019){Boquien}, {Burgarella}, {Roehlly}, {Buat}, {Ciesla}, {Corre}, {Inoue}, \& {Salas}}]{Boquien19}
{Boquien}, M., {Burgarella}, D., {Roehlly}, Y., {et~al.} 2019, \aap, 622, A103, \dodoi{10.1051/0004-6361/201834156}

\bibitem[{Boyle \& Terlevich(1998)}]{Boyle_Terlevich98}
Boyle, B.~J., \& Terlevich, R.~J. 1998, Monthly Notices of the Royal Astronomical Society, 293, L49, \dodoi{10.1111/j.1365-8711.1998.01264.x}

\bibitem[{{Brammer} {et~al.}(2012){Brammer}, {van Dokkum}, {Franx}, {Fumagalli}, {Patel}, {Rix}, {Skelton}, {Kriek}, {Nelson}, {Schmidt}, {Bezanson}, {da Cunha}, {Erb}, {Fan}, {F{\"o}rster Schreiber}, {Illingworth}, {Labb{\'e}}, {Leja}, {Lundgren}, {Magee}, {Marchesini}, {McCarthy}, {Momcheva}, {Muzzin}, {Quadri}, {Steidel}, {Tal}, {Wake}, {Whitaker}, \& {Williams}}]{brammer12}
{Brammer}, G.~B., {van Dokkum}, P.~G., {Franx}, M., {et~al.} 2012, \apjs, 200, 13, \dodoi{10.1088/0067-0049/200/2/13}

\bibitem[{{Bruzual} \& {Charlot}(2003)}]{BC03}
{Bruzual}, G., \& {Charlot}, S. 2003, \mnras, 344, 1000, \dodoi{10.1046/j.1365-8711.2003.06897.x}

\bibitem[{{Buchner} {et~al.}(2024){Buchner}, {Starck}, {Salvato}, {Netzer}, {Igo}, {Laloux}, {Georgakakis}, {Gauger}, {Olechowska}, {Lopez}, {Shankar}, {Li}, {Nandra}, \& {Merloni}}]{Buchner24}
{Buchner}, J., {Starck}, H., {Salvato}, M., {et~al.} 2024, \aap, 692, A161, \dodoi{10.1051/0004-6361/202449372}

\bibitem[{{Bunker} {et~al.}(2023){Bunker}, {Saxena}, {Cameron}, {Willott}, {Curtis-Lake}, {Jakobsen}, {Carniani}, {Smit}, {Maiolino}, {Witstok}, {Curti}, {D'Eugenio}, {Jones}, {Ferruit}, {Arribas}, {Charlot}, {Chevallard}, {Giardino}, {de Graaff}, {Looser}, {L{\"u}tzgendorf}, {Maseda}, {Rawle}, {Rix}, {Del Pino}, {Alberts}, {Egami}, {Eisenstein}, {Endsley}, {Hainline}, {Hausen}, {Johnson}, {Rieke}, {Rieke}, {Robertson}, {Shivaei}, {Stark}, {Sun}, {Tacchella}, {Tang}, {Williams}, {Willmer}, {Baker}, {Baum}, {Bhatawdekar}, {Bowler}, {Boyett}, {Chen}, {Circosta}, {Helton}, {Ji}, {Kumari}, {Lyu}, {Nelson}, {Parlanti}, {Perna}, {Sandles}, {Scholtz}, {Suess}, {Topping}, {{\"U}bler}, {Wallace}, \& {Whitler}}]{Bunker23}
{Bunker}, A.~J., {Saxena}, A., {Cameron}, A.~J., {et~al.} 2023, \aap, 677, A88, \dodoi{10.1051/0004-6361/202346159}

\bibitem[{{Caplar} {et~al.}(2018){Caplar}, {Lilly}, \& {Trakhtenbrot}}]{Caplar18}
{Caplar}, N., {Lilly}, S.~J., \& {Trakhtenbrot}, B. 2018, \apj, 867, 148, \dodoi{10.3847/1538-4357/aae691}

\bibitem[{{Casertano} {et~al.}(2000){Casertano}, {de Mello}, {Dickinson}, {Ferguson}, {Fruchter}, {Gonzalez-Lopezlira}, {Heyer}, {Hook}, {Levay}, {Lucas}, {Mack}, {Makidon}, {Mutchler}, {Smith}, {Stiavelli}, {Wiggs}, \& {Williams}}]{Casertano00}
{Casertano}, S., {de Mello}, D., {Dickinson}, M., {et~al.} 2000, \aj, 120, 2747, \dodoi{10.1086/316851}

\bibitem[{{Chabrier}(2003)}]{chabrier03}
{Chabrier}, G. 2003, Publications of the Astronomical Society of the Pacific, 115, 763, \dodoi{10.1086/376392}

\bibitem[{{Ciesla} {et~al.}(2015){Ciesla}, {Charmandaris}, {Georgakakis}, {Bernhard}, {Mitchell}, {Buat}, {Elbaz}, {LeFloc'h}, {Lacey}, {Magdis}, \& {Xilouris}}]{Ciesla_2015}
{Ciesla}, L., {Charmandaris}, V., {Georgakakis}, A., {et~al.} 2015, \aap, 576, A10, \dodoi{10.1051/0004-6361/201425252}

\bibitem[{{Dattathri} {et~al.}(2025){Dattathri}, {Natarajan}, {Porras-Valverde}, {Burke}, {Chen}, {Di Matteo}, \& {Ni}}]{Dattathri_2025}
{Dattathri}, S., {Natarajan}, P., {Porras-Valverde}, A.~J., {et~al.} 2025, \apj, 984, 122, \dodoi{10.3847/1538-4357/adbeef}

\bibitem[{{de Graaff} {et~al.}(2025){de Graaff}, {Brammer}, {Weibel}, {Lewis}, {Maseda}, {Oesch}, {Bezanson}, {Boogaard}, {Cleri}, {Cooper}, {Gottumukkala}, {Greene}, {Hirschmann}, {Hviding}, {Katz}, {Labb{\'e}}, {Leja}, {Matthee}, {McConachie}, {Miller}, {Naidu}, {Price}, {Rix}, {Setton}, {Suess}, {Wang}, {Whitaker}, \& {Williams}}]{RUBIES_2025}
{de Graaff}, A., {Brammer}, G., {Weibel}, A., {et~al.} 2025, \aap, 697, A189, \dodoi{10.1051/0004-6361/202452186}

\bibitem[{{Decarli} {et~al.}(2010){Decarli}, {Falomo}, {Treves}, {Labita}, {Kotilainen}, \& {Scarpa}}]{Decarli10}
{Decarli}, R., {Falomo}, R., {Treves}, A., {et~al.} 2010, \mnras, 402, 2453, \dodoi{10.1111/j.1365-2966.2009.16049.x}

\bibitem[{{Dekel} {et~al.}(2023){Dekel}, {Sarkar}, {Birnboim}, {Mandelker}, \& {Li}}]{Dekel_2023}
{Dekel}, A., {Sarkar}, K.~C., {Birnboim}, Y., {Mandelker}, N., \& {Li}, Z. 2023, \mnras, 523, 3201, \dodoi{10.1093/mnras/stad1557}

\bibitem[{{Dekel} {et~al.}(2025){Dekel}, {Stone}, {Chowdhury}, {Gilbaum}, {Li}, {Mandelker}, \& {van den Bosch}}]{Dekel_2025_MBH}
{Dekel}, A., {Stone}, N.~C., {Chowdhury}, D.~D., {et~al.} 2025, \aap, 695, A97, \dodoi{10.1051/0004-6361/202452393}

\bibitem[{{Dunlop} {et~al.}(2021){Dunlop}, {Abraham}, {Ashby}, {Bagley}, {Best}, {Bongiorno}, {Bouwens}, {Bowler}, {Brammer}, {Bremer}, {Calabro'}, {Carnall}, {Castellano}, {Cirasuolo}, {Conselice}, {Cullen}, {Dave}, {Dayal}, {Dekel}, {Dickinson}, {Duncan}, {Elbaz}, {Ellis}, {Ferguson}, {Ferrara}, {Finkelstein}, {Fontana}, {Furlanetto}, {Fynbo}, {Gallerani}, {Gardner}, {Giavalisco}, {Grazian}, {Grogin}, {Harikane}, {Hopkins}, {Ilbert}, {Illingworth}, {Juneau}, {Jung}, {Kartaltepe}, {Kassin}, {Kauffmann}, {Khochfar}, {Kirkpatrick}, {Kocevski}, {Koekemoer}, {Labbe}, {Laporte}, {Larson}, {Lucas}, {Magee}, {Mason}, {McCracken}, {McLeod}, {McLure}, {Merlin}, {Mesinger}, {Milvang-Jensen}, {Newman}, {Oesch}, {Ouchi}, {Pacifici}, {Papovich}, {Peacock}, {Peeples}, {Pentericci}, {Perez-Gonzalez}, {Pirzkal}, {Pope}, {Pye}, {Reddy}, {Robertson}, {Salvato}, {Santini}, {Schaerer}, {Shapley}, {Simons}, {Smit}, {Smith}, {Snyder}, {Somerville}, {Stanway}, {Stefanon}, {Tasca}, {Tikkanen}, {Tresse}, {Trump}, {Whitaker},
  {Wilkins}, {Wright}, {Wyithe}, {van Dokkum}, \& {van der Werf}}]{Dunlop21}
{Dunlop}, J.~S., {Abraham}, R.~G., {Ashby}, M. L.~N., {et~al.} 2021, {PRIMER: Public Release IMaging for Extragalactic Research}, JWST Proposal. Cycle 1, ID. \#1837

\bibitem[{{Durodola} {et~al.}(2025){Durodola}, {Pacucci}, \& {Hickox}}]{Durodola_2025}
{Durodola}, E., {Pacucci}, F., \& {Hickox}, R.~C. 2025, \apj, 985, 169, \dodoi{10.3847/1538-4357/adced2}

\bibitem[{{Eisenstein} {et~al.}(2023){Eisenstein}, {Willott}, {Alberts}, {Arribas}, {Bonaventura}, {Bunker}, {Cameron}, {Carniani}, {Charlot}, {Curtis-Lake}, {D'Eugenio}, {Endsley}, {Ferruit}, {Giardino}, {Hainline}, {Hausen}, {Jakobsen}, {Johnson}, {Maiolino}, {Rieke}, {Rieke}, {Rix}, {Robertson}, {Stark}, {Tacchella}, {Williams}, {Willmer}, {Baker}, {Baum}, {Bhatawdekar}, {Boyett}, {Chen}, {Chevallard}, {Circosta}, {Curti}, {Danhaive}, {DeCoursey}, {de Graaff}, {Dressler}, {Egami}, {Helton}, {Hviding}, {Ji}, {Jones}, {Kumari}, {L{\"u}tzgendorf}, {Laseter}, {Looser}, {Lyu}, {Maseda}, {Nelson}, {Parlanti}, {Perna}, {Pusk{\'a}s}, {Rawle}, {Rodr{\'\i}guez Del Pino}, {Sandles}, {Saxena}, {Scholtz}, {Sharpe}, {Shivaei}, {Silcock}, {Simmonds}, {Skarbinski}, {Smit}, {Stone}, {Suess}, {Sun}, {Tang}, {Topping}, {{\"U}bler}, {Villanueva}, {Wallace}, {Whitler}, {Witstok}, \& {Woodrum}}]{Eisenstein_JADES_2023}
{Eisenstein}, D.~J., {Willott}, C., {Alberts}, S., {et~al.} 2023, arXiv e-prints, arXiv:2306.02465, \dodoi{10.48550/arXiv.2306.02465}

\bibitem[{{Ferrarese} \& {Merritt}(2000)}]{Ferrarese00}
{Ferrarese}, L., \& {Merritt}, D. 2000, \apjl, 539, L9, \dodoi{10.1086/312838}

\bibitem[{{Finkelstein} {et~al.}(2024){Finkelstein}, {Leung}, {Bagley}, {Dickinson}, {Ferguson}, {Papovich}, {Akins}, {Arrabal Haro}, {Dav{\'e}}, {Dekel}, {Kartaltepe}, {Kocevski}, {Koekemoer}, {Pirzkal}, {Somerville}, {Yung}, {Amor{\'\i}n}, {Backhaus}, {Behroozi}, {Bisigello}, {Bromm}, {Casey}, {Ch{\'a}vez Ortiz}, {Cheng}, {Chworowsky}, {Cleri}, {Cooper}, {Davis}, {de la Vega}, {Elbaz}, {Franco}, {Fontana}, {Fujimoto}, {Giavalisco}, {Grogin}, {Holwerda}, {Huertas-Company}, {Hirschmann}, {Iyer}, {Jogee}, {Jung}, {Larson}, {Lucas}, {Mobasher}, {Morales}, {Morley}, {Mukherjee}, {P{\'e}rez-Gonz{\'a}lez}, {Ravindranath}, {Rodighiero}, {Rowland}, {Tacchella}, {Taylor}, {Trump}, \& {Wilkins}}]{Finkelstein24}
{Finkelstein}, S.~L., {Leung}, G. C.~K., {Bagley}, M.~B., {et~al.} 2024, \apjl, 969, L2, \dodoi{10.3847/2041-8213/ad4495}

\bibitem[{{Finkelstein} {et~al.}(2025{\natexlab{a}}){Finkelstein}, {Bagley}, {Arrabal Haro}, {Dickinson}, {Ferguson}, {Kartaltepe}, {Kocevski}, {Koekemoer}, {Lotz}, {Papovich}, {P{\'e}rez-Gonz{\'a}lez}, {Pirzkal}, {Somerville}, {Trump}, {Yang}, {Yung}, {Fontana}, {Grazian}, {Grogin}, {Kewley}, {Kirkpatrick}, {Larson}, {Pentericci}, {Ravindranath}, {Wilkins}, {Almaini}, {Amor{\'\i}n}, {Barro}, {Bhatawdekar}, {Bisigello}, {Brooks}, {Buat}, {Buitrago}, {Burgarella}, {Calabr{\`o}}, {Castellano}, {Cheng}, {Cleri}, {Cole}, {Cooper}, {Cooper}, {Costantin}, {Cox}, {Croton}, {Daddi}, {Davis}, {Dekel}, {Elbaz}, {Fern{\'a}ndez}, {Fujimoto}, {Gandolfi}, {Gardner}, {Gawiser}, {Giavalisco}, {G{\'o}mez-Guijarro}, {Guo}, {Gupta}, {Hathi}, {Harish}, {Henry}, {Hirschmann}, {Hu}, {Hutchison}, {Iyer}, {Jaskot}, {Jha}, {Jung}, {Kassin}, {Kokorev}, {Kurczynski}, {Leung}, {Llerena}, {Long}, {Lucas}, {Lu}, {McGrath}, {McIntosh}, {Merlin}, {Mobasher}, {Morales}, {Napolitano}, {Pacucci}, {Pandya}, {Rafelski}, {Rodighiero}, {Rose},
  {Santini}, {Seill{\'e}}, {Simons}, {Shen}, {Straughn}, {Tacchella}, {Taylor}, {Vanderhoof}, {Vega-Ferrero}, {Weiner}, {Willmer}, {Zhu}, {Bell}, {Wuyts}, {Holwerda}, {Wang}, {Wang}, {Zavala}, \& {CEERS Collaboration}}]{Finkelstein25}
{Finkelstein}, S.~L., {Bagley}, M.~B., {Arrabal Haro}, P., {et~al.} 2025{\natexlab{a}}, \apjl, 983, L4, \dodoi{10.3847/2041-8213/adbbd3}

\bibitem[{{Finkelstein} {et~al.}(2025{\natexlab{b}}){Finkelstein}, {Bagley}, {Arrabal Haro}, {Dickinson}, {Ferguson}, {Kartaltepe}, {Kocevski}, {Koekemoer}, {Lotz}, {Papovich}, {P{\'e}rez-Gonz{\'a}lez}, {Pirzkal}, {Somerville}, {Trump}, {Yang}, {Yung}, {Fontana}, {Grazian}, {Grogin}, {Kewley}, {Kirkpatrick}, {Larson}, {Pentericci}, {Ravindranath}, {Wilkins}, {Almaini}, {Amor{\'\i}n}, {Barro}, {Bhatawdekar}, {Bisigello}, {Brooks}, {Buat}, {Buitrago}, {Burgarella}, {Calabr{\`o}}, {Castellano}, {Cheng}, {Cleri}, {Cole}, {Cooper}, {Cooper}, {Costantin}, {Cox}, {Croton}, {Daddi}, {Davis}, {Dekel}, {Elbaz}, {Fern{\'a}ndez}, {Fujimoto}, {Gandolfi}, {Gardner}, {Gawiser}, {Giavalisco}, {G{\'o}mez-Guijarro}, {Guo}, {Gupta}, {Hathi}, {Harish}, {Henry}, {Hirschmann}, {Hu}, {Hutchison}, {Iyer}, {Jaskot}, {Jha}, {Jung}, {Kassin}, {Kokorev}, {Kurczynski}, {Leung}, {Llerena}, {Long}, {Lucas}, {Lu}, {McGrath}, {McIntosh}, {Merlin}, {Mobasher}, {Morales}, {Napolitano}, {Pacucci}, {Pandya}, {Rafelski}, {Rodighiero}, {Rose},
  {Santini}, {Seill{\'e}}, {Simons}, {Shen}, {Straughn}, {Tacchella}, {Taylor}, {Vanderhoof}, {Vega-Ferrero}, {Weiner}, {Willmer}, {Zhu}, {Bell}, {Wuyts}, {Holwerda}, {Wang}, {Wang}, {Zavala}, \& {CEERS Collaboration}}]{Finkelstein_2025}
---. 2025{\natexlab{b}}, \apjl, 983, L4, \dodoi{10.3847/2041-8213/adbbd3}

\bibitem[{{Fragione} \& {Pacucci}(2023)}]{Fragione_2023}
{Fragione}, G., \& {Pacucci}, F. 2023, \apjl, 958, L24, \dodoi{10.3847/2041-8213/ad09e5}

\bibitem[{{Fruchter} \& {Hook}(2002)}]{fruchter02}
{Fruchter}, A.~S., \& {Hook}, R.~N. 2002, \pasp, 114, 144, \dodoi{10.1086/338393}

\bibitem[{{Furtak} {et~al.}(2022){Furtak}, {Zitrin}, {Plat}, {Fujimoto}, {Wang}, {Nelson}, {Labb{\'e}}, {Bezanson}, {Brammer}, {van Dokkum}, {Endsley}, {Glazebrook}, {Greene}, {Leja}, {Price}, {Smit}, {Stark}, {Weaver}, {Whitaker}, {Atek}, {Chevallard}, {Curtis-Lake}, {Dayal}, {Franx}, {Feltre}, {Fudamoto}, {Marchesini}, {Mowla}, {Pan}, {Suess}, {Vidal-Garc{\'\i}a}, \& {Williams}}]{Furtak22}
{Furtak}, L.~J., {Zitrin}, A., {Plat}, A., {et~al.} 2022, arXiv e-prints, arXiv:2212.10531, \dodoi{10.48550/arXiv.2212.10531}

\bibitem[{{Furtak} {et~al.}(2024){Furtak}, {Labb{\'e}}, {Zitrin}, {Greene}, {Dayal}, {Chemerynska}, {Kokorev}, {Miller}, {Goulding}, {de Graaff}, {Bezanson}, {Brammer}, {Cutler}, {Leja}, {Pan}, {Price}, {Wang}, {Weaver}, {Whitaker}, {Atek}, {Bogd{\'a}n}, {Charlot}, {Curtis-Lake}, {van Dokkum}, {Endsley}, {Feldmann}, {Fudamoto}, {Fujimoto}, {Glazebrook}, {Juneau}, {Marchesini}, {Maseda}, {Nelson}, {Oesch}, {Plat}, {Setton}, {Stark}, \& {Williams}}]{Furtak24}
{Furtak}, L.~J., {Labb{\'e}}, I., {Zitrin}, A., {et~al.} 2024, \nat, 628, 57, \dodoi{10.1038/s41586-024-07184-8}

\bibitem[{{Gaia Collaboration} {et~al.}(2016){Gaia Collaboration}, {Prusti}, {de Bruijne}, {Brown}, {Vallenari}, {Babusiaux}, {Bailer-Jones}, {Bastian}, {Biermann}, {Evans}, {Eyer}, {Jansen}, {Jordi}, {Klioner}, {Lammers}, {Lindegren}, {Luri}, {Mignard}, {Milligan}, {Panem}, {Poinsignon}, {Pourbaix}, {Randich}, {Sarri}, {Sartoretti}, {Siddiqui}, {Soubiran}, {Valette}, {van Leeuwen}, {Walton}, {Aerts}, {Arenou}, {Cropper}, {Drimmel}, {H{\o}g}, {Katz}, {Lattanzi}, {O'Mullane}, {Grebel}, {Holland}, {Huc}, {Passot}, {Bramante}, {Cacciari}, {Casta{\~n}eda}, {Chaoul}, {Cheek}, {De Angeli}, {Fabricius}, {Guerra}, {Hern{\'a}ndez}, {Jean-Antoine-Piccolo}, {Masana}, {Messineo}, {Mowlavi}, {Nienartowicz}, {Ord{\'o}{\~n}ez-Blanco}, {Panuzzo}, {Portell}, {Richards}, {Riello}, {Seabroke}, {Tanga}, {Th{\'e}venin}, {Torra}, {Els}, {Gracia-Abril}, {Comoretto}, {Garcia-Reinaldos}, {Lock}, {Mercier}, {Altmann}, {Andrae}, {Astraatmadja}, {Bellas-Velidis}, {Benson}, {Berthier}, {Blomme}, {Busso}, {Carry}, {Cellino}, {Clementini},
  {Cowell}, {Creevey}, {Cuypers}, {Davidson}, {De Ridder}, {de Torres}, {Delchambre}, {Dell'Oro}, {Ducourant}, {Fr{\'e}mat}, {Garc{\'\i}a-Torres}, {Gosset}, {Halbwachs}, {Hambly}, {Harrison}, {Hauser}, {Hestroffer}, {Hodgkin}, {Huckle}, {Hutton}, {Jasniewicz}, {Jordan}, {Kontizas}, {Korn}, {Lanzafame}, {Manteiga}, {Moitinho}, {Muinonen}, {Osinde}, {Pancino}, {Pauwels}, {Petit}, {Recio-Blanco}, {Robin}, {Sarro}, {Siopis}, {Smith}, {Smith}, {Sozzetti}, {Thuillot}, {van Reeven}, {Viala}, {Abbas}, {Abreu Aramburu}, {Accart}, {Aguado}, {Allan}, {Allasia}, {Altavilla}, {{\'A}lvarez}, {Alves}, {Anderson}, {Andrei}, {Anglada Varela}, {Antiche}, {Antoja}, {Ant{\'o}n}, {Arcay}, {Atzei}, {Ayache}, {Bach}, {Baker}, {Balaguer-N{\'u}{\~n}ez}, {Barache}, {Barata}, {Barbier}, {Barblan}, {Baroni}, {Barrado y Navascu{\'e}s}, {Barros}, {Barstow}, {Becciani}, {Bellazzini}, {Bellei}, {Bello Garc{\'\i}a}, {Belokurov}, {Bendjoya}, {Berihuete}, {Bianchi}, {Bienaym{\'e}}, {Billebaud}, {Blagorodnova}, {Blanco-Cuaresma}, {Boch},
  {Bombrun}, {Borrachero}, {Bouquillon}, {Bourda}, {Bouy}, {Bragaglia}, {Breddels}, {Brouillet}, {Br{\"u}semeister}, {Bucciarelli}, {Budnik}, {Burgess}, {Burgon}, {Burlacu}, {Busonero}, {Buzzi}, {Caffau}, {Cambras}, {Campbell}, {Cancelliere}, {Cantat-Gaudin}, {Carlucci}, {Carrasco}, {Castellani}, {Charlot}, {Charnas}, {Charvet}, {Chassat}, {Chiavassa}, {Clotet}, {Cocozza}, {Collins}, {Collins}, {Costigan}, {Crifo}, {Cross}, {Crosta}, {Crowley}, {Dafonte}, {Damerdji}, {Dapergolas}, {David}, {David}, {De Cat}, {de Felice}, {de Laverny}, {De Luise}, {De March}, {de Martino}, {de Souza}, {Debosscher}, {del Pozo}, {Delbo}, {Delgado}, {Delgado}, {di Marco}, {Di Matteo}, {Diakite}, {Distefano}, {Dolding}, {Dos Anjos}, {Drazinos}, {Dur{\'a}n}, {Dzigan}, {Ecale}, {Edvardsson}, {Enke}, {Erdmann}, {Escolar}, {Espina}, {Evans}, {Eynard Bontemps}, {Fabre}, {Fabrizio}, {Faigler}, {Falc{\~a}o}, {Farr{\`a}s Casas}, {Faye}, {Federici}, {Fedorets}, {Fern{\'a}ndez-Hern{\'a}ndez}, {Fernique}, {Fienga}, {Figueras}, {Filippi},
  {Findeisen}, {Fonti}, {Fouesneau}, {Fraile}, {Fraser}, {Fuchs}, {Furnell}, {Gai}, {Galleti}, {Galluccio}, {Garabato}, {Garc{\'\i}a-Sedano}, {Gar{\'e}}, {Garofalo}, {Garralda}, {Gavras}, {Gerssen}, {Geyer}, {Gilmore}, {Girona}, {Giuffrida}, {Gomes}, {Gonz{\'a}lez-Marcos}, {Gonz{\'a}lez-N{\'u}{\~n}ez}, {Gonz{\'a}lez-Vidal}, {Granvik}, {Guerrier}, {Guillout}, {Guiraud}, {G{\'u}rpide}, {Guti{\'e}rrez-S{\'a}nchez}, {Guy}, {Haigron}, {Hatzidimitriou}, {Haywood}, {Heiter}, {Helmi}, {Hobbs}, {Hofmann}, {Holl}, {Holland}, {Hunt}, {Hypki}, {Icardi}, {Irwin}, {Jevardat de Fombelle}, {Jofr{\'e}}, {Jonker}, {Jorissen}, {Julbe}, {Karampelas}, {Kochoska}, {Kohley}, {Kolenberg}, {Kontizas}, {Koposov}, {Kordopatis}, {Koubsky}, {Kowalczyk}, {Krone-Martins}, {Kudryashova}, {Kull}, {Bachchan}, {Lacoste-Seris}, {Lanza}, {Lavigne}, {Le Poncin-Lafitte}, {Lebreton}, {Lebzelter}, {Leccia}, {Leclerc}, {Lecoeur-Taibi}, {Lemaitre}, {Lenhardt}, {Leroux}, {Liao}, {Licata}, {Lindstr{\o}m}, {Lister}, {Livanou}, {Lobel}, {L{\"o}ffler},
  {L{\'o}pez}, {Lopez-Lozano}, {Lorenz}, {Loureiro}, {MacDonald}, {Magalh{\~a}es Fernandes}, {Managau}, {Mann}, {Mantelet}, {Marchal}, {Marchant}, {Marconi}, {Marie}, {Marinoni}, {Marrese}, {Marschalk{\'o}}, {Marshall}, {Mart{\'\i}n-Fleitas}, {Martino}, {Mary}, {Matijevi{\v{c}}}, {Mazeh}, {McMillan}, {Messina}, {Mestre}, {Michalik}, {Millar}, {Miranda}, {Molina}, {Molinaro}, {Molinaro}, {Moln{\'a}r}, {Moniez}, {Montegriffo}, {Monteiro}, {Mor}, {Mora}, {Morbidelli}, {Morel}, {Morgenthaler}, {Morley}, {Morris}, {Mulone}, {Muraveva}, {Musella}, {Narbonne}, {Nelemans}, {Nicastro}, {Noval}, {Ord{\'e}novic}, {Ordieres-Mer{\'e}}, {Osborne}, {Pagani}, {Pagano}, {Pailler}, {Palacin}, {Palaversa}, {Parsons}, {Paulsen}, {Pecoraro}, {Pedrosa}, {Pentik{\"a}inen}, {Pereira}, {Pichon}, {Piersimoni}, {Pineau}, {Plachy}, {Plum}, {Poujoulet}, {Pr{\v{s}}a}, {Pulone}, {Ragaini}, {Rago}, {Rambaux}, {Ramos-Lerate}, {Ranalli}, {Rauw}, {Read}, {Regibo}, {Renk}, {Reyl{\'e}}, {Ribeiro}, {Rimoldini}, {Ripepi}, {Riva}, {Rixon},
  {Roelens}, {Romero-G{\'o}mez}, {Rowell}, {Royer}, {Rudolph}, {Ruiz-Dern}, {Sadowski}, {Sagrist{\`a} Sell{\'e}s}, {Sahlmann}, {Salgado}, {Salguero}, {Sarasso}, {Savietto}, {Schnorhk}, {Schultheis}, {Sciacca}, {Segol}, {Segovia}, {Segransan}, {Serpell}, {Shih}, {Smareglia}, {Smart}, {Smith}, {Solano}, {Solitro}, {Sordo}, {Soria Nieto}, {Souchay}, {Spagna}, {Spoto}, {Stampa}, {Steele}, {Steidelm{\"u}ller}, {Stephenson}, {Stoev}, {Suess}, {S{\"u}veges}, {Surdej}, {Szabados}, {Szegedi-Elek}, {Tapiador}, {Taris}, {Tauran}, {Taylor}, {Teixeira}, {Terrett}, {Tingley}, {Trager}, {Turon}, {Ulla}, {Utrilla}, {Valentini}, {van Elteren}, {Van Hemelryck}, {van Leeuwen}, {Varadi}, {Vecchiato}, {Veljanoski}, {Via}, {Vicente}, {Vogt}, {Voss}, {Votruba}, {Voutsinas}, {Walmsley}, {Weiler}, {Weingrill}, {Werner}, {Wevers}, {Whitehead}, {Wyrzykowski}, {Yoldas}, {{\v{Z}}erjal}, {Zucker}, {Zurbach}, {Zwitter}, {Alecu}, {Allen}, {Allende Prieto}, {Amorim}, {Anglada-Escud{\'e}}, {Arsenijevic}, {Azaz}, {Balm}, {Beck}, {Bernstein},
  {Bigot}, {Bijaoui}, {Blasco}, {Bonfigli}, {Bono}, {Boudreault}, {Bressan}, {Brown}, {Brunet}, {Bunclark}, {Buonanno}, {Butkevich}, {Carret}, {Carrion}, {Chemin}, {Ch{\'e}reau}, {Corcione}, {Darmigny}, {de Boer}, {de Teodoro}, {de Zeeuw}, {Delle Luche}, {Domingues}, {Dubath}, {Fodor}, {Fr{\'e}zouls}, {Fries}, {Fustes}, {Fyfe}, {Gallardo}, {Gallegos}, {Gardiol}, {Gebran}, {Gomboc}, {G{\'o}mez}, {Grux}, {Gueguen}, {Heyrovsky}, {Hoar}, {Iannicola}, {Isasi Parache}, {Janotto}, {Joliet}, {Jonckheere}, {Keil}, {Kim}, {Klagyivik}, {Klar}, {Knude}, {Kochukhov}, {Kolka}, {Kos}, {Kutka}, {Lainey}, {LeBouquin}, {Liu}, {Loreggia}, {Makarov}, {Marseille}, {Martayan}, {Martinez-Rubi}, {Massart}, {Meynadier}, {Mignot}, {Munari}, {Nguyen}, {Nordlander}, {Ocvirk}, {O'Flaherty}, {Olias Sanz}, {Ortiz}, {Osorio}, {Oszkiewicz}, {Ouzounis}, {Palmer}, {Park}, {Pasquato}, {Peltzer}, {Peralta}, {P{\'e}turaud}, {Pieniluoma}, {Pigozzi}, {Poels}, {Prat}, {Prod'homme}, {Raison}, {Rebordao}, {Risquez}, {Rocca-Volmerange}, {Rosen},
  {Ruiz-Fuertes}, {Russo}, {Sembay}, {Serraller Vizcaino}, {Short}, {Siebert}, {Silva}, {Sinachopoulos}, {Slezak}, {Soffel}, {Sosnowska}, {Strai{\v{z}}ys}, {ter Linden}, {Terrell}, {Theil}, {Tiede}, {Troisi}, {Tsalmantza}, {Tur}, {Vaccari}, {Vachier}, {Valles}, {Van Hamme}, {Veltz}, {Virtanen}, {Wallut}, {Wichmann}, {Wilkinson}, {Ziaeepour}, \& {Zschocke}}]{Gaia16}
{Gaia Collaboration}, {Prusti}, T., {de Bruijne}, J.~H.~J., {et~al.} 2016, \aap, 595, A1, \dodoi{10.1051/0004-6361/201629272}

\bibitem[{{Galametz} {et~al.}(2013){Galametz}, {Grazian}, {Fontana}, {Ferguson}, {Ashby}, {Barro}, {Castellano}, {Dahlen}, {Donley}, {Faber}, {Grogin}, {Guo}, {Huang}, {Kocevski}, {Koekemoer}, {Lee}, {McGrath}, {Peth}, {Willner}, {Almaini}, {Cooper}, {Cooray}, {Conselice}, {Dickinson}, {Dunlop}, {Fazio}, {Foucaud}, {Gardner}, {Giavalisco}, {Hathi}, {Hartley}, {Koo}, {Lai}, {de Mello}, {McLure}, {Lucas}, {Paris}, {Pentericci}, {Santini}, {Simpson}, {Sommariva}, {Targett}, {Weiner}, {Wuyts}, \& {CANDELS Team}}]{Galametz13}
{Galametz}, A., {Grazian}, A., {Fontana}, A., {et~al.} 2013, \apjs, 206, 10, \dodoi{10.1088/0067-0049/206/2/10}

\bibitem[{{Gebhardt} {et~al.}(2000){Gebhardt}, {Bender}, {Bower}, {Dressler}, {Faber}, {Filippenko}, {Green}, {Grillmair}, {Ho}, {Kormendy}, {Lauer}, {Magorrian}, {Pinkney}, {Richstone}, \& {Tremaine}}]{Gebhardt00}
{Gebhardt}, K., {Bender}, R., {Bower}, G., {et~al.} 2000, \apjl, 539, L13, \dodoi{10.1086/312840}

\bibitem[{{Geris} {et~al.}(2025){Geris}, {Maiolino}, {Isobe}, {Scholtz}, {D'Eugenio}, {Ji}, {Juodzbalis}, {Simmonds}, {Dayal}, {Trinca}, {Schneider}, {Arribas}, {Bhatawdekar}, {Bunker}, {Carniani}, {Charlot}, {Chevallard}, {Curtis-Lake}, {Johnson}, {Parlanti}, {Rinaldi}, {Robertson}, {Tacchella}, {Uebler}, {Venturi}, {Williams}, \& {Witstok}}]{Geris_2025}
{Geris}, S., {Maiolino}, R., {Isobe}, Y., {et~al.} 2025, arXiv e-prints, arXiv:2506.22147, \dodoi{10.48550/arXiv.2506.22147}

\bibitem[{{Greene} \& {Ho}(2005)}]{Greene_2005}
{Greene}, J.~E., \& {Ho}, L.~C. 2005, \apj, 630, 122, \dodoi{10.1086/431897}

\bibitem[{{Greene} {et~al.}(2023){Greene}, {Labbe}, {Goulding}, {Furtak}, {Chemerynska}, {Kokorev}, {Dayal}, {Williams}, {Wang}, {Setton}, {Burgasser}, {Bezanson}, {Atek}, {Brammer}, {Cutler}, {Feldmann}, {Fujimoto}, {Glazebrook}, {de Graaff}, {Leja}, {Marchesini}, {Maseda}, {Matthee}, {Miller}, {Naidu}, {Nanayakkara}, {Oesch}, {Pan}, {Papovich}, {Price}, {van Dokkum}, {Weaver}, {Whitaker}, \& {Zitrin}}]{Greene_2023}
{Greene}, J.~E., {Labbe}, I., {Goulding}, A.~D., {et~al.} 2023, arXiv e-prints, arXiv:2309.05714, \dodoi{10.48550/arXiv.2309.05714}

\bibitem[{{Greene} {et~al.}(2025){Greene}, {Setton}, {Furtak}, {Naidu}, {Volonteri}, {Dayal}, {Labbe}, {van Dokkum}, {Bezanson}, {Brammer}, {Cutler}, {Glazebrook}, {de Graaff}, {Hirschmann}, {Hviding}, {Kokorev}, {Leja}, {Liu}, {Ma}, {Matthee}, {Nanayakkara}, {Oesch}, {Pan}, {Price}, {Spilker}, {Wang}, {Weaver}, {Whitaker}, {Williams}, \& {Zitrin}}]{Greene25}
{Greene}, J.~E., {Setton}, D.~J., {Furtak}, L.~J., {et~al.} 2025, arXiv e-prints, arXiv:2509.05434, \dodoi{10.48550/arXiv.2509.05434}

\bibitem[{{Grogin} {et~al.}(2011){Grogin}, {Kocevski}, {Faber}, {Ferguson}, {Koekemoer}, {Riess}, {Acquaviva}, {Alexander}, {Almaini}, {Ashby}, {Barden}, {Bell}, {Bournaud}, {Brown}, {Caputi}, {Casertano}, {Cassata}, {Challis}, {Chary}, {Cheung}, {Cirasuolo}, {Conselice}, {Roshan Cooray}, {Croton}, {Daddi}, {Dahlen}, {Dav{\'e}}, {de Mello}, {Dekel}, {Dickinson}, {Dolch}, {Donley}, {Dunlop}, {Dutton}, {Elbaz}, {Fazio}, {Filippenko}, {Finkelstein}, {Fontana}, {Gardner}, {Garnavich}, {Gawiser}, {Giavalisco}, {Grazian}, {Guo}, {Hathi}, {H{\"a}ussler}, {Hopkins}, {Huang}, {Huang}, {Jha}, {Kartaltepe}, {Kirshner}, {Koo}, {Lai}, {Lee}, {Li}, {Lotz}, {Lucas}, {Madau}, {McCarthy}, {McGrath}, {McIntosh}, {McLure}, {Mobasher}, {Moustakas}, {Mozena}, {Nandra}, {Newman}, {Niemi}, {Noeske}, {Papovich}, {Pentericci}, {Pope}, {Primack}, {Rajan}, {Ravindranath}, {Reddy}, {Renzini}, {Rix}, {Robaina}, {Rodney}, {Rosario}, {Rosati}, {Salimbeni}, {Scarlata}, {Siana}, {Simard}, {Smidt}, {Somerville}, {Spinrad}, {Straughn},
  {Strolger}, {Telford}, {Teplitz}, {Trump}, {van der Wel}, {Villforth}, {Wechsler}, {Weiner}, {Wiklind}, {Wild}, {Wilson}, {Wuyts}, {Yan}, \& {Yun}}]{grogin11}
{Grogin}, N.~A., {Kocevski}, D.~D., {Faber}, S.~M., {et~al.} 2011, \apjs, 197, 35.
\newblock \doarXiv{1105.3753}

\bibitem[{{Guia} \& {Pacucci}(2024)}]{Guia_2024_scatter}
{Guia}, C.~A., \& {Pacucci}, F. 2024, Research Notes of the American Astronomical Society, 8, 153, \dodoi{10.3847/2515-5172/ad530c}

\bibitem[{{Guia} {et~al.}(2024){Guia}, {Pacucci}, \& {Kocevski}}]{Guia24}
{Guia}, C.~A., {Pacucci}, F., \& {Kocevski}, D.~D. 2024, Research Notes of the American Astronomical Society, 8, 207, \dodoi{10.3847/2515-5172/ad7262}

\bibitem[{{Habouzit} {et~al.}(2022){Habouzit}, {Onoue}, {Ba{\~n}ados}, {Neeleman}, {Angl{\'e}s-Alc{\'a}zar}, {Walter}, {Pillepich}, {Dav{\'e}}, {Jahnke}, \& {Dubois}}]{Habouzit22}
{Habouzit}, M., {Onoue}, M., {Ba{\~n}ados}, E., {et~al.} 2022, \mnras, 511, 3751, \dodoi{10.1093/mnras/stac225}

\bibitem[{{Harikane} {et~al.}(2023){Harikane}, {Zhang}, {Nakajima}, {Ouchi}, {Isobe}, {Ono}, {Hatano}, {Xu}, \& {Umeda}}]{Harikane23}
{Harikane}, Y., {Zhang}, Y., {Nakajima}, K., {et~al.} 2023, \apj, 959, 39, \dodoi{10.3847/1538-4357/ad029e}

\bibitem[{{Heckman} {et~al.}(2004){Heckman}, {Kauffmann}, {Brinchmann}, {Charlot}, {Tremonti}, \& {White}}]{Heckman04}
{Heckman}, T.~M., {Kauffmann}, G., {Brinchmann}, J., {et~al.} 2004, \apj, 613, 109, \dodoi{10.1086/422872}

\bibitem[{{Hirschmann}(2011)}]{Hirschmann11}
{Hirschmann}, M.~M. 2011, PhD thesis, Ludwig-Maximilians University of Munich, Germany

\bibitem[{{Hogg} {et~al.}(2010){Hogg}, {Bovy}, \& {Lang}}]{Hogg_2010}
{Hogg}, D.~W., {Bovy}, J., \& {Lang}, D. 2010, arXiv e-prints, arXiv:1008.4686, \dodoi{10.48550/arXiv.1008.4686}

\bibitem[{{Horne}(1986)}]{Horne86}
{Horne}, K. 1986, \pasp, 98, 609, \dodoi{10.1086/131801}

\bibitem[{{Hviding} {et~al.}(2025){Hviding}, {de Graaff}, {Miller}, {Setton}, {Greene}, {Labb{\'e}}, {Brammer}, {Bezanson}, {Boogaard}, {Cleri}, {Leja}, {Maseda}, {McConachie}, {Matthee}, {Naidu}, {Oesch}, {Wang}, {Whitaker}, \& {Williams}}]{Hviding25}
{Hviding}, R.~E., {de Graaff}, A., {Miller}, T.~B., {et~al.} 2025, arXiv e-prints, arXiv:2506.05459, \dodoi{10.48550/arXiv.2506.05459}

\bibitem[{{Inayoshi}(2025)}]{Inayoshi25}
{Inayoshi}, K. 2025, \apjl, 988, L22, \dodoi{10.3847/2041-8213/adea66}

\bibitem[{{Inayoshi} \& {Ichikawa}(2024)}]{Inayoshi24}
{Inayoshi}, K., \& {Ichikawa}, K. 2024, arXiv e-prints, arXiv:2402.14706, \dodoi{10.48550/arXiv.2402.14706}

\bibitem[{{Inayoshi} {et~al.}(2024){Inayoshi}, {Kimura}, \& {Noda}}]{Inayoshi24b}
{Inayoshi}, K., {Kimura}, S.~S., \& {Noda}, H. 2024, arXiv e-prints, arXiv:2412.03653, \dodoi{10.48550/arXiv.2412.03653}

\bibitem[{{Inayoshi} \& {Maiolino}(2025)}]{Inayoshi_Maiolino25}
{Inayoshi}, K., \& {Maiolino}, R. 2025, \apjl, 980, L27, \dodoi{10.3847/2041-8213/adaebd}

\bibitem[{{Inayoshi} {et~al.}(2022){Inayoshi}, {Nakatani}, {Toyouchi}, {Hosokawa}, {Kuiper}, \& {Onoue}}]{Inayoshi_2022}
{Inayoshi}, K., {Nakatani}, R., {Toyouchi}, D., {et~al.} 2022, \apj, 927, 237, \dodoi{10.3847/1538-4357/ac4751}

\bibitem[{{Izumi} {et~al.}(2021){Izumi}, {Matsuoka}, {Fujimoto}, {Onoue}, {Strauss}, {Umehata}, {Imanishi}, {Kohno}, {Kawaguchi}, {Kawamuro}, {Baba}, {Nagao}, {Toba}, {Inayoshi}, {Silverman}, {Inoue}, {Ikarashi}, {Iwasawa}, {Kashikawa}, {Hashimoto}, {Nakanishi}, {Ueda}, {Schramm}, {Lee}, \& {Suh}}]{Izumi_2021}
{Izumi}, T., {Matsuoka}, Y., {Fujimoto}, S., {et~al.} 2021, \apj, 914, 36, \dodoi{10.3847/1538-4357/abf6dc}

\bibitem[{{Jahnke} \& {Macci{\`o}}(2011{\natexlab{a}})}]{Jahnke11}
{Jahnke}, K., \& {Macci{\`o}}, A.~V. 2011{\natexlab{a}}, \apj, 734, 92, \dodoi{10.1088/0004-637X/734/2/92}

\bibitem[{{Jahnke} \& {Macci{\`o}}(2011{\natexlab{b}})}]{Jahnke_2011}
---. 2011{\natexlab{b}}, \apj, 734, 92, \dodoi{10.1088/0004-637X/734/2/92}

\bibitem[{{Jahnke} {et~al.}(2009){Jahnke}, {Bongiorno}, {Brusa}, {Capak}, {Cappelluti}, {Cisternas}, {Civano}, {Colbert}, {Comastri}, {Elvis}, {Hasinger}, {Ilbert}, {Impey}, {Inskip}, {Koekemoer}, {Lilly}, {Maier}, {Merloni}, {Riechers}, {Salvato}, {Schinnerer}, {Scoville}, {Silverman}, {Taniguchi}, {Trump}, \& {Yan}}]{Jahnke09}
{Jahnke}, K., {Bongiorno}, A., {Brusa}, M., {et~al.} 2009, \apjl, 706, L215, \dodoi{10.1088/0004-637X/706/2/L215}

\bibitem[{{Jeon} {et~al.}(2025){Jeon}, {Bromm}, {Liu}, \& {Finkelstein}}]{Jeon_2025}
{Jeon}, J., {Bromm}, V., {Liu}, B., \& {Finkelstein}, S.~L. 2025, \apj, 979, 127, \dodoi{10.3847/1538-4357/ad9f3a}

\bibitem[{{Juod{\v{z}}balis} {et~al.}(2025{\natexlab{a}}){Juod{\v{z}}balis}, {Maiolino}, {Baker}, {Lake}, {Scholtz}, {D'Eugenio}, {Trefoloni}, {Isobe}, {Tacchella}, {Bunker}, {Carniani}, {Charlot}, {Jones}, {Parlanti}, {Perna}, {Rinaldi}, {Robertson}, {{\"U}bler}, {Venturi}, \& {Willott}}]{Juodžbalis_25}
{Juod{\v{z}}balis}, I., {Maiolino}, R., {Baker}, W.~M., {et~al.} 2025{\natexlab{a}}, arXiv e-prints, arXiv:2504.03551, \dodoi{10.48550/arXiv.2504.03551}

\bibitem[{{Juod{\v{z}}balis} {et~al.}(2025{\natexlab{b}}){Juod{\v{z}}balis}, {Marconcini}, {D'Eugenio}, {Maiolino}, {Marconi}, {{\"U}bler}, {Scholtz}, {Ji}, {Arribas}, {Bennett}, {Bromm}, {Bunker}, {Carniani}, {Charlot}, {Cresci}, {Dayal}, {Egami}, {Fabian}, {Inayoshi}, {Isobe}, {Ivey}, {Jones}, {Koudmani}, {Laporte}, {Liu}, {Lyu}, {Mazzolari}, {Monty}, {Parlanti}, {P{\'e}rez-Gonz{\'a}lez}, {Perna}, {Robertson}, {Schneider}, {Sijacki}, {Tacchella}, {Trinca}, {Valiante}, {Volonteri}, {Witstok}, \& {Zhang}}]{Juodzbalis_2025}
{Juod{\v{z}}balis}, I., {Marconcini}, C., {D'Eugenio}, F., {et~al.} 2025{\natexlab{b}}, arXiv e-prints, arXiv:2508.21748, \dodoi{10.48550/arXiv.2508.21748}

\bibitem[{{Killi} {et~al.}(2023){Killi}, {Watson}, {Brammer}, {McPartland}, {Antwi-Danso}, {Newshore}, {Coe}, {Allen}, {Fynbo}, {Gould}, {Heintz}, {Rusakov}, \& {Vejlgaard}}]{Killi23}
{Killi}, M., {Watson}, D., {Brammer}, G., {et~al.} 2023, arXiv e-prints, arXiv:2312.03065, \dodoi{10.48550/arXiv.2312.03065}

\bibitem[{{Kocevski} {et~al.}(2023{\natexlab{a}}){Kocevski}, {Onoue}, {Inayoshi}, {Trump}, {Haro}, {Grazian}, {Dickinson}, {Finkelstein}, {Kartaltepe}, {Hirschmann}, {Aird}, {Holwerda}, {Fujimoto}, {Juneau}, {Amor{\'\i}n}, {Backhaus}, {Bagley}, {Barro}, {Bell}, {Bisigello}, {Calabr{\`o}}, {Cleri}, {Cooper}, {Ding}, {Grogin}, {Ho}, {Hutchison}, {Inoue}, {Jiang}, {Jones}, {Koekemoer}, {Li}, {Li}, {McGrath}, {Molina}, {Papovich}, {P{\'e}rez-Gonz{\'a}lez}, {Pirzkal}, {Wilkins}, {Yang}, \& {Yung}}]{Kocevski23b}
{Kocevski}, D.~D., {Onoue}, M., {Inayoshi}, K., {et~al.} 2023{\natexlab{a}}, \apjl, 954, L4, \dodoi{10.3847/2041-8213/ace5a0}

\bibitem[{{Kocevski} {et~al.}(2023{\natexlab{b}}){Kocevski}, {Barro}, {McGrath}, {Finkelstein}, {Bagley}, {Ferguson}, {Jogee}, {Yang}, {Dickinson}, {Hathi}, {Backhaus}, {Bell}, {Bisigello}, {Buat}, {Burgarella}, {Casey}, {Cleri}, {Cooper}, {Costantin}, {Croton}, {Daddi}, {Fontana}, {Fujimoto}, {Gardner}, {Gawiser}, {Giavalisco}, {Grazian}, {Grogin}, {Guo}, {Arrabal Haro}, {Hirschmann}, {Holwerda}, {Huertas-Company}, {Hutchison}, {Iyer}, {Jones}, {Juneau}, {Kartaltepe}, {Kewley}, {Kirkpatrick}, {Koekemoer}, {Kurczynski}, {Le Bail}, {Long}, {Lotz}, {Lucas}, {Papovich}, {Pentericci}, {P{\'e}rez-Gonz{\'a}lez}, {Pirzkal}, {Rafelski}, {Ravindranath}, {Somerville}, {Straughn}, {Tacchella}, {Trump}, {Wilkins}, {Wuyts}, {Yung}, \& {Zavala}}]{Kocevski23a}
{Kocevski}, D.~D., {Barro}, G., {McGrath}, E.~J., {et~al.} 2023{\natexlab{b}}, \apjl, 946, L14, \dodoi{10.3847/2041-8213/acad00}

\bibitem[{{Kocevski} {et~al.}(2025){Kocevski}, {Finkelstein}, {Barro}, {Taylor}, {Calabr{\`o}}, {Laloux}, {Buchner}, {Trump}, {Leung}, {Yang}, {Dickinson}, {P{\'e}rez-Gonz{\'a}lez}, {Pacucci}, {Inayoshi}, {Somerville}, {McGrath}, {Akins}, {Bagley}, {Bowler}, {Bisigello}, {Carnall}, {Casey}, {Cheng}, {Cleri}, {Costantin}, {Cullen}, {Davis}, {Donnan}, {Dunlop}, {Ellis}, {Ferguson}, {Fujimoto}, {Fontana}, {Giavalisco}, {Grazian}, {Grogin}, {Hathi}, {Hirschmann}, {Huertas-Company}, {Holwerda}, {Illingworth}, {Juneau}, {Kartaltepe}, {Koekemoer}, {Li}, {Lucas}, {Magee}, {Mason}, {McLeod}, {McLure}, {Napolitano}, {Papovich}, {Pirzkal}, {Rodighiero}, {Santini}, {Wilkins}, \& {Yung}}]{Kocevski25}
{Kocevski}, D.~D., {Finkelstein}, S.~L., {Barro}, G., {et~al.} 2025, \apj, 986, 126, \dodoi{10.3847/1538-4357/adbc7d}

\bibitem[{{Koekemoer} {et~al.}(2011){Koekemoer}, {Faber}, {Ferguson}, {Grogin}, {Kocevski}, {Koo}, {Lai}, {Lotz}, {Lucas}, {McGrath}, {Ogaz}, {Rajan}, {Riess}, {Rodney}, {Strolger}, {Casertano}, {Dahlen}, {Dickinson}, {Dolch}, {Fontana}, {Giavalisco}, {Grazian}, {Guo}, {Hathi}, {Huang}, {van der Wel}, {Yan}, {Acquaviva}, {Almaini}, {Ashby}, {Barden}, {Bell}, {Bournaud}, {Brown}, {Caputi}, {Cassata}, {Challis}, {Chary}, {Cheung}, {Cirasuolo}, {Conselice}, {Roshan Cooray}, {Croton}, {Daddi}, {Dav{\'e}}, {de Mello}, {de Ravel}, {Dekel}, {Donley}, {Dunlop}, {Dutton}, {Elbaz}, {Fazio}, {Filippenko}, {Finkelstein}, {Frazer}, {Gardner}, {Garnavich}, {Gawiser}, {Gruetzbauch}, {Hartley}, {H{\"a}ussler}, {Herrington}, {Hopkins}, {Huang}, {Jha}, {Johnson}, {Kartaltepe}, {Khostovan}, {Kirshner}, {Lani}, {Lee}, {Li}, {Madau}, {McCarthy}, {McIntosh}, {McLure}, {McPartland}, {Mobasher}, {Moreira}, {Mortlock}, {Moustakas}, {Mozena}, {Nandra}, {Newman}, {Nielsen}, {Niemi}, {Noeske}, {Papovich}, {Pentericci}, {Pope},
  {Primack}, {Ravindranath}, {Reddy}, {Renzini}, {Rix}, {Robaina}, {Rosario}, {Rosati}, {Salimbeni}, {Scarlata}, {Siana}, {Simard}, {Smidt}, {Snyder}, {Somerville}, {Spinrad}, {Straughn}, {Telford}, {Teplitz}, {Trump}, {Vargas}, {Villforth}, {Wagner}, {Wandro}, {Wechsler}, {Weiner}, {Wiklind}, {Wild}, {Wilson}, {Wuyts}, \& {Yun}}]{koekemoer11}
{Koekemoer}, A.~M., {Faber}, S.~M., {Ferguson}, H.~C., {et~al.} 2011, \apjs, 197, 36.
\newblock \doarXiv{1105.3754}

\bibitem[{{Kokorev} {et~al.}(2023){Kokorev}, {Fujimoto}, {Labbe}, {Greene}, {Bezanson}, {Dayal}, {Nelson}, {Atek}, {Brammer}, {Caputi}, {Chemerynska}, {Cutler}, {Feldmann}, {Fudamoto}, {Furtak}, {Goulding}, {de Graaff}, {Leja}, {Marchesini}, {Miller}, {Nanayakkara}, {Oesch}, {Pan}, {Price}, {Setton}, {Smit}, {Stefanon}, {Wang}, {Weaver}, {Whitaker}, {Williams}, \& {Zitrin}}]{Kokorev23}
{Kokorev}, V., {Fujimoto}, S., {Labbe}, I., {et~al.} 2023, \apjl, 957, L7, \dodoi{10.3847/2041-8213/ad037a}

\bibitem[{{Kokorev} {et~al.}(2024){Kokorev}, {Caputi}, {Greene}, {Dayal}, {Trebitsch}, {Cutler}, {Fujimoto}, {Labb{\'e}}, {Miller}, {Iani}, {Navarro-Carrera}, \& {Rinaldi}}]{Kokorev24}
{Kokorev}, V., {Caputi}, K.~I., {Greene}, J.~E., {et~al.} 2024, arXiv e-prints, arXiv:2401.09981, \dodoi{10.48550/arXiv.2401.09981}

\bibitem[{{Kormendy} \& {Ho}(2013)}]{Kormendy_Ho_2013}
{Kormendy}, J., \& {Ho}, L.~C. 2013, \araa, 51, 511, \dodoi{10.1146/annurev-astro-082708-101811}

\bibitem[{{Kormendy} \& {Richstone}(1995)}]{Kormendy1995}
{Kormendy}, J., \& {Richstone}, D. 1995, \araa, 33, 581, \dodoi{10.1146/annurev.aa.33.090195.003053}

\bibitem[{{Lambrides} {et~al.}(2024){Lambrides}, {Garofali}, {Larson}, {Ptak}, {Chiaberge}, {Long}, {Hutchison}, {Norman}, {McKinney}, {Akins}, {Berg}, {Chisholm}, {Civano}, {Cloonan}, {Endsley}, {Faisst}, {Gilli}, {Gillman}, {Hirschmann}, {Kartaltepe}, {Kocevski}, {Kokorev}, {Pacucci}, {Richardson}, {Stiavelli}, \& {Whalen}}]{Lambrides_2024}
{Lambrides}, E., {Garofali}, K., {Larson}, R., {et~al.} 2024, arXiv e-prints, arXiv:2409.13047, \dodoi{10.48550/arXiv.2409.13047}

\bibitem[{{Leung} {et~al.}(2024){Leung}, {Finkelstein}, {P{\'e}rez-Gonz{\'a}lez}, {Morales}, {Taylor}, {Barro}, {Kocevski}, {Akins}, {Carnall}, {Ch{\'a}vez Ortiz}, {Cleri}, {Cullen}, {Donnan}, {Dunlop}, {Ellis}, {Grogin}, {Hirschmann}, {Koekemoer}, {Kokorev}, {Lucas}, {McLeod}, {Papovich}, \& {Yung}}]{Leung24}
{Leung}, G. C.~K., {Finkelstein}, S.~L., {P{\'e}rez-Gonz{\'a}lez}, P.~G., {et~al.} 2024, arXiv e-prints, arXiv:2411.12005, \dodoi{10.48550/arXiv.2411.12005}

\bibitem[{{Li} {et~al.}(2025{\natexlab{a}}){Li}, {Silverman}, {Shen}, {Volonteri}, {Jahnke}, {Zhuang}, {Scoggins}, {Ding}, {Harikane}, {Onoue}, \& {Tanaka}}]{LiSilverman2025}
{Li}, J., {Silverman}, J.~D., {Shen}, Y., {et~al.} 2025{\natexlab{a}}, \apj, 981, 19, \dodoi{10.3847/1538-4357/ada603}

\bibitem[{{Li} {et~al.}(2025{\natexlab{b}}){Li}, {Silverman}, {Shen}, {Volonteri}, {Jahnke}, {Zhuang}, {Scoggins}, {Ding}, {Harikane}, {Onoue}, \& {Tanaka}}]{Li25}
---. 2025{\natexlab{b}}, \apj, 981, 19, \dodoi{10.3847/1538-4357/ada603}

\bibitem[{{Li} {et~al.}(2023){Li}, {Inayoshi}, {Onoue}, {He}, {Matsuoka}, {Pan}, {Akiyama}, {Izumi}, \& {Nagao}}]{Li_Inayoshi23}
{Li}, W., {Inayoshi}, K., {Onoue}, M., {et~al.} 2023, arXiv e-prints, arXiv:2306.06172, \dodoi{10.48550/arXiv.2306.06172}

\bibitem[{{Lupi} {et~al.}(2024){Lupi}, {Trinca}, {Volonteri}, {Dotti}, \& {Mazzucchelli}}]{Lupi_2024}
{Lupi}, A., {Trinca}, A., {Volonteri}, M., {Dotti}, M., \& {Mazzucchelli}, C. 2024, \aap, 689, A128, \dodoi{10.1051/0004-6361/202451249}

\bibitem[{{Ma} {et~al.}(2025){Ma}, {Greene}, {Setton}, {Goulding}, {Annunziatella}, {Fan}, {Kokorev}, {Labbe}, {Li}, {Lin}, {Marchesini}, {Matthee}, {Robbins}, {Sajina}, {Sawicki}, \& {Telford}}]{Ma25}
{Ma}, Y., {Greene}, J.~E., {Setton}, D.~J., {et~al.} 2025, arXiv e-prints, arXiv:2504.08032, \dodoi{10.48550/arXiv.2504.08032}

\bibitem[{{Madau} \& {Dickinson}(2014)}]{Madau_Dickinson14}
{Madau}, P., \& {Dickinson}, M. 2014, \araa, 52, 415, \dodoi{10.1146/annurev-astro-081811-125615}

\bibitem[{{Madau} \& {Haardt}(2024)}]{Madau24}
{Madau}, P., \& {Haardt}, F. 2024, \apjl, 976, L24, \dodoi{10.3847/2041-8213/ad90e1}

\bibitem[{{Magorrian} {et~al.}(1998){Magorrian}, {Tremaine}, {Richstone}, {Bender}, {Bower}, {Dressler}, {Faber}, {Gebhardt}, {Green}, {Grillmair}, {Kormendy}, \& {Lauer}}]{Magorrian98}
{Magorrian}, J., {Tremaine}, S., {Richstone}, D., {et~al.} 1998, \aj, 115, 2285, \dodoi{10.1086/300353}

\bibitem[{{Maiolino} {et~al.}(2023){Maiolino}, {Scholtz}, {Curtis-Lake}, {Carniani}, {Baker}, {de Graaff}, {Tacchella}, {{\"U}bler}, {D'Eugenio}, {Witstok}, {Curti}, {Arribas}, {Bunker}, {Charlot}, {Chevallard}, {Eisenstein}, {Egami}, {Ji}, {Jones}, {Lyu}, {Rawle}, {Robertson}, {Rujopakarn}, {Perna}, {Sun}, {Venturi}, {Williams}, \& {Willott}}]{Maiolino23}
{Maiolino}, R., {Scholtz}, J., {Curtis-Lake}, E., {et~al.} 2023, arXiv e-prints, arXiv:2308.01230, \dodoi{10.48550/arXiv.2308.01230}

\bibitem[{{Maiolino} {et~al.}(2024){Maiolino}, {Scholtz}, {Witstok}, {Carniani}, {D'Eugenio}, {de Graaff}, {{\"U}bler}, {Tacchella}, {Curtis-Lake}, {Arribas}, {Bunker}, {Charlot}, {Chevallard}, {Curti}, {Looser}, {Maseda}, {Rawle}, {Rodr{\'\i}guez del Pino}, {Willott}, {Egami}, {Eisenstein}, {Hainline}, {Robertson}, {Williams}, {Willmer}, {Baker}, {Boyett}, {DeCoursey}, {Fabian}, {Helton}, {Ji}, {Jones}, {Kumari}, {Laporte}, {Nelson}, {Perna}, {Sandles}, {Shivaei}, \& {Sun}}]{Maiolino24_gnz11}
{Maiolino}, R., {Scholtz}, J., {Witstok}, J., {et~al.} 2024, \nat, 627, 59, \dodoi{10.1038/s41586-024-07052-5}

\bibitem[{{Matthee} {et~al.}(2023){Matthee}, {Naidu}, {Brammer}, {Chisholm}, {Eilers}, {Goulding}, {Greene}, {Kashino}, {Labbe}, {Lilly}, {Mackenzie}, {Oesch}, {Weibel}, {Wuyts}, {Xiao}, {Bordoloi}, {Bouwens}, {van Dokkum}, {Illingworth}, {Kramarenko}, {Maseda}, {Mason}, {Meyer}, {Nelson}, {Reddy}, {Shivaei}, {Simcoe}, \& {Yue}}]{Matthee_2023}
{Matthee}, J., {Naidu}, R.~P., {Brammer}, G., {et~al.} 2023, arXiv e-prints, arXiv:2306.05448, \dodoi{10.48550/arXiv.2306.05448}

\bibitem[{{McConnell} \& {Ma}(2013)}]{McConnell13}
{McConnell}, N.~J., \& {Ma}, C.-P. 2013, \apj, 764, 184, \dodoi{10.1088/0004-637X/764/2/184}

\bibitem[{{Momcheva} {et~al.}(2016){Momcheva}, {Brammer}, {van Dokkum}, {Skelton}, {Whitaker}, {Nelson}, {Fumagalli}, {Maseda}, {Leja}, {Franx}, {Rix}, {Bezanson}, {Da Cunha}, {Dickey}, {F{\"o}rster Schreiber}, {Illingworth}, {Kriek}, {Labb{\'e}}, {Ulf Lange}, {Lundgren}, {Magee}, {Marchesini}, {Oesch}, {Pacifici}, {Patel}, {Price}, {Tal}, {Wake}, {van der Wel}, \& {Wuyts}}]{Momcheva16}
{Momcheva}, I.~G., {Brammer}, G.~B., {van Dokkum}, P.~G., {et~al.} 2016, \apjs, 225, 27, \dodoi{10.3847/0067-0049/225/2/27}

\bibitem[{{Naidu} {et~al.}(2025){Naidu}, {Matthee}, {Katz}, {de Graaff}, {Oesch}, {Smith}, {Greene}, {Brammer}, {Weibel}, {Hviding}, {Chisholm}, {Labb\textbackslash'e}, {Simcoe}, {Witten}, {Atek}, {Baggen}, {Belli}, {Bezanson}, {Boogaard}, {Bose}, {Covelo-Paz}, {Dayal}, {Fudamoto}, {Furtak}, {Giovinazzo}, {Goulding}, {Gronke}, {Heintz}, {Hirschmann}, {Illingworth}, {Inoue}, {Johnson}, {Leja}, {Leonova}, {McConachie}, {Maseda}, {Natarajan}, {Nelson}, {Setton}, {Shivaei}, {Sobral}, {Stefanon}, {Tacchella}, {Toft}, {Torralba}, {van Dokkum}, {van der Wel}, {Volonteri}, {Walter}, {Wang}, \& {Watson}}]{Naidu25}
{Naidu}, R.~P., {Matthee}, J., {Katz}, H., {et~al.} 2025, arXiv e-prints, arXiv:2503.16596, \dodoi{10.48550/arXiv.2503.16596}

\bibitem[{{Napolitano} {et~al.}(2025){Napolitano}, {Castellano}, {Pentericci}, {Vignali}, {Gilli}, {Fontana}, {Santini}, {Treu}, {Calabr{\`o}}, {Llerena}, {Piconcelli}, {Zappacosta}, {Mascia}, {Tripodi}, {Arrabal Haro}, {Bergamini}, {Bakx}, {Dickinson}, {Glazebrook}, {Henry}, {Leethochawalit}, {Mazzolari}, {Merlin}, {Morishita}, {Nanayakkara}, {Paris}, {Puccetti}, {Roberts-Borsani}, {Rojas Ruiz}, {Rosati}, {Vanzella}, {Vito}, {Vulcani}, {Wang}, {Yoon}, \& {Zavala}}]{Napolitano25}
{Napolitano}, L., {Castellano}, M., {Pentericci}, L., {et~al.} 2025, \apj, 989, 75, \dodoi{10.3847/1538-4357/ade706}

\bibitem[{{Ni} {et~al.}(2024){Ni}, {Chen}, {Zhou}, {Park}, {Yang}, {DiMatteo}, {Bird}, \& {Croft}}]{Ni_2024_astrid}
{Ni}, Y., {Chen}, N., {Zhou}, Y., {et~al.} 2024, arXiv e-prints, arXiv:2409.10666, \dodoi{10.48550/arXiv.2409.10666}

\bibitem[{{Onoue} {et~al.}(2023){Onoue}, {Inayoshi}, {Ding}, {Li}, {Li}, {Molina}, {Inoue}, {Jiang}, \& {Ho}}]{onoue23}
{Onoue}, M., {Inayoshi}, K., {Ding}, X., {et~al.} 2023, \apjl, 942, L17, \dodoi{10.3847/2041-8213/aca9d3}

\bibitem[{{Pacucci} {et~al.}(2025){Pacucci}, {Hernquist}, \& {Fujii}}]{Pacucci_Hernquist_2025}
{Pacucci}, F., {Hernquist}, L., \& {Fujii}, M. 2025, arXiv e-prints, arXiv:2509.02664, \dodoi{10.48550/arXiv.2509.02664}

\bibitem[{{Pacucci} \& {Loeb}(2019)}]{Pacucci_2019}
{Pacucci}, F., \& {Loeb}, A. 2019, \apjl, 870, L12, \dodoi{10.3847/2041-8213/aaf86a}

\bibitem[{Pacucci \& Loeb(2024)}]{Pacucci2024}
Pacucci, F., \& Loeb, A. 2024, The Astrophysical Journal, 964, 154, \dodoi{10.3847/1538-4357/ad3044}

\bibitem[{{Pacucci} \& {Loeb}(2025)}]{Pacucci_Loeb_2025}
{Pacucci}, F., \& {Loeb}, A. 2025, \apjl, 989, L19, \dodoi{10.3847/2041-8213/ade871}

\bibitem[{{Pacucci} \& {Narayan}(2024)}]{Pacucci24b}
{Pacucci}, F., \& {Narayan}, R. 2024, \apj, 976, 96, \dodoi{10.3847/1538-4357/ad84f7}

\bibitem[{{Pacucci} {et~al.}(2023){Pacucci}, {Nguyen}, {Carniani}, {Maiolino}, \& {Fan}}]{Pacucci_2023_overmassive}
{Pacucci}, F., {Nguyen}, B., {Carniani}, S., {Maiolino}, R., \& {Fan}, X. 2023, \apjl, 957, L3, \dodoi{10.3847/2041-8213/ad0158}

\bibitem[{{Pacucci} {et~al.}(2017){Pacucci}, {Pallottini}, {Ferrara}, \& {Gallerani}}]{Pacucci_2017}
{Pacucci}, F., {Pallottini}, A., {Ferrara}, A., \& {Gallerani}, S. 2017, \mnras, 468, L77, \dodoi{10.1093/mnrasl/slx029}

\bibitem[{{Peng}(2007)}]{Peng07}
{Peng}, C.~Y. 2007, \apj, 671, 1098, \dodoi{10.1086/522774}

\bibitem[{{Peng} {et~al.}(2002){Peng}, {Ho}, {Impey}, \& {Rix}}]{peng02}
{Peng}, C.~Y., {Ho}, L.~C., {Impey}, C.~D., \& {Rix}, H.-W. 2002, \aj, 124, 266, \dodoi{10.1086/340952}

\bibitem[{{Peng} {et~al.}(2006){Peng}, {Impey}, {Rix}, {Kochanek}, {Keeton}, {Falco}, {Leh{\'a}r}, \& {McLeod}}]{Peng06}
{Peng}, C.~Y., {Impey}, C.~D., {Rix}, H.-W., {et~al.} 2006, \apj, 649, 616, \dodoi{10.1086/506266}

\bibitem[{{P{\'e}rez-Gonz{\'a}lez} {et~al.}(2024){P{\'e}rez-Gonz{\'a}lez}, {Barro}, {Rieke}, {Lyu}, {Rieke}, {Alberts}, {Williams}, {Hainline}, {Sun}, {Puskas}, {Annunziatella}, {Baker}, {Bunker}, {Egami}, {Ji}, {Johnson}, {Robertson}, {Rodriguez Del Pino}, {Rujopakarn}, {Shivaei}, {Tacchella}, {Willmer}, \& {Willott}}]{Pablo24}
{P{\'e}rez-Gonz{\'a}lez}, P.~G., {Barro}, G., {Rieke}, G.~H., {et~al.} 2024, arXiv e-prints, arXiv:2401.08782, \dodoi{10.48550/arXiv.2401.08782}

\bibitem[{{Reines} {et~al.}(2013){Reines}, {Greene}, \& {Geha}}]{reines13}
{Reines}, A.~E., {Greene}, J.~E., \& {Geha}, M. 2013, \apj, 775, 116, \dodoi{10.1088/0004-637X/775/2/116}

\bibitem[{{Reines} \& {Volonteri}(2015)}]{Reines_Volonteri_2015}
{Reines}, A.~E., \& {Volonteri}, M. 2015, \apj, 813, 82, \dodoi{10.1088/0004-637X/813/2/82}

\bibitem[{{Rinaldi} {et~al.}(2025){Rinaldi}, {Rieke}, {Wu}, {Gilbert}, {Pacucci}, {Barchiesi}, {Alberts}, {Carniani}, {Bunker}, {Bhatawdekar}, {D'Eugenio}, {Ji}, {Johnson}, {Hainline}, {Kokorev}, {Kumari}, {Iani}, {Lyu}, {Maiolino}, {Parlanti}, {Robertson}, {Sun}, {Vignali}, {Williams}, {Willmer}, \& {Zhu}}]{Rinaldi_2025}
{Rinaldi}, P., {Rieke}, G.~H., {Wu}, Z., {et~al.} 2025, arXiv e-prints, arXiv:2507.17738, \dodoi{10.48550/arXiv.2507.17738}

\bibitem[{{Rusakov} {et~al.}(2025){Rusakov}, {Watson}, {Nikopoulos}, {Brammer}, {Gottumukkala}, {Harvey}, {Heintz}, {Nielsen}, {Sim}, {Sneppen}, {Vijayan}, {Adams}, {Austin}, {Conselice}, {Goolsby}, {Toft}, \& {Witstok}}]{Rusakov25}
{Rusakov}, V., {Watson}, D., {Nikopoulos}, G.~P., {et~al.} 2025, arXiv e-prints, arXiv:2503.16595, \dodoi{10.48550/arXiv.2503.16595}

\bibitem[{{Scoggins} {et~al.}(2023){Scoggins}, {Haiman}, \& {Wise}}]{Scoggins_2023}
{Scoggins}, M.~T., {Haiman}, Z., \& {Wise}, J.~H. 2023, \mnras, 519, 2155, \dodoi{10.1093/mnras/stac3715}

\bibitem[{{Silverman} {et~al.}(2025){Silverman}, {Li}, {Ding}, {Onoue}, {Strauss}, {Matsuoka}, {Izumi}, {Jahnke}, {Treu}, {Volonteri}, {Phillips}, {Andika}, {Aoki}, {Arita}, {Baba}, {Bosman}, {Eilers}, {Fan}, {Fujimoto}, {Habouzit}, {Haiman}, {Imanishi}, {Inayoshi}, {Iwasawa}, {Kashikawa}, {Kawaguchi}, {Lee}, {Lupi}, {Nagao}, {Schindler}, {Schramm}, {Shimasaku}, {Toba}, {Trakhtenbrot}, {Umehata}, {Vestergaard}, {Walter}, {Wang}, \& {Yang}}]{Silverman25}
{Silverman}, J., {Li}, J., {Ding}, X., {et~al.} 2025, arXiv e-prints, arXiv:2507.23066, \dodoi{10.48550/arXiv.2507.23066}

\bibitem[{{Silverman} {et~al.}(2009){Silverman}, {Lamareille}, {Maier}, {Lilly}, {Mainieri}, {Brusa}, {Cappelluti}, {Hasinger}, {Zamorani}, {Scodeggio}, {Bolzonella}, {Contini}, {Carollo}, {Jahnke}, {Kneib}, {Le F{\`e}vre}, {Merloni}, {Bardelli}, {Bongiorno}, {Brunner}, {Caputi}, {Civano}, {Comastri}, {Coppa}, {Cucciati}, {de la Torre}, {de Ravel}, {Elvis}, {Finoguenov}, {Fiore}, {Franzetti}, {Garilli}, {Gilli}, {Iovino}, {Kampczyk}, {Knobel}, {Kova{\v{c}}}, {Le Borgne}, {Le Brun}, {Mignoli}, {Pello}, {Peng}, {Perez Montero}, {Ricciardelli}, {Tanaka}, {Tasca}, {Tresse}, {Vergani}, {Vignali}, {Zucca}, {Bottini}, {Cappi}, {Cassata}, {Fumana}, {Griffiths}, {Kartaltepe}, {Koekemoer}, {Marinoni}, {McCracken}, {Memeo}, {Meneux}, {Oesch}, {Porciani}, \& {Salvato}}]{Silverman09}
{Silverman}, J.~D., {Lamareille}, F., {Maier}, C., {et~al.} 2009, \apj, 696, 396, \dodoi{10.1088/0004-637X/696/1/396}

\bibitem[{{Song} {et~al.}(2016){Song}, {Finkelstein}, {Ashby}, {Grazian}, {Lu}, {Papovich}, {Salmon}, {Somerville}, {Dickinson}, {Duncan}, {Faber}, {Fazio}, {Ferguson}, {Fontana}, {Guo}, {Hathi}, {Lee}, {Merlin}, \& {Willner}}]{Song_2016}
{Song}, M., {Finkelstein}, S.~L., {Ashby}, M. L.~N., {et~al.} 2016, \apj, 825, 5, \dodoi{10.3847/0004-637X/825/1/5}

\bibitem[{{Stern} \& {Laor}(2012)}]{Stern_Laor12}
{Stern}, J., \& {Laor}, A. 2012, \mnras, 423, 600, \dodoi{10.1111/j.1365-2966.2012.20901.x}

\bibitem[{{Suh} {et~al.}(2020){Suh}, {Civano}, {Trakhtenbrot}, {Shankar}, {Hasinger}, {Sanders}, \& {Allevato}}]{Suh20}
{Suh}, H., {Civano}, F., {Trakhtenbrot}, B., {et~al.} 2020, \apj, 889, 32, \dodoi{10.3847/1538-4357/ab5f5f}

\bibitem[{{Sun} {et~al.}(2015){Sun}, {Trump}, {Brandt}, {Luo}, {Alexander}, {Jahnke}, {Rosario}, {Wang}, \& {Xue}}]{Sun15}
{Sun}, M., {Trump}, J.~R., {Brandt}, W.~N., {et~al.} 2015, \apj, 802, 14, \dodoi{10.1088/0004-637X/802/1/14}

\bibitem[{Sun {et~al.}(2024)Sun, Lyu, Rieke, Ji, Sun, Zhu, Bunker, Cargile, Circosta, D’Eugenio, Egami, Hainline, Helton, Rinaldi, Robertson, Scholtz, Shivaei, Stone, Tacchella, Williams, Willmer, \& Willott}]{Sun2024}
Sun, Y., Lyu, J., Rieke, G.~H., {et~al.} 2024, The Astrophysical Journal, 978, 98, \dodoi{10.3847/1538-4357/ad973b}

\bibitem[{{Tanaka} {et~al.}(2025){Tanaka}, {Silverman}, {Ding}, {Jahnke}, {Trakhtenbrot}, {Lambrides}, {Onoue}, {Andika}, {Bongiorno}, {Faisst}, {Gillman}, {Hayward}, {Hirschmann}, {Koekemoer}, {Kokorev}, {Liu}, {Magdis}, {Renzini}, {Casey}, {Drakos}, {Franco}, {Gozaliasl}, {Kartaltepe}, {Liu}, {McCracken}, {Rhodes}, {Robertson}, \& {Toft}}]{Tanaka25}
{Tanaka}, T.~S., {Silverman}, J.~D., {Ding}, X., {et~al.} 2025, \apj, 979, 215, \dodoi{10.3847/1538-4357/ad9d0a}

\bibitem[{{Taylor} {et~al.}(2025{\natexlab{a}}){Taylor}, {Kokorev}, {Kocevski}, {Akins}, {Cullen}, {Dickinson}, {Finkelstein}, {Arrabal Haro}, {Bromm}, {Giavalisco}, {Inayoshi}, {Juneau}, {Leung}, {P{\'e}rez-Gonz{\'a}lez}, {Somerville}, {Trump}, {Amor{\'\i}n}, {Barro}, {Burgarella}, {Brooks}, {Carnall}, {Casey}, {Cheng}, {Chisholm}, {Chworowsky}, {Davis}, {Donnan}, {Dunlop}, {Ellis}, {Fern{\'a}ndez}, {Fujimoto}, {Grogin}, {Gupta}, {Hathi}, {Jung}, {Hirschmann}, {Kartaltepe}, {Koekemoer}, {Larson}, {Leung}, {Llerena}, {Lucas}, {McLeod}, {McLure}, {Napolitano}, {Papovich}, {Stanton}, {Tripodi}, {Wang}, {Wilkins}, {Yung}, \& {Zavala}}]{Taylor25b}
{Taylor}, A.~J., {Kokorev}, V., {Kocevski}, D.~D., {et~al.} 2025{\natexlab{a}}, \apjl, 989, L7, \dodoi{10.3847/2041-8213/ade789}

\bibitem[{{Taylor} {et~al.}(2025{\natexlab{b}}){Taylor}, {Finkelstein}, {Kocevski}, {Jeon}, {Bromm}, {Amor{\'\i}n}, {Arrabal Haro}, {Backhaus}, {Bagley}, {Banados}, {Bhatawdekar}, {Brooks}, {Calabr{\`o}}, {Ch{\'a}vez Ortiz}, {Cheng}, {Cleri}, {Cole}, {Davis}, {Dickinson}, {Donnan}, {Dunlop}, {Ellis}, {Fern{\'a}ndez}, {Fontana}, {Fujimoto}, {Giavalisco}, {Grazian}, {Guo}, {Hathi}, {Holwerda}, {Hirschmann}, {Inayoshi}, {Kartaltepe}, {Khusanova}, {Koekemoer}, {Kokorev}, {Larson}, {Leung}, {Lucas}, {McLeod}, {Napolitano}, {Onoue}, {Pacucci}, {Papovich}, {P{\'e}rez-Gonz{\'a}lez}, {Pirzkal}, {Somerville}, {Trump}, {Wilkins}, {Yung}, \& {Zhang}}]{Taylor25}
{Taylor}, A.~J., {Finkelstein}, S.~L., {Kocevski}, D.~D., {et~al.} 2025{\natexlab{b}}, \apj, 986, 165, \dodoi{10.3847/1538-4357/add15b}

\bibitem[{{{\"U}bler} {et~al.}(2023{\natexlab{a}}){{\"U}bler}, {Maiolino}, {Curtis-Lake}, {P{\'e}rez-Gonz{\'a}lez}, {Curti}, {Perna}, {Arribas}, {Charlot}, {Marshall}, {D'Eugenio}, {Scholtz}, {Bunker}, {Carniani}, {Ferruit}, {Jakobsen}, {Rix}, {Rodr{\'\i}guez Del Pino}, {Willott}, {Boeker}, {Cresci}, {Jones}, {Kumari}, \& {Rawle}}]{Ubler23}
{{\"U}bler}, H., {Maiolino}, R., {Curtis-Lake}, E., {et~al.} 2023{\natexlab{a}}, \aap, 677, A145, \dodoi{10.1051/0004-6361/202346137}

\bibitem[{{{\"U}bler} {et~al.}(2023{\natexlab{b}}){{\"U}bler}, {Maiolino}, {Curtis-Lake}, {P{\'e}rez-Gonz{\'a}lez}, {Curti}, {Arribas}, {Charlot}, {Perna}, {Marshall}, {D'Eugenio}, {Scholtz}, {Bunker}, {Carniani}, {Ferruit}, {Jakobsen}, {Rix}, {Rodr{\'\i}guez Del Pino}, {Willott}, {B{\"o}ker}, {Cresci}, {Jones}, {Kumari}, \& {Rawle}}]{Uebler23}
---. 2023{\natexlab{b}}, arXiv e-prints, arXiv:2302.06647, \dodoi{10.48550/arXiv.2302.06647}

\bibitem[{{Villa-V{\'e}lez} {et~al.}(2021){Villa-V{\'e}lez}, {Buat}, {Theul{\'e}}, {Boquien}, \& {Burgarella}}]{VillaV21}
{Villa-V{\'e}lez}, J.~A., {Buat}, V., {Theul{\'e}}, P., {Boquien}, M., \& {Burgarella}, D. 2021, \aap, 654, A153, \dodoi{10.1051/0004-6361/202140890}

\bibitem[{{Weller} {et~al.}(2023){Weller}, {Pacucci}, {Natarajan}, \& {Di Matteo}}]{Weller_2023_overmassive}
{Weller}, E.~J., {Pacucci}, F., {Natarajan}, P., \& {Di Matteo}, T. 2023, \mnras, 522, 4963, \dodoi{10.1093/mnras/stad1362}

\bibitem[{{Williams} {et~al.}(2023){Williams}, {Alberts}, {Ji}, {Hainline}, {Lyu}, {Rieke}, {Endsley}, {Suess}, {Johnson}, {Florian}, {Shivaei}, {Rujopakarn}, {Baker}, {Bhatawdekar}, {Boyett}, {Bunker}, {Carniani}, {Charlot}, {Curtis-Lake}, {DeCoursey}, {de Graaff}, {Egami}, {Eisenstein}, {Gibson}, {Hausen}, {Helton}, {Maiolino}, {Maseda}, {Nelson}, {Perez-Gonzalez}, {Rieke}, {Robertson}, {Sun}, {Tacchella}, {Willmer}, \& {Willott}}]{Williams23}
{Williams}, C.~C., {Alberts}, S., {Ji}, Z., {et~al.} 2023, arXiv e-prints, arXiv:2311.07483, \dodoi{10.48550/arXiv.2311.07483}

\bibitem[{{Wyithe} \& {Loeb}(2003)}]{Wyithe_Loeb03}
{Wyithe}, J. S.~B., \& {Loeb}, A. 2003, \apj, 595, 614, \dodoi{10.1086/377475}

\bibitem[{Yang {et~al.}(2017)Yang, Brandt, Vito, Chen, Trump, Luo, Sun, Xue, Koekemoer, Schneider, Vignali, \& Wang}]{Yang18}
Yang, G., Brandt, W.~N., Vito, F., {et~al.} 2017, Monthly Notices of the Royal Astronomical Society, 475, 1887, \dodoi{10.1093/mnras/stx2805}

\bibitem[{{Yang} {et~al.}(2022){Yang}, {Boquien}, {Brandt}, {Buat}, {Burgarella}, {Ciesla}, {Lehmer}, {Ma{\l}ek}, {Mountrichas}, {Papovich}, {Pons}, {Stalevski}, {Theul{\'e}}, \& {Zhu}}]{Yang22}
{Yang}, G., {Boquien}, M., {Brandt}, W.~N., {et~al.} 2022, \apj, 927, 192, \dodoi{10.3847/1538-4357/ac4971}

\bibitem[{{Yang} {et~al.}(2020){Yang}, {Wang}, {Fan}, {Hennawi}, {Davies}, {Yue}, {Eilers}, {Farina}, {Wu}, {Bian}, {Pacucci}, \& {Lee}}]{Yang_2020}
{Yang}, J., {Wang}, F., {Fan}, X., {et~al.} 2020, \apj, 904, 26, \dodoi{10.3847/1538-4357/abbc1b}

\bibitem[{{Yue} {et~al.}(2024){Yue}, {Eilers}, {Simcoe}, {Mackenzie}, {Matthee}, {Kashino}, {Bordoloi}, {Lilly}, \& {Naidu}}]{Yue24}
{Yue}, M., {Eilers}, A.-C., {Simcoe}, R.~A., {et~al.} 2024, \apj, 966, 176, \dodoi{10.3847/1538-4357/ad3914}

\bibitem[{{Yung} {et~al.}(2021){Yung}, {Somerville}, {Finkelstein}, {Hirschmann}, {Dav{\'e}}, {Popping}, {Gardner}, \& {Venkatesan}}]{Yung2021}
{Yung}, L.~Y.~A., {Somerville}, R.~S., {Finkelstein}, S.~L., {et~al.} 2021, \mnras, 508, 2706, \dodoi{10.1093/mnras/stab2761}

\bibitem[{{Zhuang} {et~al.}(2025){Zhuang}, {Li}, {Shen}, {Lin}, {Shapley}, {Wang}, {Wu}, \& {Yang}}]{Zhuang_2025}
{Zhuang}, M.-Y., {Li}, J., {Shen}, Y., {et~al.} 2025, arXiv e-prints, arXiv:2505.20393, \dodoi{10.48550/arXiv.2505.20393}

\end{thebibliography}


\end{document}